\documentclass[acmsmall,10pt]{acmart}
\AtBeginDocument{%
  \providecommand\BibTeX{{%
    \normalfont B\kern-0.5em{\scshape i\kern-0.25em b}\kern-0.8em\TeX}}}

\usepackage{amsmath}
\usepackage{amsthm}
\usepackage{bbm}
\usepackage{stmaryrd}
\usepackage{algorithm,algpseudocode}
\usepackage{xspace}
\usepackage{multirow}

\newtheorem{theorem}{Theorem}[section]
\newtheorem{definition}[theorem]{Definition}

\definecolor{rednew}{rgb}{1,0.5,0}
\definecolor{bluenew}{rgb}{0,0.25,0.75}
\definecolor{greennew}{rgb}{0,0.75,0.25}
\newcommand{\toolname}{\textsc{StatCoder}\xspace}
\newcommand{\approxin}{\;\tilde{\in}\;}

\newcommand{\blue}[1]{{\color{bluenew}#1}}
\newcommand{\green}[1]{{\color{greennew}#1}}

\begin{document}

\title[Synthesizing Machine Learning Programs]{Synthesizing Machine Learning Programs with PAC Guarantees via Statistical Sketching}

\author{Osbert Bastani}
\email{obastani@seas.upenn.edu}
\affiliation{%
  \institution{University of Pennsylvania}
  \country{USA}}

\renewcommand{\shortauthors}{Bastani}

\begin{abstract}
We study the problem of synthesizing programs that include machine learning components such as deep neural networks (DNNs). We focus on statistical properties, which are properties expected to hold with high probability---e.g., that an image classification model correctly identifies people in images with high probability. We propose novel algorithms for sketching and synthesizing such programs by leveraging ideas from statistical learning theory to provide statistical soundness guarantees. We evaluate our approach on synthesizing list processing programs that include DNN components used to process image inputs, as well as case studies on image classification and on precision medicine. Our results demonstrate that our approach can be used to synthesize programs with probabilistic guarantees.
\end{abstract}




\maketitle

\section{Introduction}

Machine learning has recently become a powerful tool for solving challenging problems in artificial intelligence. As a consequence, there has been a great deal of interest in incorporating \emph{machine learning components} such as deep neural networks (DNNs) into real-world systems, ranging from healthcare decision-making~\cite{gulshan2016development,esteva2017dermatologist,komorowski2018artificial}, to robotics perception and control~\cite{ren2016faster,levine2016end}, to improving performance of software systems~\cite{kraska2018case,lee2018accelerating,kraska2019sagedb,chen2019relational,chen2020program}.

In these domains, there is often a need to ensure correctness properties of the overall system. To reason about such properties, we need to reason about properties of the incorporated machine learning components. However, it is in general impossible to absolutely guarantee correctness of a machine learning component---e.g., we can never guarantee that a DNN correctly detects every single image containing a pedestrian. Instead, we consider \emph{statistical properties}, which are properties that hold with high probability with respect to the distribution of inputs---e.g., we may want to ensure that the DNN detects 95\% of pedestrians encountered by an autonomous car.

We propose a framework for synthesizing programs that incorporate machine learning components while satisfying  statistical correctness properties. Our framework consists of two components.\footnote{For completeness, our framework also includes a third component for statistical verification of machine learning programs~\cite{younes2002probabilistic,sen2004statistical}, which is described in Appendix~\ref{sec:verification}.}
First, it includes a novel \emph{statistical sketching} algorithm, which builds on the concept of sketching~\cite{solar2008program} to provide statistical guarantees. At a high level, it takes as input a \emph{sketch} annotated with specifications encoding statistical properties that are expected to hold, as well as holes corresponding to real-valued thresholds for making decisions (e.g., the confidence level at which to label an image as containing a pedestrian or to diagnose a patient with a disease). Since statistical properties depend on the data distribution, it additionally takes as input a labeled dataset of training examples (separate from those used to train the DNNs). Then, our algorithm selects values to fill the holes in the sketch so all the given specifications are satisfied.

Second, our framework uses this sketching algorithm in conjunction with a syntax-guided synthesizer~\cite{alur2013syntax} to synthesize programs in a specific domain that provably satisfy statistical guarantees. Our strategy is to first synthesize a sketch whose specifications encode the overall statistical correctness property, and then apply our sketching algorithm to fill the holes in the sketch so these specifications are satisfied with high probability.

The key challenge is providing statistical guarantees for programs using DNNs. To do so, we leverage generalization bounds from statistical learning theory~\cite{valiant1984theory,kearns1994introduction,vapnik2013nature}. These bounds can be thought of as a variant of concentration inequalities to apply to parameters that are estimated based on a dataset. Traditionally, it is hard to apply generalization bounds to obtain useful guarantees on the performance of machine learning components. One reason is that modern machine learning models such as DNNs do not satisfy assumptions in learning theory. However, a deeper issue is that learning theory can only prove bounds with respect to the best model in a given family, not the ``true'' model. More precisely, given a model family $\mathcal{F}$ and a training dataset $\vec{z}$, learning theory provides bounds on the performance of the model $\hat{f}$ learned using $\vec{z}$ in the following form:
\begin{align*}
L(\hat{f})\le L(f^*)+\mathcal{G}(\mathcal{F},n),
\end{align*}
where $L$ a loss function (e.g., the accuracy of a model $f$) and $\mathcal{G}$ is a measure of the complexity of $\mathcal{F}$ in the context of the amount of available training data $n=|\vec{z}|$ (e.g., VC dimension~\cite{kearns1994introduction} or Rademacher complexity~\cite{bartlett2002rademacher}). In particular, learning theory provides no tools for bounding the error $L(f^*)$ of the optimal model given infinite training examples. In other words, learning theory cannot guarantee that $\hat{f}$ is good at detecting pedestrians; at best, that given enough data, it is as good at detecting pedestrians as the best possible DNN $f^*\in\mathcal{F}$.

However, we can provide guarantees for loss functions $L$ where we know that there exists \emph{some} solution $f^*$ with zero loss $L(f^*)=0$. In an analogy with program verification, we cannot in general devise verification algorithms that are both sound and complete. Instead, the goal is to devise algorithms that are as precise as possible subject to a soundness constraint. Similarly, our goal is to learn models that perform as well as possible while satisfying a statistical property---i.e., we want a model that empirically minimizes the number of false alarms while still satisfying the correctness guarantee. For instance, this approach satisfies the above condition since $L(f^*)=0$ if $f^*$ predicts there is a pedestrian in every image. Thus, we might learn a model that is guaranteed to detect 95\% of pedestrians but reports many false alarms (but in practice, we often achieve good performance).

In this context, we show how to use learning theory to sketch programs with statistical guarantees. The machine learning components (e.g., DNNs) in the given sketch have already been trained before our sketching algorithm is applied. In particular, the only task that must be performed by our sketching algorithm is to choose threshold values to fill the holes in the sketch in a way that satisfies the given specifications while maximizing performance---e.g., choose the confidence level of the DNN above which an image contains a pedestrian so we detect 95\% of pedestrians. Then, generalization bounds can give us formal guarantees because (i) we are only synthesize a handful of parameters, so the generalization error $\mathcal{G}(\mathcal{F},n)$ is small, and (ii) we can always choose the thresholds to make conservative decisions, so the error $L(f^*)$ of the best possible model is small (we choose the loss $L$ to measure whether the given specifications are satisfied, not the performance).

Next, we propose an algorithm for synthesizing machine learning programs that leverages our statistical sketching algorithm as a subroutine. We consider a specification saying that with high probability, the synthesized program should either return the correct answer or return ``unknown''. This specification is consistent with the above discussion since we can na\"{i}vely ensure correctness by always returning ``unknown''. Then, our goal is to synthesize a program that satisfies the desired statistical specification but returns ``unknown'' as rarely as possible. To achieve this goal, our synthesis algorithm first uses a standard enumerative synthesis algorithm to identify a program that is correct when given access to ground truth labels. When ground truth labels are unavailable, this program must instead use labels predicted by machine learning components; to satisfy the specification, these components include holes corresponding to predicted confidences below which the component returns ``unknown''. Then, our algorithm uses statistical sketching to fill these holds with thresholds in a way that satisfies the statistical property encoded by the given specification.

Our sketching algorithm requires the holes in the sketch to be annotated with local correctness properties; then, it fill sthe holes in a way that satisfies these annotations. Thus, the main challenge for our synthesizer is how to label holes in the sketch with these annotations so that if the annotations hold true, the given specification is satisfied. We instantiate such a strategy in the context of list processing programs where the components can include learned DNNs such as object classifiers or detectors. In particular, our algorithm analyzes the sketch to allocate allowable errors to each hole in a way that the overall error of the program is bounded by the desired amount.

We have implemented our approach in a tool called \toolname, and evaluate it in two ways. First, we evaluate its ability to synthesize list processing programs satisfying statistical properties where the program inputs are images, and DNN components are used to classify or detect objects in these images. Our results show that our algorithm for error allocation outperforms a na\"{i}ve baseline, and that our novel statistical learning theory bound outperforms using a more traditional bound. Second, we perform two case studies of our sketching algorithm: one on ImageNet classification and another on a medical prediction task, which demonstrate additional interesting applications of our sketching algorithm. In summary, our contributions are:
\begin{itemize}
\item A sketching language for writing programs that incorporate machine learning components in a way that ensures correctness guarantees (Section~\ref{sec:language}).
\item Algorithms for sketching (Section~\ref{sec:sketching}) and synthesizing (Section~\ref{sec:synthesis}) such programs.
\item An empirical evaluation (Section~\ref{sec:evaluation}) validating our approach in the context of our list processing domain (for synthesis) as well as two case studies (for sketching).
\end{itemize}

\section{Overview}
\label{sec:motivating}

We describe how our statistical sketching algorithm can construct a subroutine for detecting whether an image contains a person, guaranteeing that if the image contains a person, then it returns ``true'' with high probability. Then, we describe how our synthesizer uses the sketching algorithm to synthesize a program that counts the number of people in a sequence of images.

\paragraph{Statistical sketching.}

We assume given a DNN component $f:\mathcal{X}\to[0,1]$ that, given an image $x\in\mathcal{X}$, predicts whether $x$ contains a person. In particular, $f(x)$ is a score indicating its confidence that $x$ contains a person; higher score means more likely to contain a person. We do not assume the the scores are reliable---e.g., they may be overconfident. We assume that the ground truth label $y^*\in\mathcal{Y}=\{0,1\}$ indicates whether $x$ contains a person. For example, $f(x)$ may be the probability that an image contains a person according to a pretrained DNN such as ResNet~\cite{he2016deep}; then, the goal is to tailor this DNN to the current task in a way that provides correctness guarantees.

In particular, our goal is to choose a threshold $c\in[0,1]$ such that the program returns that the given image $x$ contains a person if $f(x)$ has confidence at least $1-c$---i.e., $f(x)\ge1-c$, or equivalently, $1-f(x)\le c$. Furthermore, we want $c$ to be correct in the following sense:
\begin{align*}
(y^*=1)\Rightarrow(1-f(x)\le c)
\end{align*}
That is, if the image contains a person (i.e., $y^*=1$), then the classifier should say so (i.e., $1-f(x)\le c$). Note that we do not require the converse---i.e., the program may incorrectly conclude that an image contains a person even if it does not. That is, we want soundness (i.e., no false negatives) but not necessarily completeness (i.e., no false positives). However, we cannot guarantee that soundness holds for every image $x$; instead, we want to guarantee it holds with high probability. There are two ways to formulate probabilistic correctness. First, we can say $c$ is \emph{$\epsilon$-approximately correct} if
\begin{align}
\label{eqn:g1}
\mathbb{P}_{p(x,y^*)}\big(y^*=1\Rightarrow 1-f(x)\le c\big)\ge1-\epsilon,
\end{align}
where $p(x,y^*)$ is the data distribution and $\epsilon\in\mathbb{R}_{>0}$ is a user-provided confidence level---i.e., $f$ is correct for $1-\epsilon$ fraction of images sampled from $p(x,y^*)$ that contain a person. Alternatively, we can say $c$ is $\epsilon$-approximately correct if
\begin{align}
\label{eqn:g2}
\mathbb{P}_{p(x,y^*)}\big(1-f(x)\le c\mid y^*=1\big)\ge1-\epsilon.
\end{align}
We refer to (\ref{eqn:g1}) as an \emph{implication guarantee} and (\ref{eqn:g2}) as a \emph{conditional guarantee}. The difference is how ``irrelevant examples'' (i.e., $y^*=0$) are counted: (\ref{eqn:g1}) counts them as being correctly handled, whereas (\ref{eqn:g2}) omits them from consideration. In our example, (\ref{eqn:g1}) says we can count all images without people as being correctly handled. If most images do not contain people, then we can make a large number of mistakes on images that contain people and still achieve $\ge1-\epsilon$ correctness overall. In contrast, (\ref{eqn:g2}) ignores images without people, so we must obtain $\ge1-\epsilon$ correctness rate on images with people alone. However, using (\ref{eqn:g2}), if there are very few images with people, then estimates of the error rate can be very noisy, making our algorithm very conservative. We allow the user choose which guarantee to use; intuitively, (\ref{eqn:g1}) can be used if the goal is to bound the overall error rate, whereas (\ref{eqn:g2}) should be used if it is to bound the error rate among relevant examples. Our syntax for expressing the specification in our example is
\begin{align*}
1-f(x)\le c~~\{y^*=1\}_{0.05}^\mid.
\end{align*}
The syntax $\mid$ indicates the conditional guarantee (\ref{eqn:g2}); we use $\Rightarrow$ to indicate (\ref{eqn:g1}).

We need to slightly weaken our guarantees in an additional way. The reason is that our algorithm relies on training examples $\vec{z}=\{(x_1,y_1^*),...,(x_n,y_n^*)\}$ to choose $c$, where $(x_i,y_i^*)\sim p$ are i.i.d. samples from $p(x,y^*)$. Thus, as with probably approximately correct (PAC) bounds from statistical learning theory~\cite{valiant1984theory,haussler1991equivalence}, we need to additional allow a possibility that our algorithm fails altogether due to the randomness in our training examples $\vec{z}$. In particular, consider an algorithm $A$ that chooses $c=A(\vec{z})$; then, we say $A$ is \emph{$(\epsilon,\delta)$-PAC} if
\begin{align*}
\mathbb{P}_{p(\vec{z})}\big(A(\vec{z})\text{ is }\epsilon\text{-approximately correct}\big)\ge1-\delta
\end{align*}
where $p(\vec{z})$ is the distribution over the training examples $\vec{z}$, and $\delta\in\mathbb{R}_{>0}$ is another user-provided confidence level. Then, given the sketch, a value $\delta\in\mathbb{R}_{>0}$, and a dataset $\vec{z}$, our algorithm synthesizes a value of $c$ to fill \texttt{??1} in a way that ensures that the specification holds (i.e., $c$ is $\epsilon$-approximately correct) with probability at least $1-\delta$.

\paragraph{Synthesis algorithm.}

Next, suppose we want to synthesize a program that counts the number of people in a list of images $\ell=(x_1,...,x_n)$. Intuitively, we can do so by writing a simple list processing program around our DNN for detecting people. In particular, letting
\begin{align*}
(\textsf{predict}_{\textsf{person}}~x)=\mathbbm{1}(1-f(x)\le\;??)
\end{align*}
be our DNN component, where the detection threshold has been left as a hole, then the sketch
\begin{align*}
\tilde{P}_{\text{ex}}=(\textsf{fold}~+~(\textsf{map}~\textsf{predict}_{\textsf{person}}~\ell)~0)
\end{align*}
counts the number of people in $\ell$. Given a few input-output examples along with the ground truth labels for each image, we can use a standard enumerative synthesizer to compute the sketch $\tilde{P}_{\text{ex}}$, assuming $\textsf{predict}_{\textsf{person}}$ returns the ground truth label. In particular, this sketch has a single hole in the DNN component $\textsf{predict}_{\textsf{person}}$ that remains to be filled.

Note that $\tilde{P}_{\text{ex}}$ evaluates correctly if $\textsf{predict}_{\textsf{person}}$ returns the ground truth label, but in general, it may make mistakes. Thus, the correctness property for the synthesized program $P_{\text{ex}}$ needs to account for the possibility that $\textsf{predict}_{\textsf{person}}$ may return incorrectly. Mirroring the correctness property for a single prediction, suppose we want a program $P_{\text{ex}}$ that conservatively overestimates the number of people in $\ell$.\footnote{In Section~\ref{sec:synthesis}, our synthesis algorithm is presented for the case where it returns the correct answer or ``unknown'' with high probability, but as we discuss in Section~\ref{sec:discussion}, it can easily be modified to return an overestimate of the correct answer.}
In particular, given confidence levels $\epsilon,\delta\in\mathbb{R}_{>0}$, we say a completion $P_{\text{ex}}$ of $\tilde{P}_{\text{ex}}$ is \emph{$\epsilon$-approximately correct} if
\begin{align*}
\mathbb{P}_{p(\alpha)}(\llbracket P_{\text{ex}}\rrbracket_{\ell}\ge y^*)\ge1-\epsilon,
\end{align*}
where $\alpha=(\ell,y^*)$ is an example, and $\llbracket P\rrbracket_{\ell}$ denotes the output of running program $P$ on input $\ell$. Then, we say our synthesis algorithm is \emph{$(\epsilon,\delta)$-probably approximately correct (PAC)} if
\begin{align*}
\mathbb{P}_{p(\vec{\alpha})}(A(\tilde{P}_{\text{ex}},\vec{\alpha})\text{ is $\epsilon$-approximately correct})\ge1-\delta,
\end{align*}
where $P_{\text{ex}}=A(\tilde{P}_{\text{ex}},\vec{\alpha})$ is the program synthesized using our algorithm and training examples $\vec{\alpha}$.

Using our statistical sketching algorithm, we can provide $(\epsilon',\delta')$-PAC guarantees on $\textsf{predict}_{\textsf{person}}$ for any $\epsilon',\delta'\in\mathbb{R}_{>0}$; thus, the question is how to choose (i) the appropriate specification, (ii) the parameters of this specification, and (iii) the confidence levels $\epsilon',\delta'$. These choices depend on the specification that we want to ensure for the synthesized program $P_{\text{ex}}$. In our example, we can use the specification above---i.e., that $\textsf{predict}_{\textsf{person}}$ returns $1$ with high probability if there is a person:
\begin{align*}
(\textsf{predict}_{\textsf{person}}~x)=\mathbbm{1}(1-f(x)\le??)~\{y^*=1\}_{\epsilon'}^|.
\end{align*}
In general, the specification on $\textsf{predict}_{\textsf{person}}$ may have additional parameters (in particular, for real-valued predictions, an error tolerance $e$).

Next, we need to choose $\epsilon',\delta'$. While there is only one hole, $\textsf{predict}_{\textsf{person}}$ is executed multiple times (assuming $\textsf{length}(\ell)>1$). We need to choose $\epsilon'$ and $\delta'$ so that with high probability, $\textsf{predict}_{\textsf{person}}$ is correct for \emph{all} applications. For simplicity, we assume given an upper bound $N\in\mathbb{N}$ on the maximum possible length of $\ell$ (we discuss how we might remove this assumption in Section~\ref{sec:discussion}). Given $N$, we take $\epsilon'=\epsilon/N$ and $\delta'=\delta/N$; then, we use our sketching algorithm to synthesize $c$ to fill the hole in $\textsf{predict}_{\textsf{person}}$. By a union bound, for a given list $\ell$, all applications of $\textsf{predict}_{\textsf{person}}$ are correct with probability at least $1-\epsilon$, and this property holds with probability at least $1-\delta$. Under this event, $P_{\text{ex}}$ returns correctly---i.e., $P_{\text{ex}}$ satisfies the desired $(\epsilon,\delta)$-PAC guarantee.

\section{Sketch Language}
\label{sec:language}

In this section, we describe the syntax and semantics of our sketch language, as well as the desired correctness properties we expect that synthesized programs should satisfy.

\paragraph{Syntax.}

Our sketch language is shown in Figure~\ref{fig:semantics}. Intuitively, in the expression $\phi(P,c)~\{Q\}_{\epsilon}^\omega$, $Q$ is a specification that we want to ensure holds, $P$ is a score (intuitively, it should indicate the likelihood that $Q$ holds, but we make no assumptions about it), $c$ is a threshold below which we consider $Q$ to be satisfied, $\epsilon$ is the allowed failure probability, and $\omega$ indicates whether we want a \emph{conditional guarantee} (i.e., $\omega=\;\mid$, the guarantee (\ref{eqn:g2})) or \emph{implication guarantee} (i.e., $\omega=\;\Rightarrow$, the guarantee (\ref{eqn:g1})). We assume that $P$ evaluates to a value in $\mathbb{R}$, $c\in\mathbb{R}$, and $Q$ evaluates to a value in $\{0,1\}$. Note that $Q$ is itself a program; unlike programs $P$, it can use \emph{ground truth inputs} $y$. Finally, either $c$ and $\epsilon$ in this expression can be left as a hole $??$ (but not both simultaneously).

We say $P$ is \emph{complete} if it contains no holes and \emph{partial} otherwise. We use $\mathcal{P}$ to denote the space of programs, $\bar{\mathcal{P}}\subseteq\mathcal{P}$ to denote the space of complete programs, and $\bar{P}\in\bar{\mathcal{P}}$ to denote a complete program. For $P\in\mathcal{P}$, we use $\Phi(P)$ to denote the expressions $\phi(P',c)~\{Q\}_{\epsilon}^\omega$ in $P$ (including cases where $c$ or $\epsilon$ is a hole), $\Phi_{??}^c(P)\subseteq\Phi(P)$ to denote the expressions $\phi(P',??)~\{Q\}_{\epsilon}^\omega$ in $P$, $\Phi_{??}^{\epsilon}(P)\subseteq\Phi(P)$ to denote the expressions $\phi(P',c)~\{Q\}_{??}^\omega$ in $P$, and $\Phi_{??}(P)=\Phi_{??}^c(P)\cup\Phi_{??}^{\epsilon}(P)$.

\paragraph{Semantics.}

We define two semantics for programs $P$, shown in Figure~\ref{fig:semantics}:
\begin{itemize}
\item {\bf Train semantics:} Given a \emph{training valuation} $\alpha\in\mathcal{A}$, where $\alpha:\mathcal{X}\cup\mathcal{Y}\to\mathcal{C}$ maps both inputs and ground truth inputs $y$ to values, the \emph{train semantics} $\llbracket\cdot\rrbracket_{\alpha}^*$ evaluate $Q$ instead of $\phi(P,c)$. Since they ignore $\phi$, they can be applied to both partial and complete programs.
\item {\bf Test semantics:} Given a \emph{test valuation} $\beta\in\mathcal{B}$, where $\beta:\mathcal{X}\to\mathcal{C}$ maps inputs to values, the \emph{test semantics} $\llbracket\cdot\rrbracket_{\beta}$ evaluate $\phi(P,c)$ instead of $Q$. They only apply to complete programs.
\end{itemize}

\begin{figure}
\centering
\scriptsize
\begin{minipage}{0.48\textwidth}
\begin{align*}
P~::=~&c\mid x\mid f(P,...,P) \\
&\mid\phi(P,c)~\{Q\}_{\epsilon}^\omega\mid\phi(P,??)~\{Q\}_{\epsilon}^\omega\mid\phi(P,c)~\{Q\}_{??}^\omega
\\
Q~::=~&c\mid x\mid y\mid f(Q,...,Q)
\end{align*}
\end{minipage}
\begin{minipage}{0.5\textwidth}
\begin{minipage}{0.2\textwidth}
\begin{align*}
\llbracket c\rrbracket_{\alpha}^*&=c \\
\llbracket x\rrbracket_{\alpha}^*&=\alpha(x) \\
\llbracket y\rrbracket_{\alpha}^*&=\alpha(y)
\end{align*}
\end{minipage}
\vspace{5pt}
\begin{minipage}{0.8\textwidth}
\begin{align*}
\llbracket f(P,...,P)\rrbracket_{\alpha}^*&=f(\llbracket P\rrbracket_{\alpha}^*,...,\llbracket P\rrbracket_{\alpha}^*) \\
\llbracket f(Q,...,Q)\rrbracket_{\alpha}^*&=f(\llbracket Q\rrbracket_{\alpha}^*,...,\llbracket Q\rrbracket_{\alpha}^*) \\
\llbracket\phi(P,c)~\{Q\}_{\epsilon}^\omega\rrbracket_{\alpha}^*&=\llbracket Q\rrbracket_{\alpha}^*
\end{align*}
\end{minipage}
\begin{minipage}{0.2\textwidth}
\begin{align*}
\llbracket c\rrbracket_{\beta}&=c \\
\llbracket x\rrbracket_{\beta}&=\beta(x)
\end{align*}
\end{minipage}
\begin{minipage}{0.8\textwidth}
\begin{align*}
\llbracket f(P,...,P)\rrbracket_{\beta}&=f(\llbracket P\rrbracket_{\beta},...,\llbracket P\rrbracket_{\beta}) \\
\llbracket\phi(P,v)~\{Q\}_{\epsilon}^\omega\rrbracket_{\beta}&=\mathbbm{1}(\llbracket P\rrbracket_{\beta}>c)
\end{align*}
\end{minipage}
\end{minipage}
\caption{Syntax (left), train semantics (right, top), and test semantics (right, bottom). The production rules in the syntax are implicitly universally quantified over constant values $c\in\mathcal{C}$, input variables $x\in\mathcal{X}$, \emph{ground truth input variables} $y\in\mathcal{Y}$, components $f\in\mathcal{F}$ where $f:\mathcal{C}^k\to\mathcal{C}$, $\epsilon\in\mathbb{R}_{>0}$, and $\omega\in\{\mid,\Rightarrow\}$. The distinguished component $\phi\in\mathcal{F}$ is a function $\phi:\mathbb{R}^2\to\mathbb{R}$ defined by $\phi(z,t)=\mathbbm{1}(z\le t)$.}
\label{fig:semantics}
\end{figure}

\paragraph{Correctness properties.}

We define what it means for a complete program to be correct---i.e., satisfies its specifications. We begin with correctness of a single specification.
\begin{definition}
\rm
Given a distribution $p(\alpha)$ over test valuations $\alpha\in\mathcal{A}$, $\phi(\bar{P},c)~\{Q\}_{\epsilon}^\mid$ is \emph{approximately sound} if it satisfies the \emph{conditional guarantee}\footnote{Note that since $\alpha$ includes valuations of $x\in\mathcal{X}$, we can use it in conjunction both train semantics and test semantics.}
\begin{align*}
\mathbb{P}_{p(\alpha)}\big(\llbracket\phi(\bar{P},c)~\{Q\}_{\epsilon}^\omega\rrbracket_{\alpha}\bigm\vert\llbracket\phi(\bar{P},c)~\{Q\}_{\epsilon}^\omega\rrbracket_{\alpha}^*\big)\ge1-\epsilon,
\end{align*}
and $\{Q\}_{\epsilon}^\Rightarrow$ is \emph{approximately sound} if it satisfies the \emph{implication guarantee}
\begin{align*}
\mathbb{P}_{p(\alpha)}\big(\llbracket\phi(\bar{P},c)~\{Q\}_{\epsilon}^\omega\rrbracket_{\alpha}^*\Rightarrow\llbracket\phi(\bar{P},c)~\{Q\}_{\epsilon}^\omega\rrbracket_{\alpha}\big)\ge1-\epsilon.
\end{align*}
\end{definition}
This property can be thought of as probabilistic soundness; it says that we should have $\phi(\bar{P},c)\Rightarrow Q$ with high probability, which means that $\phi(\bar{P},c)$ is a sound overapproximation of $Q$.
\begin{definition}
\rm
A complete program $\bar{P}$ is \emph{approximately correct} (denoted $\bar{P}\in\bar{\mathcal{P}}^*$) if every expression $\phi(\bar{P}',c)~\{Q\}_{\epsilon}^\omega$ in $\bar{P}$ is approximately sound.
\end{definition}

\section{Statistical Sketching}
\label{sec:sketching}

Next, we describe our algorithm for synthesizing values $c$ and $\epsilon$ to fill holes in a given sketch. Our algorithm, shown in Figure~\ref{alg:sketch}, takes as input a sketch $P$, training valuations $\vec{\alpha}=(\alpha_1,...,\alpha_n)$, where $\alpha_1,...,\alpha_n\sim p$ are i.i.d. samples, and a confidence level $\delta\in\mathbb{R}_{>0}$, and outputs a complete program $A(\bar{P},\vec{\alpha})\in\bar{\mathcal{P}}^*$ that is approximately correct with probability at least $1-\delta$ with respect to $p(\vec{\alpha})$.

Our algorithm synthesizes $c$ and $\epsilon$ in a bottom-up fashion, so that all subtrees of the current expression are complete. Our sketching algorithm uses probabilistic bounds in conjunction with the given samples $\vec{\alpha}$ to provide guarantees. Intuitively, since we are estimating parameters from data, our problem is a statistical learning problem~\cite{valiant1984theory}, so we can leverage techniques from statistical learning theory to provide guarantees on the synthesized sketch.

For synthesizing $c$---i.e., an expression $E=\phi(P,c)~\{Q\}_{\epsilon}^\omega$. Letting $z_{\alpha}=\llbracket P\rrbracket_{\alpha}\in\mathbb{R}$ and $z_{\alpha}^*=\llbracket E\rrbracket$, then $c$ is $\epsilon$-approximately correct if $z_{\alpha}\le c$ conditioned on $z_{\alpha}^*=1$ (if $\omega=\;\mid$) or whenever $z_{\alpha}^*=1$ (if $\omega=\;\Rightarrow$) with probability at least $1-\epsilon$ with respect to $p(\alpha)$. In either case, synthesizing $c$ is equivalent to a binary classification problem with labels $z_{\alpha}^*$, with a one-dimensional hypothesis space $c\in\mathbb{R}$ and a one-dimensional feature space $z_{\alpha}\in\mathbb{R}$. Furthermore, this problem is simple---$c$ is a linear classifier. Thus, we could use standard learning theory results to provide guarantees.

However, we can obtain sharper guarantees using a learning theory bound specialized to our setting. We build on a bound based on~\cite{haussler1991equivalence} (Section~\ref{sec:learningtheory}) tailored to the \emph{realizable} setting, where there exists a classifier that makes zero mistakes. Our setting is realizable, since $c=\infty$ always makes zero mistakes. The main difference is that their bound always chooses a classifier that makes zero mistakes, which can be overly conservative. We prove a novel generalization bound that allows for some number $k$ of mistakes that is a function of $\epsilon$, $\delta$, and $n$.

Synthesizing a value $\epsilon$ is a bit different, since we are not classifying examples that depend on a single $\alpha$, but examples that depend on $\vec{\alpha}$. Thus, we can formulate it as a learning problem where the examples are $\vec{\alpha}$; however, this approach is complicated due to the need to figure out how to divide our given samples $\vec{\alpha}$ into multiple sub-examples $\vec{\alpha}_1,...\vec{\alpha}_n$. Instead, we use an approach based on Hoeffding's inequality~\cite{hoeffding1994probability} (Section~\ref{sec:concentration2}) to infer $\epsilon$. In particular, Hoeffding's inequality gives us a lower bound on the correctness rate $\mathbb{P}_{p(\alpha)}(z_{\alpha}\mid z_{\alpha}^*)\ge1-\epsilon$ (if $\omega=\;\mid$) or $\mathbb{P}_{p(\alpha)}(z_{\alpha}^*\Rightarrow z_{\alpha})\ge1-\epsilon$ (if $\omega=\;\Rightarrow$), and we can simply use this $\epsilon$.

Finally, our sketching algorithm uses the above two approaches to synthesize $c$ and $\epsilon$ (Section~\ref{sec:synthesisalgo}).

\subsection{A Learning Theory Bound}
\label{sec:learningtheory}

\paragraph{Problem formulation.}

We consider a unary classification problem with one-dimensional feature and hypothesis spaces. In particular, given a probability distribution $p(z)$ over $z\in\mathbb{R}$ (the feature), the goal is to select the smallest possible threshold $t\in\mathbb{R}$ (the hypothesis) such that
\begin{align}
\label{eqn:epssynthesize}
\mathbb{P}_{p(z)}(z\le t)\ge1-\epsilon
\end{align}
for a given $\epsilon\in\mathbb{R}_{>0}$. That is, we want the smallest possible $t$ such that $z\in(-\infty,t]$ with probability at least $1-\epsilon$ according to $p(z)$. We denote the subset of $t$ that satisfies (\ref{eqn:epssynthesize}) by
\begin{align*}
\mathcal{T}_{\epsilon}&=\left\{t\in\mathbb{R}\mid\mathbb{P}_{p(z)}(z\le t)\ge1-\epsilon\right\}.
\end{align*}
To compute such a $t$, we are given a training set of examples $\vec{z}=(z_1,...,z_n)\in\mathbb{R}^n$, where $z_1,...,z_n\sim p$ are $n$ i.i.d. samples from $p$. An \emph{estimator} $\hat{t}$ is a mapping $\hat{t}:\mathbb{R}^n\to\mathbb{R}$. Then, the constraint (\ref{eqn:epssynthesize}) is $\hat{t}(\vec{z})\in\mathcal{T}_{\epsilon}$; we say such a $\hat{t}$ is \emph{$\epsilon$-approximately correct}---i.e., it is correct for ``most'' samples $z\sim p$.

In general, we are unable to guarantee that $\hat{t}$ is approximately correct due to the randomness in the training examples $\vec{z}$. Thus, we additionally allow for a small probability $\delta\in\mathbb{R}_{>0}$ that $\hat{t}$ is not approximately correct.
\begin{definition}
\rm
Given $\epsilon,\delta\in\mathbb{R}_{>0}$, $\hat{t}$ is \emph{$(\epsilon,\delta)$-PAC} if $\mathbb{P}_{p(\vec{z})}(\hat{t}(\vec{z})\in\mathcal{T}_{\epsilon})\ge1-\delta$.
\end{definition}
That is, $\hat{t}(\vec{z})$ is approximately correct with probability at least $1-\delta$ according to $p(\vec{z})$. Our goal is to construct an $(\epsilon,\delta)$-PAC estimator $\hat{t}(\vec{z})$ that tries to minimize $\hat{t}(\vec{z})$.

\paragraph{Estimator.}

Given $\epsilon,\delta\in\mathbb{R}_{>0}$, consider the estimator
\begin{align}
\label{eqn:that}
\hat{t}(\vec{z})
=\inf_{t\in\mathbb{R}}\left\{t\in\mathbb{R}\bigm\vert L(t;\vec{z})\le k\right\}+\gamma(\vec{z})
\quad\text{where}\quad
k=\max\left\{h\in\mathbb{N}\Biggm\vert\sum_{i=0}^h\binom{n}{i}\epsilon^i(1-\epsilon)^{n-i}\le\delta\right\}
\end{align}
where the \emph{empirical loss} is $L(t;\vec{z})=\sum_{z\in\vec{z}}\mathbbm{1}(z>t)$, and where $\gamma(\vec{z})>0$ is an arbitrary positive function. Intuitively, the empirical loss counts the number of mistakes that $t$ makes on the training data---i.e., $z\in\vec{z}$ such that $z\not\in(-\infty,t]$. To compute the solution $k$ in (\ref{eqn:that}), we start with $h=0$ and increment it until it no longer satisfies the condition. To ensure numerical stability, this computation is performed using logarithms. Note that $k$ does not exist if the set inside the maximum in (\ref{eqn:that}) is empty; in this case, we choose $\hat{\psi}(\vec{z})=0$, which trivially satisfies the PAC property. To compute $\hat{t}(\vec{z})$, we sort the training examples $z_1,...,z_n$ by magnitude, so $z_1\ge z_2\ge...\ge z_n$. Finally, $z_{k+1}$ solves the minimization problem in (\ref{eqn:that}), so $\hat{t}(\vec{z})=z_{k+1}+\gamma(\vec{z})$. If $k$ does not exist, then we choose $\hat{t}(\vec{z})=\infty$, which trivially satisfies the PAC property. We have the following; see Appendix~\ref{sec:thmpacproof} for a proof:
\begin{theorem}
\label{thm:pac}
The estimator $\hat{t}(\vec{z})$ in (\ref{eqn:that}) is $(\epsilon,\delta)$-PAC.
\end{theorem}

\begin{algorithm}[t]
\begin{algorithmic}
\Procedure{Sketch}{$P,\vec{\alpha},\delta$}
\State $m\gets|\Phi_{??}(P)|$
\For{$E\in\textsf{BottomUp}(\Phi_{??}(P))$}
\If{$E=\phi(\bar{P}',??)~\{Q\}_{\epsilon}^\omega$}
\State Compute $\vec{z}_{\vec{\alpha}}$ according to (\ref{eqn:synthesis})
\State Compute $\hat{t}(\vec{z}_{\vec{\alpha}})$ according to (\ref{eqn:that}) with $(\epsilon,\delta/m)$
\State Fill the hole $??$ with $\hat{t}(\vec{z}_{\alpha})$
\ElsIf{$E=\phi(\bar{P}',c)~\{Q\}_{??}^\omega$}
\State Compute $\vec{z}_{\vec{\alpha}}$ according to (\ref{eqn:synthesis})
\State Compute $\hat{\nu}(\vec{z}_{\vec{\alpha}})$ according to (\ref{eqn:nuhat}) with $\delta/m$
\State Fill the hole $??$ with $1-\hat{\nu}(\vec{z}_{\vec{\alpha}})$
\EndIf
\EndFor
\State\Return {\bf true}
\EndProcedure
\end{algorithmic}
\caption{Use learning theory to sketch $\bar{P}$ that is approximately correct.}
\label{alg:sketch}
\end{algorithm}

\subsection{A Concentration Bound}
\label{sec:concentration2}

\paragraph{Problem formulation.}

Consider a Bernoulli distribution $p=\text{Bernoulli}(\mu)$ with unknown mean $\mu\in[0,1]$. Our goal is to compute a lower bound $\nu\in[0,1]$ of $\mu$---i.e., $\mu\ge\nu$. For example, if $\mu$ is the error rate of a classifier, then $\nu$ is a lower bound on this rate. To compute $\nu$, we are given a training set $\vec{z}=(z_1,...,z_n)\in\{0,1\}^n$, where $z_1,...,z_n\sim p$ are $n$ i.i.d. samples from $p$. An \emph{estimator} is a mapping $\hat{\nu}:\mathbb{R}^n\to\mathbb{R}$. We say $\hat{\nu}$ is \emph{correct} if it satisfies $\mu\ge\hat{\nu}(\vec{z})$. We are unable to guarantee that $\hat{\nu}(\vec{z})$ is correct due to the randomness in the training examples $\vec{z}$. Thus, we additionally allow for a small probability $\delta\in\mathbb{R}_{>0}$ that $\hat{\psi}(\vec{z})$ is not correct---i.e., it is \emph{probably correct (PC)}.
\begin{definition}
\rm
Given $\delta\in\mathbb{R}_{>0}$, $\hat{\nu}$ is \emph{$\delta$-PC} if $\mathbb{P}_{p(\vec{z})}\big(\mu\ge\hat{\nu}(\vec{z})\big)\ge1-\delta$.
\end{definition}
In other words, $\hat{\nu}(\vec{z})$ is correct with probability at least $1-\delta$ according to the randomness in $p(\vec{z})$. Our goal is to construct an $\delta$-PC estimator $\hat{\nu}(\vec{z})$.

\paragraph{Estimator.}

Given $\delta\in\mathbb{R}_{>0}$, consider the estimator
\begin{align}
\label{eqn:nuhat}
\hat{\nu}(\vec{z})=\hat{\mu}(\vec{z})-\sqrt{\frac{\log(1/\delta)}{2n}},
\end{align}
where $\hat{\mu}(\vec{z})=n^{-1}\sum_{z\in\vec{z}}z$ is an estimate of $\mu$ based on the samples $\vec{z}$; we take $\hat{\nu}(\vec{z})=0$ if (\ref{eqn:nuhat}) is negative. Intuitively, the second term in $\hat{\nu}(\vec{z})$ is a correction to $\hat{\mu}(\vec{z})$ to ensure it is $(\epsilon,\delta)$-PC, based on Hoeffding's inequality~\cite{hoeffding1994probability}. We have the following; see Appendix~\ref{sec:binommeanproof} for a proof:
\begin{theorem}
\label{thm:binommean}
The estimator $\hat{\nu}$ is $\delta$-PC.
\end{theorem}

\subsection{Sketching Algorithm}
\label{sec:synthesisalgo}

\paragraph{Problem formulation.}

A sketching algorithm $A:\mathcal{P}\times\mathcal{A}^n\to\bar{\mathcal{P}}$ takes as input a partial program $P\in\mathcal{P}$, together with a set of test valuations $\vec{\alpha}=(\alpha_1,...,\alpha_n)\in\mathcal{A}^n$, where $\alpha_1,...,\alpha_n\sim p$ are i.i.d. samples from an underlying distribution $p(\alpha)$. Then, $\bar{P}=A(P,\vec{\alpha})$ should be a complete program that is approximately correct by filling each hole in expressions $\phi(P',??)~\{Q\}_{\epsilon}^\omega\in\Phi_{??}^c(P)$ with a value $c\in\mathbb{R}$ and each hole in expressions $\phi(P',c)~\{Q\}_{??}^\omega\in\phi_{??}^{\epsilon}(P)$ with a value $\epsilon\in\mathbb{R}_{>0}$. We assume that every expression in $\Phi(P)$ has a hole---i.e., $\Phi(P)=\Phi_{??}(P)$; otherwise, we cannot guarantee that the existing thresholds in these expressions are approximately sound.
\begin{definition}
\rm
A partial program $P\in\mathcal{P}$ is a \emph{full sketch}, denoted $P\in\mathcal{P}^0$, if $\Phi_{??}(P)=\Phi(P)$.
\end{definition}
Then, we say $A$ is \emph{correct} if $A(P,\vec{\alpha})\in\bar{\mathcal{P}}^*$. We cannot guarantee this property; instead, given $\delta\in\mathbb{R}_{>0}$, we want it to hold with probability at least $1-\delta$ according to $p(\vec{\alpha})$.
\begin{definition}
\rm
A sketching algorithm $A:\mathcal{P}^0\times\mathcal{A}^n\to\bar{\mathcal{P}}$ is \emph{$\delta$-probably approximately correct (PAC)} if for all $P\in\mathcal{P}^0$, we have $\mathbb{P}_{p(\vec{\alpha})}\big(A(P,\vec{\alpha})\in\bar{\mathcal{P}}^*\big)\ge1-\delta$.
\end{definition}
Note that this definition does not include $\epsilon$ since these values are provide in the given sketch.

\paragraph{Algorithm.}

Our sketching algorithm is shown in Algorithm~\ref{alg:sketch}. At a high level, it fills each hole so that the resulting expressions $\phi(\bar{P}',c)~\{Q\}_{\epsilon}^\omega$ are all approximately sound. The order in which these expressions are processed is important; a expression cannot be processed until all its descendants have been processed. This order ensures that $\bar{P}'$ is complete, so it can be evaluated. In Algorithm~\ref{alg:sketch}, the function $\textsf{BottomUp}$ ensures that the expressions in $\Phi_{??}(P)$ is processed in such an order. The algorithm allocates a $\delta/m$ probability of failure for each expression, where $m=|\Phi_{??}(P)|$.

\paragraph{Synthesizing $c$.}

We describe how our algorithm synthesizes a threshold $c$ for an expression $E=\phi(\bar{P}',??)~\{Q\}_{\epsilon}^\omega$. Given a single test valuation $\alpha\sim p$, consider the values
\begin{align*}
z_{\alpha}=\llbracket\bar{P}'\rrbracket_{\alpha}
\qquad\text{and}\qquad
z_{\alpha}^*=\llbracket\phi(\bar{P}',??)~\{Q\}_{\epsilon}^\omega\rrbracket_{\alpha}^*
\end{align*}
Given $c\in\mathbb{R}$, it follows by definition of $\llbracket\cdot\rrbracket_{\alpha}$ that
\begin{align*}
\llbracket\phi(\bar{P}',c)~\{Q\}_{\epsilon}^\omega\rrbracket_{\alpha}=\mathbbm{1}(z_{\alpha}\le c).
\end{align*}
Thus, $E$ is approximately sound for some $c\in\mathbb{R}$ if and only if
\begin{align*}
\mathbb{P}_{p(\alpha)}(z_{\alpha}\le c\mid z_{\alpha}^*)\ge1-\epsilon&\quad\text{if}~\omega=\;\mid
\qquad\text{or}\qquad
\mathbb{P}_{p(\alpha)}(z_{\alpha}^*\Rightarrow z_{\alpha}\le c)\ge1-\epsilon\quad\text{if}~\omega=\;\Rightarrow.
\end{align*}
Given $\vec{\alpha}=(\alpha_1,...,\alpha_n)$, where $\alpha_1,...,\alpha_n\sim p$ i.i.d.,
\begin{align}
\label{eqn:synthesis}
\vec{z}_{\vec{\alpha}}=
\begin{cases}
\{z_{\alpha}\mid\alpha\in\vec{\alpha}\wedge z_{\alpha}^*\}&\text{if}~\omega=\;\mid \\
\{z_{\alpha}^*\Rightarrow z_{\alpha}\mid\alpha\in\vec{\alpha}\}&\text{if}~\omega=\;\Rightarrow
\end{cases}
\end{align}
is a vector of i.i.d. samples. The estimator $\hat{t}(\vec{z}_{\vec{\alpha}})$ in (\ref{eqn:that}) with parameters $(\epsilon,\delta/m)$ ensures approximate soundness with high probability---i.e.,
\begin{align*}
\mathbb{P}_{p(\alpha)}\big(z_{\alpha}\le\hat{t}(\vec{z}_{\vec{\alpha}})\mid z_{\alpha}^*\big)\ge1-\epsilon&\quad\text{if}~\omega=\;\mid
\quad\text{or}\quad
\mathbb{P}_{p(\alpha)}\big(z_{\alpha}^*\Rightarrow z_{\alpha}\le\hat{t}(\vec{z}_{\vec{\alpha}})\big)\ge1-\epsilon\quad\text{if}~\omega=\;\Rightarrow.
\end{align*}
holds with probability at least $1-\delta/m$ according to $p(\vec{\alpha})$.

\paragraph{Synthesizing $\epsilon$.}

We describe how our algorithm synthesizes a confidence level $\epsilon$ for an expression $E=\phi(\bar{P}',c)~\{Q\}_{??}^\omega$. Given a single test valuation $\alpha\sim p$, consider the values
\begin{align*}
z_{\alpha}=\llbracket\phi(\bar{P}',c)~\{Q\}_{??}^\omega\rrbracket_{\alpha}
\qquad\text{and}\qquad
z_{\alpha}^*=\llbracket\phi(\bar{P}',c)~\{Q\}_{??}^\omega\rrbracket_{\alpha}^*.
\end{align*}
Note that we compute these values even though the $\epsilon$ is a hole, since $\llbracket\cdot\rrbracket_{\alpha}$ and $\llbracket\cdot\rrbracket_{\alpha}^*$ do not depend on $\epsilon$. Also, note that unlike the case of synthesizing $c$, where $z_{\alpha}\in\mathbb{R}$ is a score, in this case, $z_{\alpha}\in\{0,1\}$ is a binary value. Given $\epsilon\in\mathbb{R}_{>0}$, $E$ is $\epsilon$-approximately sound for $\epsilon$ if and only if
\begin{align*}
\mathbb{P}_{p(\alpha)}(z_{\alpha}\mid z_{\alpha}^*)\ge1-\epsilon&\quad\text{if}~\omega=\;\mid
\qquad\text{or}\qquad
\mathbb{P}_{p(\alpha)}(z_{\alpha}^*\Rightarrow z_{\alpha})\ge1-\epsilon\quad\text{if}~\omega=\;\Rightarrow.
\end{align*}
Given $\vec{\alpha}=(\alpha_1,...,\alpha_n)$, where $\alpha_1,...,\alpha_n\sim p$ are i.i.d. samples, $\vec{z}_{\vec{\alpha}}$ defined in (\ref{eqn:synthesis}) is a vector of i.i.d. samples from $\text{Bernoulli}(\mu)$. Then, the estimator $\hat{\nu}(\vec{z}_{\vec{\alpha}})$ in (\ref{eqn:nuhat}) with parameter $\delta/m$ is a lower bound on $\mu$ with high probability---i.e.,
\begin{align*}
\mathbb{P}_{p(\alpha)}(z_{\alpha}\mid z_{\alpha}^*)\ge\hat{\nu}(\vec{z}_{\vec{\alpha}})&\quad\text{if}~\omega=\;\mid
\qquad\text{or}\qquad
\mathbb{P}_{p(\alpha)}(z_{\alpha}^*\Rightarrow z_{\alpha})\ge\hat{\nu}(\vec{z}_{\vec{\alpha}})\quad\text{if}~\omega=\;\Rightarrow.
\end{align*}
holds with probability at least $1-\delta/m$ according to $p(\vec{\alpha})$. Thus, it suffices to choose $1-\epsilon=\hat{\nu}(\vec{z}_{\vec{\alpha}})$.

The following guarantee follows from Theorems~\ref{thm:pac} \&~\ref{thm:binommean} by a union bound over $\Phi(\bar{P})$:
\begin{theorem}
\label{thm:sketch}
Algorithm~\ref{alg:sketch} is $\delta$-PAC.
\end{theorem}

\section{Synthesis Algorithm}
\label{sec:synthesis}

We now describe a syntax-guided synthesizer that uses our sketching algorithm to identify programs with machine learning components while satisfying a desired error guarantee. In general, to design such a synthesizer, we need to design a space of specifications along with a domain-specific language (DSL) of programs. For clarity, we focus on a specific set of design choices; as we discuss in Section~\ref{sec:discussion}, our approach straightforwardly generalizes in several ways. We consider the following choices:
\begin{itemize}
\item {\bf Specifications:} We consider specifications $\tilde{\psi}=(\psi,\epsilon,e)$, consisting of both a traditional part $\psi$ indicating the logical property that the train semantics of the program should satisfy (provided either as a logical formula or input-output examples), and a statistical part $(\epsilon,e)$ indicating that the program should have error at most $e$ with probability at least $1-\epsilon$ with respect to $p(\alpha)$, or else return $\varnothing$.
\item {\bf DSL:} We consider a DSL (shown in Figure~\ref{fig:dsl}) of list processing programs where the inputs are images of integers. Our DSL includes components designed to predict the integer represented by a given image. These components return the predicted value if its confidence is above a certain threshold, and return $\varnothing$ otherwise. Values $\varnothing$ are propagated as $\varnothing$ by all components in our DSL---i.e., if any input to a function is $\varnothing$, then its output is also $\varnothing$.
\end{itemize}
For clarity, we refer to specifications $\tilde{\psi}$ as \emph{task specifications} and specifications on DSL components as \emph{component specifications}. As a running example, consider the program in Figure~\ref{fig:synthexample}. This program predicts the value $x$ of the image $\textsf{input}^1$ (as an integer) and values $\ell$ of the images in the list $\textsf{input}^2$ (as real values), and then sums the values in $\ell$ that are greater than equal to $x$. It contains three components that have component specifications: the two machine learning components $\textsf{predict}_{\textsf{int}}$ and $\textsf{predict}_{\textsf{float}}$, along with the inequality \textsf{cond-$\le$}. The first two component specifications ensure that the corresponding machine learning model returns correctly (or $\varnothing$) with high probability. For the last one, note that in the expression $y_1\le y_2$, the inputs $y_1$ and $y_2$ may have a small amount of prediction error, so if they are to close together (i.e., $|y_1-y_2|\le c$ for some $c\in\mathbb{R}_{\le0}$), then $y_1\le y_2$ might be incorrect. Thus, to ensure $\le$ returns correctly, \textsf{cond-$\le$} returns $\varnothing$ if $|y_1-y_2|\le c$.

Finally, note that we use $\omega=\;\Rightarrow$, indicating that our goal is to synthesize $\bar{P}$ such that the the overall success rate is bounded---i.e., $\mathbb{P}_{p(\alpha)}\big(\llbracket\bar{P}\rrbracket_{\alpha}=\varnothing\vee|\llbracket\bar{P}\rrbracket_{\alpha}-\llbracket\bar{P}\rrbracket_{\alpha}^*|>e\big)\ge1-\epsilon$. We could use $\omega=\;\mid$ here if we instead wanted to bound the probability of failure conditioned on $\llbracket\bar{P}\rrbracket_{\alpha}\neq\varnothing$.

\begin{figure*}
\centering\scriptsize
\begin{minipage}{0.4\textwidth}
\begin{align*}
P_{\tau}::=\;&\textsf{input}_{\tau}^1\mid\cdots\mid\textsf{input}_{\tau}^{k_\tau} \\
&\mid(P_{\tau'\to\tau}~P_{\tau'}) \\
&\mid(\textsf{fold}~P_{\tau'\to\tau\to\tau}~P_{\textsf{list}(\tau')}~P_{\tau}) \\
P_{\textsf{list}(\tau)}::=\;&(\textsf{map}~P_{\tau'\to\tau}~P_{\textsf{list}(\tau')}) \\
&\mid(\textsf{filter}~P_{\tau\to\textsf{bool}}~P_{\textsf{list}(\tau)}) \\
&\mid(\textsf{slice}~P_{\textsf{list}(\tau)}~P_{\textsf{int}}~P_{\textsf{int}}) \\
P_{\textsf{int}}::=\;&(\textsf{length}~P_{\textsf{list}(\tau)})
\end{align*}
\end{minipage}
\begin{minipage}{0.45\textwidth}
\begin{align*}
P_{\sigma\to\sigma\to\sigma}::=\;&+\mid- \\
P_{\textsf{int}\to\textsf{int}\to\textsf{bool}}::=&\;\le\;\mid\;=\;\mid\;\ge \\
P_{\textsf{float}\to\textsf{float}\to\textsf{bool}}::=&\;\textsf{cond-$\le$}\mid\textsf{cond-$\ge$} \\
P_{\textsf{image}\to\sigma}::=&\;\textsf{predict}_{\sigma} \\
P_{\textsf{image}\to\textsf{image}}::=&\;\textsf{cond-flip}
\end{align*}
\end{minipage}
\begin{minipage}{\textwidth}
\begin{align*}
(\textsf{predict}_{\textsf{int}}~x)&=(\textsf{if}~\hat{p}(x,\hat{f}(x))\ge\;??_c~\{\hat{f}(x)=y^*\}_{??_\epsilon}^\Rightarrow~\textsf{then}~\hat{f}(x)~\textsf{else}~\varnothing) \\
(\textsf{predict}_{\textsf{float}}~x)&=(\textsf{if}~\hat{p}(x,\hat{f}(x))\ge\;??_c~\{|\hat{f}(x)-y^*|\le\;??_e\}_{??_\epsilon}^\Rightarrow~\textsf{then}~\hat{f}(x)~\textsf{else}~\varnothing) \\
(\textsf{cond-flip}~x)&=(\textsf{if}~\hat{p}_{\textsf{flip}}(x,\hat{f}_{\textsf{flip}}(x))\ge\;??_c~\{\hat{f}_{\textsf{flip}}(x)=y_{\textsf{flip}}^*\}_{??_\epsilon}^\Rightarrow~\textsf{then}~(\textsf{cond-flip}_0~x)~\textsf{else}~\varnothing) \\
(\textsf{cond-flip}_0~x)&=(\textsf{if}~\hat{f}_{\textsf{flip}}(x)~\textsf{then}~\textsf{flip}(x)~\textsf{else}~x) \\
(\textsf{cond-$\le$}~y_1~y_2)&=(\textsf{if}~|y_1-y_2|\ge\;??_c~\{y_1^*\le y_2^*\}_{??_\epsilon}^\Rightarrow~\text{then}~y_1\ge y_2~\textsf{else}~\varnothing) \\
(\textsf{cond-$\ge$}~y_1~y_2)&=(\textsf{if}~|y_1-y_2|\ge\;??_c~\{y_1^*\ge y_2^*\}_{??_\epsilon}^\Rightarrow~\text{then}~y_1\ge y_2~\textsf{else}~\varnothing)
\end{align*}
\end{minipage}
\caption{This figure shows our domain-specific language (DSL) of list processing programs over images of inputs. The top half shows the production rules; these rules are implicitly universally quantified over the type variables $\tau$ and $\sigma$, where $\tau::=\;\textsf{bool}\mid\textsf{int}\mid\textsf{float}\mid\textsf{image}\mid\textsf{list}(\tau)\mid\tau\to\tau$ and $\sigma::=\;\textsf{int}\mid\textsf{float}\mid\textsf{image}$. The bottom half shows the semantics of functions in our language that have statistical specifications.}
\label{fig:dsl}
\end{figure*}

\begin{algorithm}[t]
\begin{algorithmic}
\Procedure{Synthesize}{$\vec{\alpha},\psi,\epsilon,e,N,\delta$}
\State $\tilde{P}\gets\textsf{SynthesizePartialSketch}(\psi)$
\State $\vec{\alpha}_{\text{synth}},\vec{\alpha}_{\text{sketch}}\gets\textsf{Split}(\vec{\alpha})$
\State $P\gets\operatorname*{\arg\max}_{P'\in\textsf{FillAll}(\tilde{P},\epsilon,e)}\textsf{Score}(\textsf{Sketch}(P',\vec{\alpha}_{\textsf{synth}},\delta))$
\State \textbf{return} $\textsf{Sketch}(P,\vec{\alpha}_{\text{sketch}},\delta)$
\EndProcedure
\end{algorithmic}
\caption{Use learning theory to synthesize $\bar{P}$ that is approximately correct.}
\label{alg:synthesis}
\end{algorithm}

Given labeled training examples $\vec{\alpha}$, a task specification $\tilde{\psi}$, a maximum list length $N$, and a confidence level $\delta$, our algorithm shown in Algorithm~\ref{alg:synthesis} synthesizes a complete program $\bar{P}$ that satisfies $\tilde{\psi}$ with probability at least $1-\delta$. At a high level, this algorithm proceeds in three steps:
\begin{itemize}
\item \textbf{Step 1:} First, our algorithm uses the logical specification $\psi$ to identify a sketch $\tilde{P}$ whose train semantics is consistent with $\psi$. Note that the train semantics for sketches in our DSL in Figure~\ref{fig:dsl} are well-defined even when the holes left unfilled. We refer to $\tilde{P}$ as a \emph{partial sketch}, since it has additional holes that cannot be filled by our sketching algorithm.
\item \textbf{Step 2:} While our algorithm uses our sketching algorithm described in Algorithm~\ref{alg:sketch} to fill holes $??_c$ in $\tilde{P}$, it must first fill the holes $??_\epsilon$ and $??_e$ (described below), which cannot be handled by this algorithm. To this end, it analyzes the program to identify constraints on the values of $\epsilon$ and $e$ that can be assigned to each hole $??_\epsilon$ and $??_e$, respectively and satisfy the desired task specification $(\epsilon,e)$. Given candidate values $\vec{e}$ and $\vec{\epsilon}$, it constructs the sketch $P=\textsf{Fill}(\tilde{P},\vec{\epsilon},\vec{e})$, and evaluates the success rate $\textsf{Score}(P)$ (i.e., how often $\llbracket P\rrbracket_{\alpha}\neq\varnothing$). It chooses the sketch $P$ that maximizes this objective over a finite set of choices of $\vec{\epsilon}$ and $\vec{e}$.
\item \textbf{Step 3:} Finally, it uses a held-out set of labeled examples $\vec{\alpha}_{\text{sketch}}$ in conjunction with our sketching algorithm in Algorithm~\ref{alg:sketch} to synthesize 
We use a held-out set since Theorem~\ref{thm:sketch} only holds if the examples $\vec{\alpha}_{\text{sketch}}$ are not used to construct the sketch $P$.
\end{itemize}
In Figure~\ref{fig:synthexample}, we show the partial sketch $\tilde{P}_{\text{ex}}$ along with two analyses which are used to help compute the search space over $\vec{\epsilon}$ and $\vec{e}$. Below, we describe our DSL and synthesis algorithm in more detail.

\begin{figure*}
\centering\scriptsize
\begin{tabular}{lc}
\toprule \\
\textbf{Task Specification} & $\tilde{\psi}_{\text{ex}}=\big(\psi=\{[1,2,3]\mapsto3,~[2,4,2]\mapsto4\},~\epsilon=0.05,~e=6,~N=3,~\delta=0.05\big)$ \\\\
\textbf{Partial Sketch} & $\tilde{P}_{\text{ex}}=(\textsf{fold}~+~(\textsf{filter}~(\underbrace{\textsf{cond-$\le$}}_{f_1}~(\underbrace{\textsf{predict}_{\textsf{int}}}_{f_2}~\textsf{input}^1))~(\textsf{map}~\underbrace{\textsf{predict}_{\textsf{float}}}_{f_3}~\textsf{input}^2)~0))$ \\\\
$\arraycolsep=0pt\begin{array}{l}\textbf{Components}\\\textbf{with Holes}\end{array}$ &
$\arraycolsep=1pt\begin{array}{rl}
f_1&=(\lambda y_1~(\lambda y_2~(\textsf{if}~|y_1-y_2|\ge\;??_c~\{y_1^*\le y_2^*\}_{??_\epsilon}^\Rightarrow~\text{then}~y_1\le y_2~\textsf{else}~\varnothing))) \\
f_2&=(\lambda x~(\textsf{if}~\hat{p}(x,\hat{f}(x))\ge\;??_c~\{\hat{f}(x)=y^*\}_{??_\epsilon}^\Rightarrow~\textsf{then}~\hat{f}(x)~\textsf{else}~\varnothing)) \\
f_3&=(\lambda x~(\textsf{if}~\hat{p}(x,\hat{f}(x))\ge\;??_c~\{|\hat{f}(x)-y^*|\le\;??_e\}_{??_\epsilon}^\Rightarrow~\textsf{then}~\hat{f}(x)~\textsf{else}~\varnothing))
\end{array}$
\\\\
\bottomrule
\end{tabular}
\caption{Example of a task in our list processing domain. Given $\tilde{\psi}_{\text{ex}}$, the goal is to synthesize a program $\bar{P}$ whose train semantics satisfies $\psi$, and whose test semantics satisfy $\mathbb{P}_{p(\alpha)}\big(\llbracket\bar{P}\rrbracket_{\alpha}=\varnothing\vee|\llbracket\bar{P}\rrbracket_{\alpha}-\llbracket\bar{P}\rrbracket_{\alpha}^*|\le e\big)\ge1-\epsilon$.}
\label{fig:synthexample}
\end{figure*}

\subsection{Domain-Specific Language}

Our DSL is summarized in Figure~\ref{fig:dsl}. To be precise, this figure shows sketches in our language; filling holes in these sketches produces a program in our language. At a high level, the language consists of standard list processing operators such as map, filter, and fold, along with a set of functions that can be applied to individual integers, real numbers, or images.

\paragraph{Machine learning components.}

Our DSL has three machine learning components: $\textsf{predict}_{\text{int}}$, $\textsf{predict}_{\text{float}}$, and $\textsf{cond-flip}$. The first two predict the value in a given image. They are identical except for their component specification; whereas the integer predictions must be exactly correct, the real-valued predictions are allowed to have bounded error. We describe these specifications below. This difference gives the user flexibility in terms of what kind of guarantees they want to provide.

The third machine learning component checks if the input image is flipped along the vertical axis. We include it to demonstrate how our approach can combine multiple machine learning components. It only returns an image if it is confident about its prediction; otherwise, it returns $\varnothing$.

\paragraph{Component specifications.}

Intuitively, there are two kinds of component specifications in our language: (i) require that the output is exactly correct, and (ii) require that the error of the output is bounded. There are four components in (i): $\textsf{predict}_{\textsf{int}}$, $\textsf{cond-flip}$, \textsf{cond-$\le$}, and \textsf{cond-$\ge$}. The first two are straightforward---they consist of a machine learning component, and return the predicted value if the prediction confidence is a threshold to be synthesized, and return $\varnothing$ otherwise.

The latter two are result from challenges handling inequalities on real-valued predictions. In particular, real-valued predictions (i.e., by $\textsf{predict}_{\textsf{float}}$) can be wrong by a bounded amount, yet the return value of $\le$ and $\ge$ is a Boolean value that must be exactly correct. Thus, these components include a component specification indicating that their output must be correct with high probability. Note that the scoring function used in the condition is $|y_1-y_2|$; intuitively, if the inputs $y_1$ and $y_2$ are far apart (i.e., $|y_1-y_2|$ is large), then the predicted result is less likely to be an error.

The $\textsf{predict}_{\textsf{float}}$ component is the only one in (ii). The only difference from $\textsf{predict}_{\textsf{int}}$ is that it only requires that the prediction is correct to within some bounded amount of error---i.e., $|\hat{f}(x)-y^*|\le e$, for some $e\in\mathbb{R}_{\ge0}$. Note that $e$ is left as a hole to be filled.

\paragraph{Holes.}

Our language has three kinds of holes. The first two are holes $??_c$ and $??_\epsilon$; these are in our sketch DSL in Figure~\ref{fig:dsl}. Note that in that DSL, each component specification could only have either $c$ or $\epsilon$ as a hole, but here we allow both to be left as holes; our algorithm searches over choices of $\epsilon$ to fill holes $??_\epsilon$, and uses our sketching algorithm in Algorithm~\ref{alg:sketch} to fill holes $??_c$. The third kind of hole is the hole $??_e$ in the component specification $|\hat{f}(x)-y^*|\le\;??_e$ for $\textsf{predict}_{\textsf{float}}$, which indicates the magnitude of error allowed by the prediction of that component. As with $??_\epsilon$ holes, the $??_e$ holes are filled by our algorithm before our sketching algorithm is applied. Intuitively, $??_\epsilon$ (resp., $??_e$) holes must be filled in a way that satisfies the overall $\epsilon$ failure probability guarantee (resp., $e$ error guarantee) in the user-provided task specification $\tilde{\psi}$.

\subsection{Synthesis Algorithm}

Our algorithm (Algorithm~\ref{alg:synthesis}) takes as input labeled training examples $\vec{\alpha}$, a task specification $\tilde{\psi}=(\psi,\epsilon,e)$, and $\delta\in\mathbb{R}_{>0}$, and returns a program $\bar{P}$ that satisfies $\tilde{\psi}$ with probability $\ge1-\delta$.

\paragraph{Step 1: Syntax-guided synthesis.}

Our algorithm first synthesizes a partial sketch $\tilde{P}$ in our DSL whose train semantics satisfies $\psi$---i.e., $\textsf{UNSAT}_{\alpha,y}\big(y=\llbracket\tilde{P}\rrbracket_{\alpha}^*\wedge\neg\psi(\alpha,y)\big)$. Importantly, note that $\llbracket\tilde{P}\rrbracket_{\alpha}^*$ is well-defined even though there are holes in $\tilde{P}$. We can compute $\tilde{P}$ using any standard synthesizer.

\paragraph{Step 2: Sketching $\epsilon$ and $e$.}

Next, our algorithm fills the holes $??_\epsilon$ in $\tilde{P}$ with values $\vec{\epsilon}$ and holes $??_e$ with values $\vec{e}$ to obtain a sketch $P=\textsc{Fill}(\tilde{P},\vec{\epsilon},\vec{e})$. Since $P$ only has holes $??_\epsilon$, we can use Algorithm~\ref{alg:sketch} to fill these holes in a way that guarantees correctness for the given values $\vec{\epsilon}$ and $\vec{e}$---i.e.,
\begin{align}
\label{eqn:synthguarantee}
\mathbb{P}_{p(\alpha)}\big(|\llbracket \bar{P}\rrbracket_{\alpha}-\llbracket P\rrbracket_{\alpha}^*|\le e\big)\ge1-\epsilon,
\end{align}
where $\bar{P}$ is a completion of $P$ where the holes $??_c$ in $P$ have been filled with values $\vec{c}$. We need to use $\bar{P}$ since the test semantics are not well-defined for sketches $P$. In particular, we need to choose values $\vec{\epsilon}$ and $\vec{e}$ that ensure that (\ref{eqn:synthguarantee}) holds for \emph{all} possible completions $\bar{P}$ of $P$.

Furthermore, we not only want to choose $\vec{\epsilon}$ and $\vec{e}$ to ensure correctness, but also to maximize a quantitative property of $\bar{P}$. In particular, we want to choose it in a way that maximizes the probability that $P$ does not return $\varnothing$---i.e., maximize the score
\begin{align*}
\textsf{Score}(P)=\mathbb{P}_{p(\alpha)}\big(\llbracket\bar{P}\rrbracket_{\alpha}\neq\varnothing\big)
\end{align*}
Note that the score depends critically on the choice of thresholds $\vec{c}$ used to fill holes $??_c$ in $P$. Thus, given a set of candidate choices $\vec{\epsilon}$ and $\vec{e}$, our algorithm constructs the corresponding sketch $P'=\textsf{Fill}(\tilde{P},\vec{\epsilon},\vec{e})$, uses our sketching algorithm to fill the holes $??_c$ in $P'$ to obtain $\bar{P}'=\textsf{Sketch}(P',\vec{\alpha},\delta)$, and finally scores $\bar{P}'$. Then, our algorithm chooses $P'$ with the highest score. In Algorithm~\ref{alg:synthesis}, we let $\textsf{FillAll}(\tilde{P},\epsilon,e)$ denote the set of all sketches $P'$ constructed from candidates $\vec{\epsilon}$ and $\vec{e}$.

One important detail is that Algorithm~\ref{alg:sketch} requires that $P$ is a straight-line program---i.e., it cannot handle loops. For now, we assume that we are given a bound $N\in\mathbb{N}$ on the maximum length of any input list. Then, we can unroll list operations such as map, filter, and fold into straight-line code. Algorithm~\ref{alg:synthesis} uses this strategy to apply Algorithm~\ref{alg:sketch} to sketches $P$. We describe how we can remove the assumption that we have an upper bound $N$ in Section~\ref{sec:discussion}.

\paragraph{Step 3: Sketching $c$.}

Finally, we use Algorithm~\ref{alg:sketch} to choose values $\vec{c}$ to fill holes $??_c$ in the highest scoring sketch $P$ from the previous step, and return the result $\bar{P}=\textsf{Sketch}(P,\vec{\alpha}_{\text{sketch}},\delta)$. Importantly, in the previous step, $P$ is chosen based on a subset $\vec{\alpha}_{\textsf{synth}}$ of the training examples $\vec{\alpha}$, whereas in this step, $\bar{P}$ is constructed based on a disjoint subset $\vec{\alpha}_{\textsf{sketch}}$. We choose these two subsets to be of equal size since Algorithm~\ref{alg:sketch} is sensitive to the number of examples in $\vec{\alpha}$. This strategy ensures that $P$ does not depend on the random variable $\vec{\alpha}_{\textsf{sketch}}$, thereby ensuring that Theorem~\ref{thm:sketch} holds.

\subsection{Search Space Over $\vec{\epsilon}$ and $\vec{e}$}
\label{sec:searchspace}

Here, we describe how we choose candidates $\vec{\epsilon}$ and $\vec{e}$ in Step 2 so that the candidate sketches $P'=\textsf{Fill}(\tilde{P},\vec{\epsilon},\vec{e})$ satisfy (\ref{eqn:synthguarantee}). At a high level, for $\vec{\epsilon}$, for each component $f$ of $\tilde{P}$ with an $??_\epsilon$ hole, we compute $\llbracket\tilde{P}\rrbracket_f^\#$, which is the number of times $f$ occurs in the unrolled version of $\tilde{P}$; then, we consider $\vec{\epsilon}=(\epsilon_{f_1},...,\epsilon_{f_d})$ such that $\sum_f\epsilon_f\le\epsilon$. For $\vec{e}$, for each component $f$ of $\tilde{P}$ with an $??_e$ hole, we compute $\llbracket\tilde{P}\rrbracket^{\textsf{err}}:\vec{e}\mapsto e'$, which is a linear function mapping $\vec{e}$ to an upper bound $e'$ on the error of the output; then, we consider $\vec{e}$ such that $\llbracket\tilde{P}\rrbracket^{\textsf{err}}(\vec{e})\le e$. We provide details below.

\begin{figure*}
\centering\scriptsize
\begin{minipage}{0.43\textwidth}
\begin{align*}
\llbracket(F~L)\rrbracket_f^\#&=\llbracket F\rrbracket_f^\#+\llbracket L\rrbracket_f^\#
\\
\llbracket(\textsf{fold}~F~L~B)\rrbracket_f^\#&=N\cdot\llbracket F\rrbracket_f^\#+\llbracket L\rrbracket_f^\#+\llbracket B\rrbracket_f^\# \\
\llbracket(\textsf{map}~F~L)\rrbracket_f^\#&=N\cdot\llbracket F\rrbracket_f^\#+\llbracket L\rrbracket_f^\# \\
\llbracket(\textsf{filter}~F~L)\rrbracket_f^\#&=N\cdot\llbracket F\rrbracket_f^\#+\llbracket L\rrbracket_f^\# \\
\llbracket(\textsf{slice}~L~I_1~I_2)\rrbracket_f^\#&=\llbracket L\rrbracket_f^\#+\llbracket I_1\rrbracket_f^\#+\llbracket I_2\rrbracket_f^\# \\
\llbracket(\textsf{length}~L)\rrbracket_f^\#&=\llbracket L\rrbracket_f^\# \\
\llbracket f'\rrbracket_f^\#&=\mathbbm{1}(f'=f) \\
\llbracket\textsf{input}_\tau^i\rrbracket_f^\#&=0
\end{align*}
\end{minipage}
\begin{minipage}{0.53\textwidth}
\begin{align*}
\llbracket(F~L)\rrbracket^{\textsf{err}}&=\llbracket F\rrbracket^{\textsf{err}}(\llbracket L\rrbracket^{\textsf{err}})
\\
\llbracket(\textsf{fold}~F~L~B)\rrbracket^{\textsf{err}}&=\max_{n\in\{0,1,...,N\}}(\llbracket F\rrbracket^{\textsf{err}})^n(\llbracket L\rrbracket^{\textsf{err}},\llbracket B\rrbracket^{\textsf{err}}) \\
\llbracket(\textsf{map}~F~L)\rrbracket^{\textsf{err}}&=\llbracket F\rrbracket^{\textsf{err}}(\llbracket L\rrbracket^{\textsf{err}}) \\
\llbracket(\textsf{filter}~F~L)\rrbracket^{\textsf{err}}&=\llbracket L\rrbracket^{\textsf{err}} \\
\llbracket(\textsf{slice}~L~I_1~I_2)\rrbracket^{\textsf{err}}&=\llbracket L\rrbracket^{\textsf{err}} \\
\llbracket(\textsf{length}~L)\rrbracket^{\textsf{err}}&=0 \\
\llbracket f\rrbracket^{\textsf{err}}&=\begin{cases}
\lambda\eta.e_f&\text{if}~f=\textsf{predict}_{\textsf{float}} \\
\lambda\eta.\lambda\eta'.\eta+\eta'&\text{if}~f\in\{+,-\} \\
\lambda\eta.\eta&\text{otherwise}
\end{cases} \\
\llbracket\textsf{input}_\tau^i\rrbracket^{\textsf{err}}&=0
\end{align*}
\end{minipage}
\caption{Rules Algorithm~\ref{alg:synthesis} uses to compute the search space over $\vec{\epsilon}$ (left) and $\vec{e}$ (right). In the rule of $\llbracket\cdot\rrbracket^{\textsf{err}}$ for \textsf{fold}, $f^n(\ell,b)=f(\ell,f^{n-1}(\ell,b))$ (and $f^0(\ell,b)=b$) is the function $f$ iterated $n$ times in its second argument. The definitions of $\eta$, $\eta'$, and $\eta+\eta'$ in the rule for $\llbracket f\rrbracket^{\textsf{err}}$ are given in Section~\ref{sec:searchspace}.}
\label{fig:synthanalysis}
\end{figure*}

\paragraph{Search space over $\vec{\epsilon}$.}

First, we describe our search space over parameter values $\vec{\epsilon}$ used to fill holes $??_\epsilon$ so that the overall failure rate is at most $\epsilon$. Note that here, $\vec{\epsilon}=(\epsilon_{f_1},...,\epsilon_{f_k})$, where $\mathcal{F}_{\tilde{P}}=\{f_1,...,f_k\}$ are subexpressions of $\tilde{P}$ of the form $\textsf{predict}_{\text{int}}$, $\textsf{predict}_{\textsf{float}}$, $\textsf{cond-flip}$, \textsf{cond-$\le$}, or \textsf{cond-$\ge$}, since each of these subexpressions contains exactly one hole of the form $??_\epsilon$.

Intuitively, we can ensure correctness via a union bound---i.e., if the sum of the $\epsilon_f$ is bounded by $\epsilon$, then the overall failure probability is also bounded by $\epsilon$. The key caveat is that to apply Algorithm~\ref{alg:sketch}, we need to unroll the sketch $P=\textsf{Fill}(\tilde{P},\vec{\epsilon},\vec{e})$. Thus, we need to count a value $\epsilon_f$ multiple times if the corresponding subexpression $f$ occurs multiple times in the unrolled version of $P$.

In particular, the rules $\llbracket P\rrbracket_{f'}^\#$ shown in Figure~\ref{fig:synthanalysis} are designed to count the number of occurrences of the subexpression $f'$ in the \emph{unrolled} version of $P$. Note that in these rules, $f'$ refers to a \emph{specific} subexpression, and $\mathbbm{1}(f=f')$ refers to whether $f$ is that specific subexpression; multiple uses of the same construct (e.g., a program with two uses of $\textsf{predict}_{\textsf{int}}$) are counted separately. These rules are straightforward; for instance, when unrolling the \textsf{fold} operator, the expressions for the list $L$ and the initial value $B$ are included exactly once, whereas the function expression $F$ occurs $N$ times. Then, to ensure that the failure probability is at most $\epsilon$, it suffices for $\vec{\epsilon}$ to satisfy
\begin{align}
\label{eqn:epsconstraint}
\sum_{f\in\mathcal{F}_{\tilde{P}}}\llbracket\tilde{P}\rrbracket_f^\#\cdot\epsilon_f\le\epsilon.
\end{align}
Now, let $\Delta_{\mathcal{F}_{\tilde{P}}}=\{\vec{x}\in\mathbb{R}^{|\mathcal{F}_{\tilde{P}}|}\mid\forall f\;.\;0\le x_f\le 1\wedge\sum_{f\in\mathcal{F}_{\tilde{P}}}x_f=1\}$ be the regular simplex in $\mathbb{R}^{|\mathcal{F}_{\tilde{P}}|}$. Now, given any $\vec{x}\in\Delta_{\mathcal{F}_{\tilde{P}}}$, letting $\epsilon_f=x_f\cdot\epsilon/\llbracket\tilde{P}\rrbracket_f^\#$, then (\ref{eqn:epsconstraint}) is satisfied. In our algorithm, we search over a finite set of points from $\Delta_{\mathcal{F}_{\tilde{p}}}$, and construct the corresponding set of values $\vec{\epsilon}$.

In Figure~\ref{fig:example}, the rule for $\textsf{filter}$ applies $f_1=$\textsf{cond-$\le$} and $f_2=\textsf{predict}_{\textsf{int}}$ each $N=3$ times (where $N$ is the given bound on the list length), so we have $\llbracket\tilde{P}_{\text{ex}}\rrbracket_{f_1}^\#=\llbracket\tilde{P}_{\text{ex}}\rrbracket_{f_2}^\#=3$. Similarly, \textsf{map} applies $f_3=\textsf{predict}_{\textsf{float}}$ a total of $N=3$ times, so $\llbracket\tilde{P}_{\text{ex}}\rrbracket_{f_1}^\#=3$. As an example of a point in our search space, taking $\vec{x}=(1/3,1/3,1/3)$ yields $\vec{\epsilon}=(1/9,1/9,1/9)$.

\paragraph{Search space over $\vec{e}$.}

Next, we describe our search space over parameter values $\vec{e}$ used to fill holes $??_e$ so the overall error is at most $e$. Similar to  before, $\vec{e}=(e_{f_1},...,e_{f_h})$, but this time $\mathcal{G}_{\tilde{P}}=\{f_1,...,f_h\}$ are subexpressions of $\tilde{P}$ of the form $\textsf{predict}_{\textsf{float}}$, which each contain exactly one hole of the form $??_e$. In this case, we define an analysis that bounds the overall error of the output of $\bar{P}=\textsf{Fill}(P,\vec{\epsilon},\vec{e})$ for any $\vec{\epsilon}$ as a function of $\vec{e}$. More precisely, $\llbracket P\rrbracket^{\textsf{err}}$ satisfies the following property:
\begin{align}
\label{eqn:erranalysis}
\big\|\llbracket\textsf{Fill}(P,\vec{\epsilon},\vec{e})\rrbracket_{\alpha}-\llbracket P\rrbracket_{\alpha}^*\big\|_{\infty}\le\llbracket P\rrbracket^{\textsf{err}}(\vec{e})
\end{align}
for all $\vec{\epsilon}$ and $\vec{e}$, and for all $\alpha$ such that all component specifications in $\textsf{Fill}(P,\vec{\epsilon},\vec{e}$ hold for $\vec{\alpha}$. In other words, (\ref{eqn:erranalysis}) bounds the error of the output for examples $\alpha$ such that predictions fall within the desired error bounds (failures happen with probability at most $\epsilon$ according to our choices of $\vec{\epsilon}$).

Note that (\ref{eqn:erranalysis}) uses the $L_{\infty}$ norm. For scalar outputs, we have $\|x-x'\|_{\infty}=|x-x'|$. For list outputs, for the $L_{\infty}$ norm to be well-defined, we need to ensure that $x=\llbracket\textsf{Fill}(P,\vec{\epsilon},\vec{e})\rrbracket_{\alpha}$ and $x'=\llbracket\bar{P}\rrbracket_{\alpha}^*$ are of the same length (at least, when all component specifications are satisfied). In particular, the only potential case where $x$ and $x'$ have unequal lengths is if $\bar{P}$ contains a \textsf{filter} operator. We focus on filtering real-valued lists; filtering integer-valued lists is similar (and there are no operations to filter list-valued lists or image-valued lists). In the real-valued case, the filter function must be either \textsf{cond-$\le$} and \textsf{cond-$\ge$}. Assuming the component specifications on \textsf{cond-$\le$} and \textsf{cond-$\ge$} are satisfied, then their (Boolean) outputs are guaranteed to be equal, so the outputs of the \textsf{filter} operator have equal length under train and test semantics. Thus, $\|x-x'\|_{\infty}$ is well-defined.

Given $\llbracket P\rrbracket^{\textsf{err}}$, our goal is to compute $\vec{e}$ satisfying
\begin{align}
\label{eqn:errconstraint}
\llbracket P\rrbracket^{\textsf{err}}(\vec{e})\le e.
\end{align}
As with $\vec{\epsilon}$, we can construct a candidate $\vec{e}$ for any point in $x\in\Delta_{\mathcal{G}_{\tilde{P}}}$ by taking $e_f=x_f\cdot e/a_f$, where $\llbracket\tilde{P}\rrbracket^{\textsf{err}}=\sum_{f\in\mathcal{G}_{\tilde{P}}}a_f\cdot e_f$. In Figure~\ref{fig:synthexample}, we have $\llbracket\tilde{P}\rrbracket^{\textsf{err}}=3\cdot e_{f_3}$, so there is a single candidate $e_{f_3}=e/3$.

Next, we describe the rules $\llbracket P\rrbracket^{\textsf{err}}$, which are shown in Figure~\ref{fig:synthanalysis} (right). They compute an symbolic expression of the form $\eta=\sum_{f\in\mathcal{G}_{\tilde{P}}}a_f\cdot e_f\in\mathcal{E}_{\tilde{P}}$, where $a_f\in\mathbb{R}_{\ge0}$ and $e_f$ is a symbol. Given $\vec{e}$, an expression $\eta$ can be evaluated by substituting $\vec{e}$ for the symbols $e_f$ in $\eta$. Now, the rule for function application assumes given a function abstraction $\llbracket F\rrbracket^{\textsf{err}}:\mathcal{E}_{\tilde{P}}\to\mathcal{E}_{\tilde{P}}$. In particular, $\llbracket F\rrbracket^{\textsf{err}}$ is the identity function except for $\textsf{predict}_{\textsf{float}}$, $+$, and $-$. The case $\textsf{predict}_{\textsf{float}}$ follows since we have assumed that the component specification holes, and the component specification for $f=\textsf{predict}_{\textsf{float}}$ says exactly that $|\llbracket\bar{f}\rrbracket_{\alpha}-\llbracket f\rrbracket_{\alpha}^*|\le e_f$ for any completion $\bar{f}$ of $f$. For $+$ and $-$, letting $\eta=\sum_{f\in\mathcal{G}_{\tilde{P}}}a_f\cdot e_f$ and $\eta'=\sum_{f\in\mathcal{G}_{\tilde{P}}}a_f'\cdot e_f$, we define $\eta+\eta'=\sum_{f\in\mathcal{G}_{\tilde{P}}}(a_f+a_f')\cdot e_f$. The rule for map follows since we are using the $L_{\infty}$ norm, so the bound is applied elementwise. The remaining rules are straightforward.

In Figure~\ref{fig:synthexample}, the rule for $\textsf{predict}_{\text{float}}$ returns $e_{f_3}$, so the rule for $\textsf{map}$ returns $3\cdot e_{f_3}$ (since the given bound on the list length is $N=3$). The remaining rules propagate this value, so $\llbracket\tilde{P}_{\textsf{ex}}\rrbracket^{\textsf{err}}=3\cdot e_{f_3}$.

Finally, the fact that $\llbracket\cdot\rrbracket^{\text{err}}$ is a linear function follows by structural induction. Additional components (e.g., multiplication) can result in nonlinear expressions, but a similar approach applies.

\paragraph{Overall search space.}

Our overall search space consists of pairs $\vec{\epsilon}$ and $\vec{e}$ such that $\vec{\epsilon}$ satisfies (\ref{eqn:epsconstraint}) and $\vec{e}$ satisfies (\ref{eqn:errconstraint}); given such a pair, $\textsf{FillAll}(\tilde{P},\epsilon,e)$ includes the program $P=\textsf{Fill}(\tilde{P},\vec{\epsilon},\vec{e})$. Together, (\ref{eqn:epsconstraint}) and (\ref{eqn:errconstraint}) ensure the desired property (\ref{eqn:synthguarantee}). In particular, for any completion $\bar{P}$ of $P$, (\ref{eqn:errconstraint}) ensures that $|\llbracket\bar{P}\rrbracket_{\alpha}-\llbracket P\rrbracket_{\alpha}^*|\le e$ as long as $\alpha$ satisfies all the component specifications, and (\ref{eqn:epsconstraint}) ensures that $\alpha$ satisfies the component specifications with probability at least $1-\epsilon$ over $p(\alpha)$.

\section{Discussion}
\label{sec:discussion}

\paragraph{Generality.}

In Section~\ref{sec:synthesis}, we described a synthesizer tailored to the language in Figure~\ref{fig:dsl}. Our approach generalizes straightforwardly in several ways. First, we note that the $\textsf{predict}_{\textsf{int}}$ and $\textsf{predict}_{\textsf{float}}$ machine learning components are not specific to images of integers, and represent general classification and regression problems, respectively. Furthermore, we can also include additional list processing components as long as we provide the abstract semantics $\llbracket\cdot\rrbracket^\#$ and $\llbracket\cdot\rrbracket^{\textsf{err}}$. Thus, our algorithm can be viewed as a general algorithm for synthesizing list processing programs with DNNs for classification and regression, where the specification is that with high probability, the program should return the either the correct answer (within some given error tolerance) or $\varnothing$.

We can also modify the specification in certain ways; for instance, we can ignore certain kinds of errors by modifying the annotations on $\textsf{predict}_{\textsf{int}}$ and $\textsf{predict}_{\textsf{float}}$.
For instance, to allow for one-sided errors in regression problems (e.g., it is fine to say ``person'' when there isn't one but not vice versa), we can simply drop the absolute values from the task specification $\psi$ and from the annotations on $\textsf{predict}_{\textsf{float}}$.
For this case, the algorithm for allocating errors $e$ works as is, but in general, it may need to be modified to ensure the annotations imply the specification.

\paragraph{Bound on examples.}

In Section~\ref{sec:synthesis}, we assumed given a bound $N$ on the maximum length of any list observed during program execution. Intuitively, we can circumvent this assumption by computing a high probability bound $N$; the error probability can be included in the user-provided allowable error rate $\epsilon$. In particular, let $\llbracket\tilde{P}\rrbracket^{\textsf{len}}_{\alpha}$ denote the maximum list length observed while executing $\tilde{P}$ on input $\alpha$. Then, suppose we can obtain $N$ such that
\begin{align*}
\mathbb{P}_{p(\alpha)}\big(\llbracket\tilde{P}\rrbracket^{\textsf{len}}_{\alpha}\le N)\ge1-\frac{\epsilon}{2}.
\end{align*}
Now, if we synthesize a completion $\bar{P}$ of $\tilde{P}$ with overall error rate $\le\epsilon/2$, then by a union bound, the total error rate is $\le\epsilon$. Finally, to obtain such an $N$, we can use the specification
\begin{align*}
\llbracket\tilde{P}\rrbracket_{\alpha}^{\textsf{len}}\le??~\{\textsf{true}\}_{\epsilon/2}^\Rightarrow.
\end{align*}
Letting $c$ be the synthesized value used to fill the hole, the specification says that $\llbracket\tilde{P}\rrbracket_{\alpha}^{\textsf{len}}\le c$ with probability at least $\epsilon/2$ according to $p(\alpha)$, which is exactly the desired condition on $N$; thus, we can take $N=c$. Note that since the specification is \textsf{true}, we can use either $\mid$ or $\Rightarrow$.

\section{Evaluation}
\label{sec:evaluation}

We describe our evaluation on synthesizing list processing programs, as well as on two case studies: (i) a state-of-the-art image classifier, and (ii) a random forest trained to predict Warfarin drug dosage. In addition, we describe an extension of (i) to object detection in Appendix~\ref{sec:appendixdetection}.

\subsection{Synthesizing List Processing Programs with Image Classification}
\label{sec:expsynthesis}

\begin{table*}
\centering\scriptsize
\begin{tabular}{llrrrrrr}
\toprule
\multicolumn{1}{c}{\multirow{2}{*}{\textbf{DSL Variant}}} & \multicolumn{1}{c}{\multirow{2}{*}{\textbf{Task}}} & \multicolumn{3}{c}{\textbf{$\varnothing$ Rate}} & \multicolumn{3}{c}{\textbf{Failure Rate}} \\
& & \toolname & No Search & $k=0$ & \toolname & No Search & $k=0$ \\
\midrule
\multirow{4}{*}{int} & sum $x\in\ell$ & 0.000 & 0.000 & 0.177 & 0.018 & 0.018 & 0.001 \\
& max $x\in\ell$ & 0.000 & 0.000 & 0.177 & 0.008 & 0.008 & 0.001 \\
& sum $x\in\ell$ that are $\le k$ & 0.001 & 0.022 & 0.206 & 0.016 & 0.010 & 0.001 \\
& max first $k$ elements $x\in\ell$ & 0.000 & 0.008 & 0.195 & 0.007 & 0.007 & 0.000 \\
& count $x\in\ell$ that are $\le k$ & 0.001 & 0.022 & 0.206 & 0.000 & 0.000 & 0.000 \\
\midrule
average & \multicolumn{1}{c}{--} & 0.000 & 0.010 & 0.192 & 0.010 & 0.009 & 0.001 \\
\midrule
\multirow{4}{*}{float} & sum $x\in\ell$ & 0.000 & 0.000 & 0.000 & 0.001 & 0.001 & 0.001 \\
& max $x\in\ell$ & 0.000 & 0.000 & 0.000 & 0.000 & 0.000 & 0.000 \\
& sum $x\in\ell$ that are $\le k$ & 0.000 & 1.000 & 1.000 & 0.010 & 0.000 & 0.000 \\
& max first $k$ elements $x\in\ell$ & 0.000 & 0.005 & 0.177 & 0.000 & 0.000 & 0.000 \\
& count $x\in\ell$ that are $\le k$ & 0.000 & 1.000 & 1.000 & 0.000 & 0.000 & 0.000 \\
\midrule
average & \multicolumn{1}{c}{--} & 0.000 & 0.401 & 0.435 & 0.002 & 0.000 & 0.000 \\
\midrule
\multirow{4}{*}{flip} & sum $x\in\ell$ & 0.015 & 0.016 & 0.230 & 0.012 & 0.012 & 0.001 \\
& max $x\in\ell$ & 0.015 & 0.016 & 0.230 & 0.006 & 0.006 & 0.001 \\
& sum $x\in\ell$ that are $\le k$ & 0.025 & 0.085 & 0.265 & 0.012 & 0.004 & 0.001 \\
& max first $k$ elements $x\in\ell$ & 0.063 & 0.046 & 0.258 & 0.005 & 0.004 & 0.000 \\
& count $x\in\ell$ that are $\le k$ & 0.025 & 0.085 & 0.265 & 0.000 & 0.000 & 0.000 \\
\midrule
average & \multicolumn{1}{c}{--} & 0.029 & 0.050 & 0.250 & 0.007 & 0.005 & 0.001 \\
\midrule
\multirow{4}{*}{fast} & sum $x\in\ell$ & 0.033 & 0.033 & 0.706 & 0.026 & 0.026 & 0.000 \\
& max $x\in\ell$ & 0.033 & 0.033 & 0.706 & 0.008 & 0.008 & 0.000 \\
& sum $x\in\ell$ that are $\le k$ & 0.039 & 0.127 & 0.755 & 0.023 & 0.005 & 0.000 \\
& max first $k$ elements $x\in\ell$ & 0.035 & 0.061 & 1.000 & 0.010 & 0.007 & 0.000 \\
& count $x\in\ell$ that are $\le k$ & 0.039 & 0.127 & 0.755 & 0.000 & 0.000 & 0.000 \\
\midrule
average & \multicolumn{1}{c}{--} & 0.036 & 0.076 & 0.784 & 0.013 & 0.009 & 0.000 \\
\midrule
overall & \multicolumn{1}{c}{--} & 0.016 & 0.134 & 0.415 & 0.008 & 0.006 & 0.000 \\
\bottomrule \\
\end{tabular}
\caption{We show results on synthesizing list processing programs, for both our approach (\toolname) and the baseline that does not search over $\vec{\epsilon}$ and $\vec{e}$ (``No Search''). For each DSL variant and each task, we show the ``$\varnothing$ Rate'' $\mathbb{P}_{p(\alpha)}(\llbracket\bar{P}\rrbracket_{\alpha}=\varnothing)$, and the ``Failure Rate'' $\mathbb{P}_{p(\alpha)}(\llbracket\bar{P}\rrbracket_{\alpha}\neq\varnothing\wedge|\llbracket\bar{P}\rrbracket_{\alpha}-\llbracket\bar{P}\rrbracket_{\alpha}^*=\varnothing|)>e$.}
\label{tab:mnistresults}
\end{table*}

\paragraph{Experimental setup.}

We evaluate our synthesis algorithm on our list processing domain in Section~\ref{sec:synthesis}. Inputs are lists of MNIST digits~\cite{lecun1998gradient}. We use a convolutional DNN (two convolutional layers followed by two fully connected layers, with ReLU activations)~\cite{krizhevsky2012imagenet} to predict the integer in an image, trained on the MNIST training set; it achieves 99.2\% accuracy. We also train a single layer DNN, which is 4.04$\times$ faster but only 98.5\% accurate. Finally, for inputs with the flip component, with consider input images flipped along their horizontal axis. We train a DNN to predict whether a given image is flipped; it achieves 99.6\% accuracy.

For the synthesizer, we use a standard enumerative synthesizer that returns the smallest program in terms of depth (but chooses arbitrarily among equal depth programs). We give it 5 labeled input-output examples as a specification $\psi$, along with the type of the function to be synthesized~\cite{feser2015synthesizing,osera2015type}. For the search space over each $\vec{\epsilon}$ and $\vec{e}$, we consider values $\vec{x}_0\in\{1,3,5\}^d$, where $d=|\mathcal{F}_{\tilde{P}}|$ or $d=|\mathcal{G}_{\tilde{P}}|$, and then take $\vec{x}=\vec{x}_0/\|\vec{x}_0\|_1$ to normalize it to $\Delta^d$. We also compare to (i) a baseline ``No Search'', which only considers a single $\vec{x}_0=(1,...,1)$, and (ii) a baseline ``$k=0$'', which uses a variant of our generalization bound that uses either $k=0$ (or $k=\varnothing$, if there are insufficient samples); this strategy captures the guarantees provided by traditional generalization bounds from statistical learning theory~\cite{haussler1991equivalence,kearns1994introduction,vapnik2013nature}. We use our algorithm with parameters $\epsilon=\delta=0.05$, $e=6$, and $N=3$. We use 2500 MNIST test set images for each $\alpha_{\text{synth}}$ and $\alpha_{\text{sketch}}$, and the remaining 5000 for evaluation. Next, we consider four variants of our DSL:
\begin{itemize}
\item \textbf{Int:} Restrict to components with integer type and omit the \textsf{cond-flip} component 
\item \textbf{Float:} Same as ``int'', but include components with real types
\item \textbf{Flip:} Same as ``int'', but include the flip component
\item \textbf{Fast:} Same as ``int'', but use the fast neural network.
\end{itemize}
For each variant, we consider five list processing tasks, which are designed to exercise different kinds of components. These programs all take as input a list $\ell$ of images $x\in\ell$; in addition, several of them take as input a second image $k$ that encodes some information relevant to task. Then, they output an integer or real value (as specified by $\psi$). The tasks are shared across the different DSL variants, but specific programs change based on the available components.

\begin{figure*}
\centering
\begin{tabular}{ccc}
\includegraphics[width=0.3\textwidth]{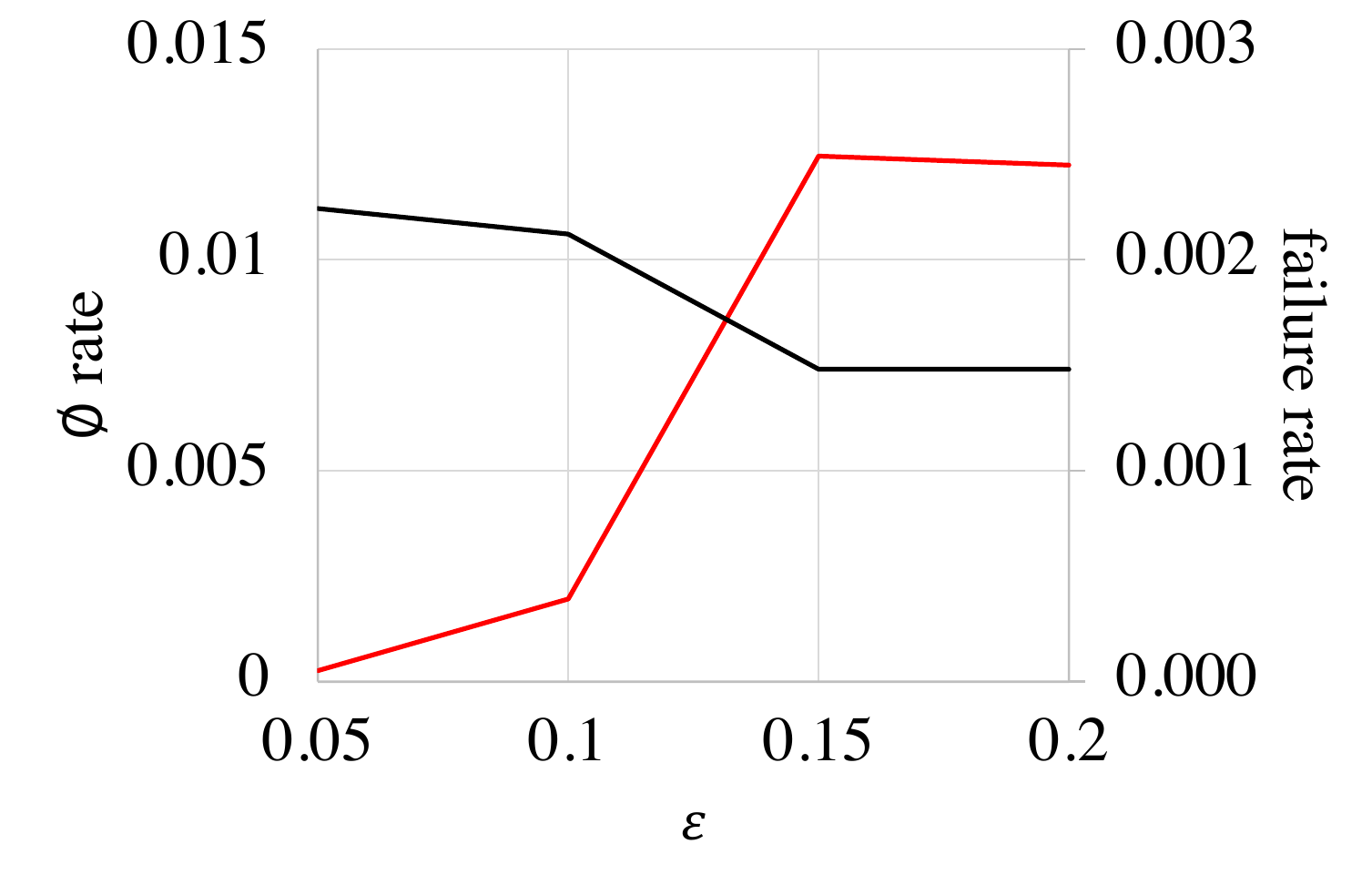} &
\includegraphics[width=0.3\textwidth]{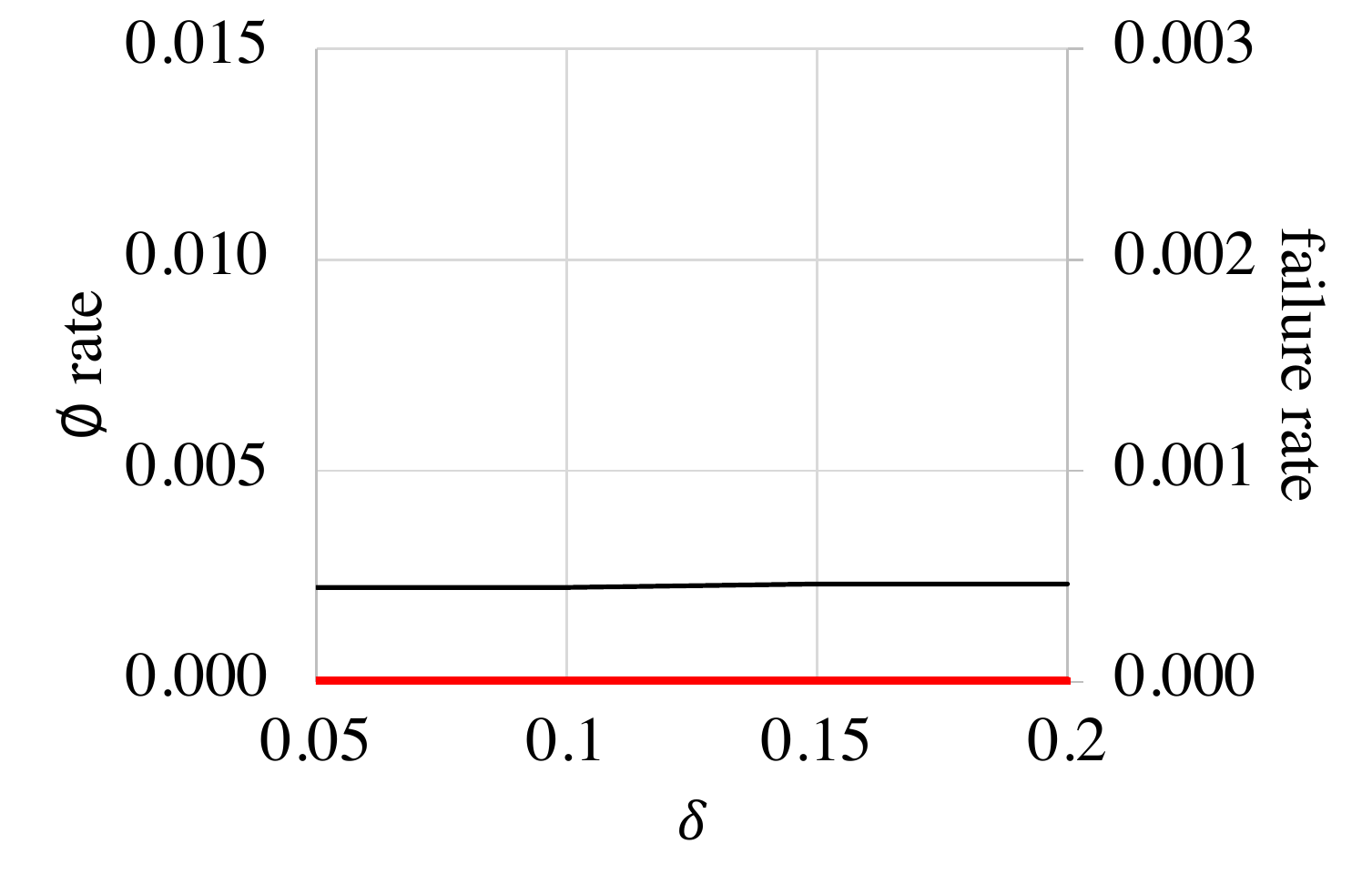} &
\includegraphics[width=0.3\textwidth]{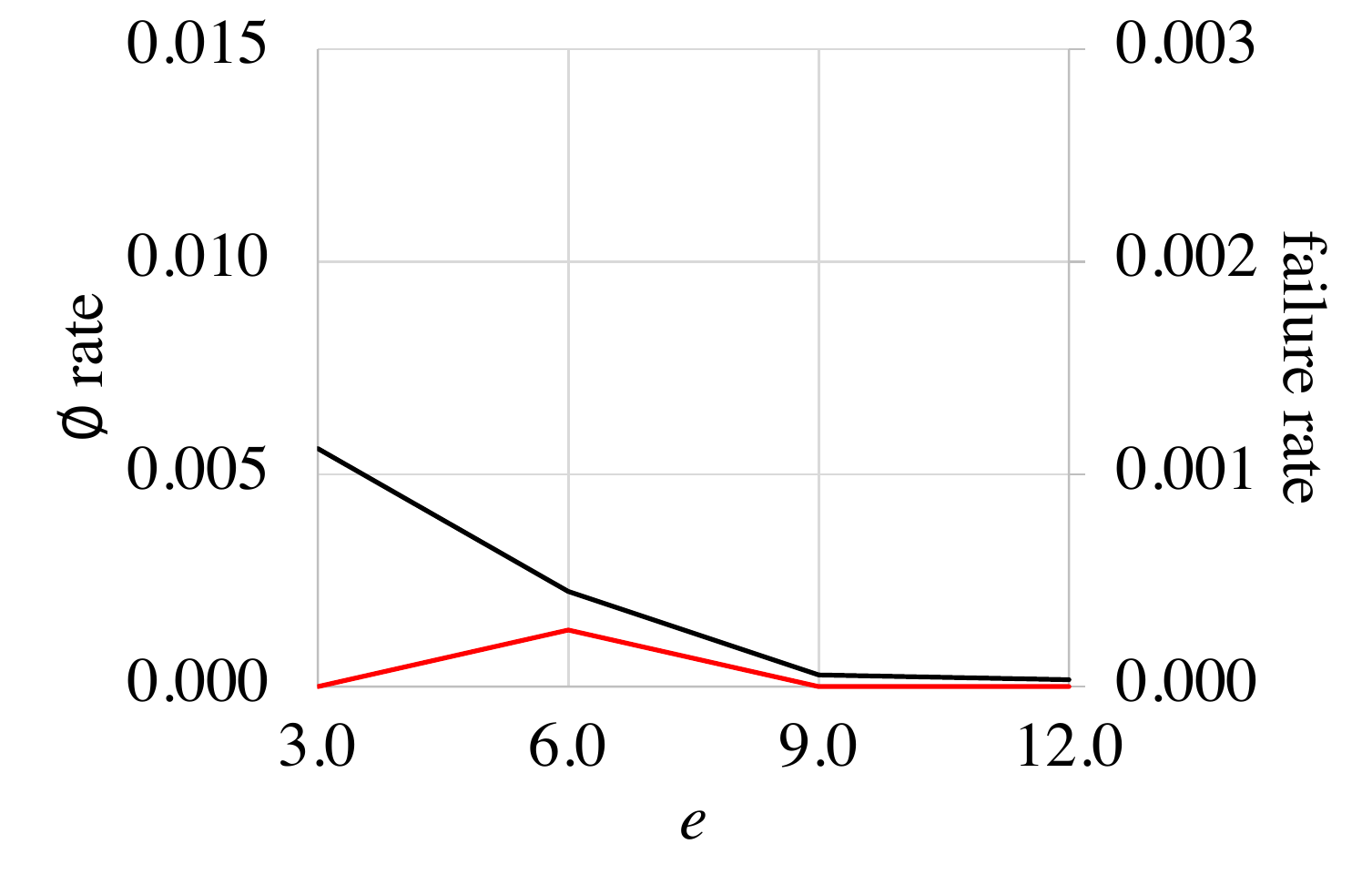}
\\
(a) & (b) & (c) \\
\end{tabular}
\caption{For list processing programs, we show $\varnothing$ rate (black) and failure rate (red) as a function of (a) $\epsilon$, (b) $\delta$, and (c) $e$, on average for (a,b) ``Int'' programs and (c) ``Float'' programs. Defaults are $\epsilon=\delta=0.05$ and $e=6.0$.}
\label{fig:mnistplots}
\end{figure*}

\paragraph{Results.}

We show results in Table~\ref{tab:mnistresults}. For the program $\bar{P}$ synthesized using each our approach $\toolname$ and our baseline that does not search over $\vec{\epsilon}$ and $\vec{e}$, we show the following metrics:
\begin{itemize}
\item \textbf{$\varnothing$ Rate:} The rate at which $\bar{P}$ returns $\varnothing$---i.e., $\mathbb{P}_{p(\alpha)}(\llbracket\bar{P}\rrbracket_{\alpha}=\varnothing)$.
\item \textbf{Failure Rate:} The rate at which $\bar{P}$ makes mistakes---i.e.,
\begin{align*}
\mathbb{P}_{p(\alpha)}\big(\llbracket\bar{P}\rrbracket_{\alpha}\neq\varnothing\wedge|\llbracket\bar{P}\rrbracket_{\alpha}-\llbracket\bar{P}\rrbracket_{\alpha}^*=\varnothing|\big)>e.
\end{align*}
\end{itemize}
As can be seen, both \toolname and the baseline always achieve the desired failure rate bound of $\epsilon=0.05$. Furthermore, by searching over candidates $\vec{\epsilon}$ and $\vec{e}$, \toolname substantially outperforms the baseline, achieving an $8\times$ reduction in $\varnothing$ rate on average. For simpler programs (i.e., sum and max), the two perform similarly since there is only a single hole, so the search space only contains one candidate. However, for larger programs, the search improves performance by up to an order of magnitude. There is a single case where the baseline performs better (the fourth program in the ``flip'' DSL), due to random chance since the dataset $\alpha_{\text{sketch}}$ used to synthesize the final program $\bar{P}$ from $\tilde{P}$ differs from the dataset $\alpha_{\text{synth}}$ used to choose $\vec{\epsilon}$ and $\vec{e}$. \toolname outperforms the ``$k=0$'' baseline by an even larger margin, due to the fact that the generalization bound is overly conservative; these results demonstrate the importance of using a generalization bound specialized to our setting rather than a more traditional generalization bound that minimizes the empirical risk.

Next, in Figure~\ref{fig:mnistplots}, we show how these results vary as a function of the specification parameters $\epsilon$, $\delta$, and $e$. As can be seen, $\epsilon$ has the largest effect on $\varnothing$ and failure rates, followed by $e$; as expected, $\delta$ has almost no effect since the dependence of our bound on $\delta$ is logarithmic.

Finally, we note that the failure rates for the ``fast'' DSL are very low. Thus, we could use our technique to chain together the fast program with the slow one, along the same lines as discussed in our case study in Section~\ref{sec:evalclassification}; we estimate that doing so results in a $3\times$ speedup on average.

\begin{table*}
\centering\scriptsize
\begin{tabular}{lrrrrrr}
\toprule
\multicolumn{1}{c}{\multirow{2}{*}{\textbf{Task}}} & \multicolumn{3}{c}{\textbf{$\varnothing$ Rate}} & \multicolumn{3}{c}{\textbf{Failure Rate}} \\
& \toolname & No Search & $k=0$ & \toolname & No Search & $k=0$ \\
\midrule
count the number of people in $x$ & 0.054 & 0.054 & 0.901 & 0.124 & 0.124 & 0.003 \\
check if $x$ contains a person & 0.054 & 0.054 & 0.901 & 0.124 & 0.124 & 0.003 \\
count people near the center of $x$ & 0.290 & 0.290 & 0.901 & 0.032 & 0.032 & 0.003 \\
find people near a car & 0.901 & 0.901 & 1.000 & 0.003 & 0.003 & 0.000 \\
minimum distance from a person to the center of $x$ & 0.149 & 0.149 & 0.901 & 0.023 & 0.023 & 0.000 \\
\midrule
average & 0.290 & 0.290 & 0.921 & 0.061 & 0.061 & 0.002 \\
\bottomrule \\
\end{tabular}
\caption{We show results on synthesizing list processing programs over object detection, for our approach \toolname. For each DSL variant and each task, we show the ``$\varnothing$ Rate'' $\mathbb{P}_{p(\alpha)}(\llbracket\bar{P}\rrbracket_{\alpha}=\varnothing)$, and the ``Failure Rate'' $\mathbb{P}_{p(\alpha)}(\llbracket\bar{P}\rrbracket_{\alpha}\neq\varnothing\wedge|\llbracket\bar{P}\rrbracket_{\alpha}-\llbracket\bar{P}\rrbracket_{\alpha}^*=\varnothing|)>e$. Parameters are $\epsilon=\delta=0.2$ and $e=20.0$.}
\label{tab:cocoresults}
\end{table*}

\subsection{Synthesizing List Processing Programs with Object Detection}

\paragraph{Experimental setup.}

Next, we consider synthesizing programs that operate over the predictions made by a state-of-the-art DNN for object detection. We assume given a DNN component $\hat{f}$ that given an image $x$, is designed to detect people and cars in $x$. We use a pretrained state-of-the-art object detector called Faster R-CNN~\cite{ren2016faster} available in PyTorch~\cite{paszke2017automatic}, tailored to the COCO dataset~\cite{lin2014microsoft}, which is a dataset of real-world images containing people, cars, and other objects. There are multiple variants of Faster R-CNN; we use the most accurate one, X101-FPN with $3\times$ learning rate schedule.

We represent this DNN as a component $\hat{f}:\mathcal{X}\to\mathcal{Y}=\mathcal{D}^*\times\mathbb{R}$, where $\hat{f}(x)=(\hat{y}(x),\hat{p}(x))$ consists of a list of \emph{detections} $d\in\hat{y}(x)$ along with a correctness score $\hat{p}(x)$ that the prediction is correct. Each detection $d\in\mathcal{D}=\mathbb{R}^2\times\mathcal{Z}$ is itself a tuple $d=(b,z)$ including the position $b$ and predicted category of the object. The ground truth label $y^*$ for an image $x$ is a list of detections $d\in y^*$. In general, we cannot expect to get a perfect match between the predicted bounding boxes and the ground truth ones. Typically, two detections $d,d^*$ \emph{match}, denoted $\|d-d^*\|\le e$, where $e$ is a specified error tolerance, if the distance between their centers satisfies $\|b-b^*\|_{\infty}\le e$. Furthermore, we write $\|\hat{y}(x)-y^*\|\le e$ if $|\hat{y}(x)|=|y^*|$ and there exists a one-to-one correspondence between $d\in\hat{f}(x)$ and $d^*\in y^*$ such that $\|d-d^*\|\le e$. Then, we define $\textsf{predict}:\mathcal{X}\to(\mathcal{Y}\cup\varnothing)$ by
\begin{align*}
(\textsf{predict}~x)=(\textsf{if}~\hat{p}(x)\ge??_c~\{\|\hat{y}(x)-y^*\|\le??_e\}_{??_\epsilon}^\Rightarrow~\textsf{then}~\hat{y}(x)~\textsf{else}~\varnothing).
\end{align*}
In other words, the specification says that a correct prediction is if the error tolerance is below a level $??_e$ to be specified. Thus, given $e$ and $\epsilon$ to fill $??_e$ and $??_\epsilon$, respectively, our sketching algorithm synthesizes a threshold $c$ to fill $??_c$ in a way that guarantees that this specification holds. Then, \textsf{predict} returns $\hat{y}(x)$ if the DNN is sufficiently confident in its prediction, and $\varnothing$ otherwise.

We can use this component in conjunction with our synthesis algorithm in the same way that it uses $\textsf{predict}_{\textsf{float}}$. In particular, we define the abstract semantics
\begin{align*}
\llbracket(\textsf{predict}~x)\rrbracket^{\textsf{err}}=\lambda\eta.e_{\textsf{predict}}.
\end{align*}
These semantics enable it to select the error tolerance $e$ to fill $??_e$. The remainder of the synthesis algorithm proceeds as in Section~\ref{sec:expsynthesis}. We use parameters $\epsilon=\delta=0.2$, $e=20.0$, and $N=3$, and use $n=1000$ COCO validation set images for each $\alpha_{\text{synth}}$ and $\alpha_{\text{sketch}}$ and the remaining $1503$ for evaluation. We use larger $\epsilon$ and $\delta$ since the accuracy of the object detector is significantly lower than that of the image classifier, so the $\varnothing$ rates are very high for smaller choices.

We evaluate our approach on synthesizing five programs, which include additional list processing components: (i) $(\textsf{product}~L~L')$, which returns the list of all pairs $(x,x')$ such that $x\in\llbracket L\rrbracket$ and $x'\in\llbracket L'\rrbracket$, (ii) $(\textsf{compose}~f~f')$, which returns the composition $\lambda x.f(f'(x))$, (iii) $(\text{is}_{z'}~D)$, which returns $\mathbbm{1}(z=z')$, where $\llbracket D\rrbracket=(b,z)$ is a detection and $z'\in\mathcal{Z}$ is an object category, and (iv) $(\textsf{distance}~D~D')$, which returns the distance $\|b-b'\|_{\infty}$ between two detections $\llbracket D\rrbracket=(b,z)$ and $\llbracket D'\rrbracket=(b',z')$. Their abstract semantics are straightforward: for $\llbracket\cdot\rrbracket^\#$, they each evaluate each of their arguments once, and for $\llbracket\cdot\rrbracket^{\textsf{err}}$, the only one that propagates errors is \textsf{distance}, for which
\begin{align*}
\llbracket(\textsf{distance}~D~D')\rrbracket^{\textsf{err}}=\llbracket D\rrbracket^{\textsf{err}}+\llbracket D'\rrbracket^{\textsf{err}}.
\end{align*}

\paragraph{Results.}

We provide results in Table~\ref{tab:cocoresults}. The trends are similar to Section~\ref{sec:expsynthesis}; the main difference is that search does not help in this case, likely because there is only a single machine learning component so optimizing the allocation does not significantly affect performance. Finally, we can chain these programs with a faster object detector to reduce running time; see Appendix~\ref{sec:appendixdetection}.

\begin{figure}
\tiny
\begin{minipage}{0.48\textwidth}
$\begin{array}{l}
\texttt{def is\_person(x, y\_true=None):} \\
\qquad\texttt{if \blue{1.0 - f(x) <= ??1} \green{\{y\_true\} [|, 0.05]}:} \\
\qquad\qquad\texttt{return True} \\
\qquad\texttt{else:} \\
\qquad\qquad\texttt{return False}  \vspace{5pt} \\
\texttt{def is\_person\_fast(x):} \\
\qquad\texttt{if \blue{1.0 - f\_fast(x) <= ??2} \green{\{is\_person(x)\} [|, 0.05]}:} \\
\qquad\qquad\texttt{return is\_person(x)} \\
\qquad\texttt{else:}\\
\qquad\qquad\texttt{return False}
\end{array}$
\end{minipage}
\begin{minipage}{0.48\textwidth}
$\begin{array}{l}
\texttt{def monitor\_correctness(x):} \\
\qquad\texttt{if np.random.uniform() <= 0.99:} \\
\qquad\qquad\texttt{return} \\
\qquad\texttt{passert \blue{1.0 - f\_fast(x) <= ??2} \green{\{is\_person(x)\} [|, 0.05]}} \vspace{5pt} \\
\texttt{def monitor\_speed(x):} \\
\qquad\texttt{passert \blue{1.0 - f\_fast(x) > ??2} \green{\{true\} [|, ??3]}}
\end{array}$
\end{minipage}
\caption{A program used to predict whether an image $x$ contains a person. Specifications are shown in green; curly brackets is the specification and square brackets is the value of $\epsilon$. The corresponding inequality with a hole in blue. Holes with the same number are filled with the same value.}
\label{fig:example}
\end{figure}

\begin{figure*}
\centering
\begin{tabular}{ccc}
\includegraphics[width=0.3\textwidth]{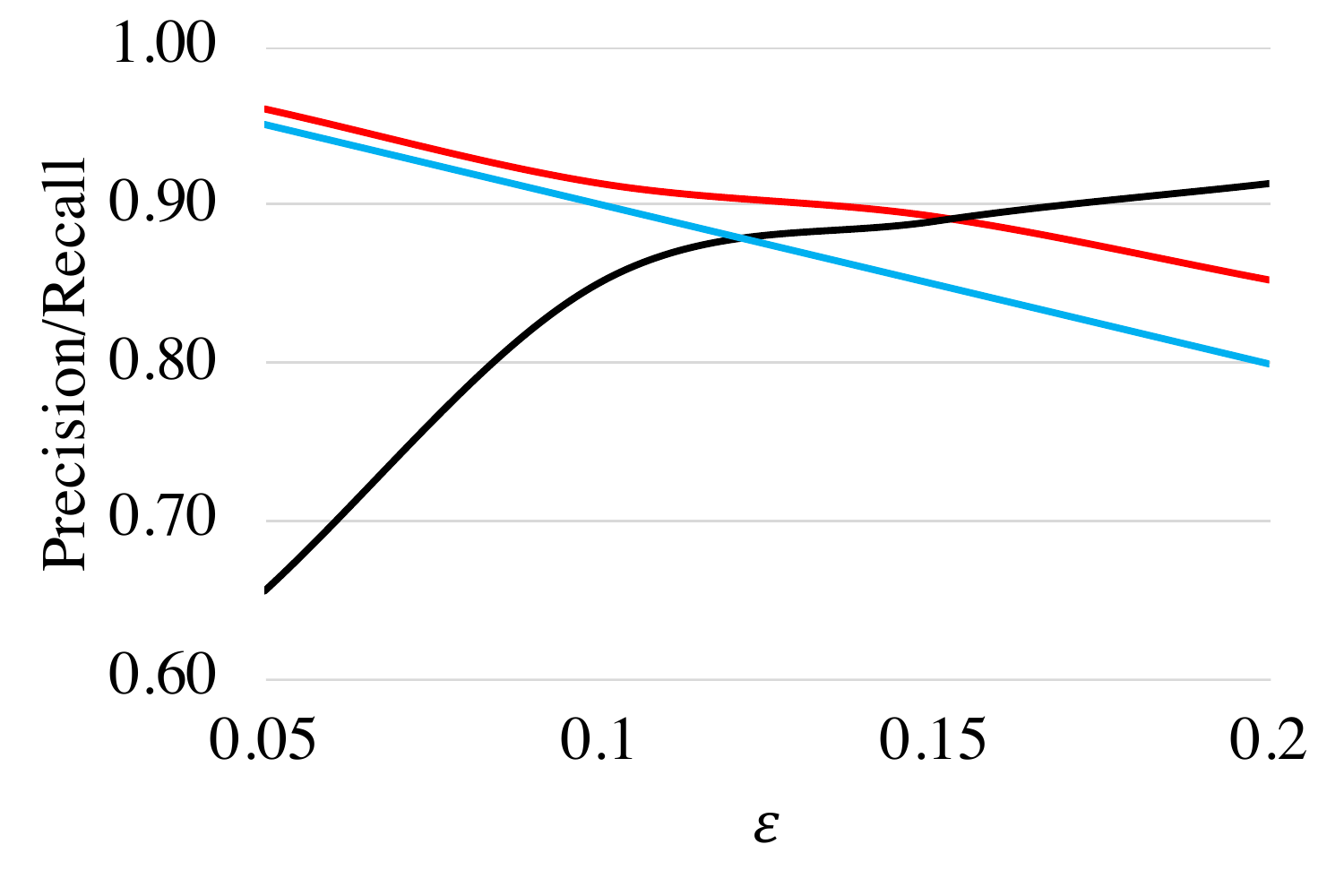} &
\includegraphics[width=0.3\textwidth]{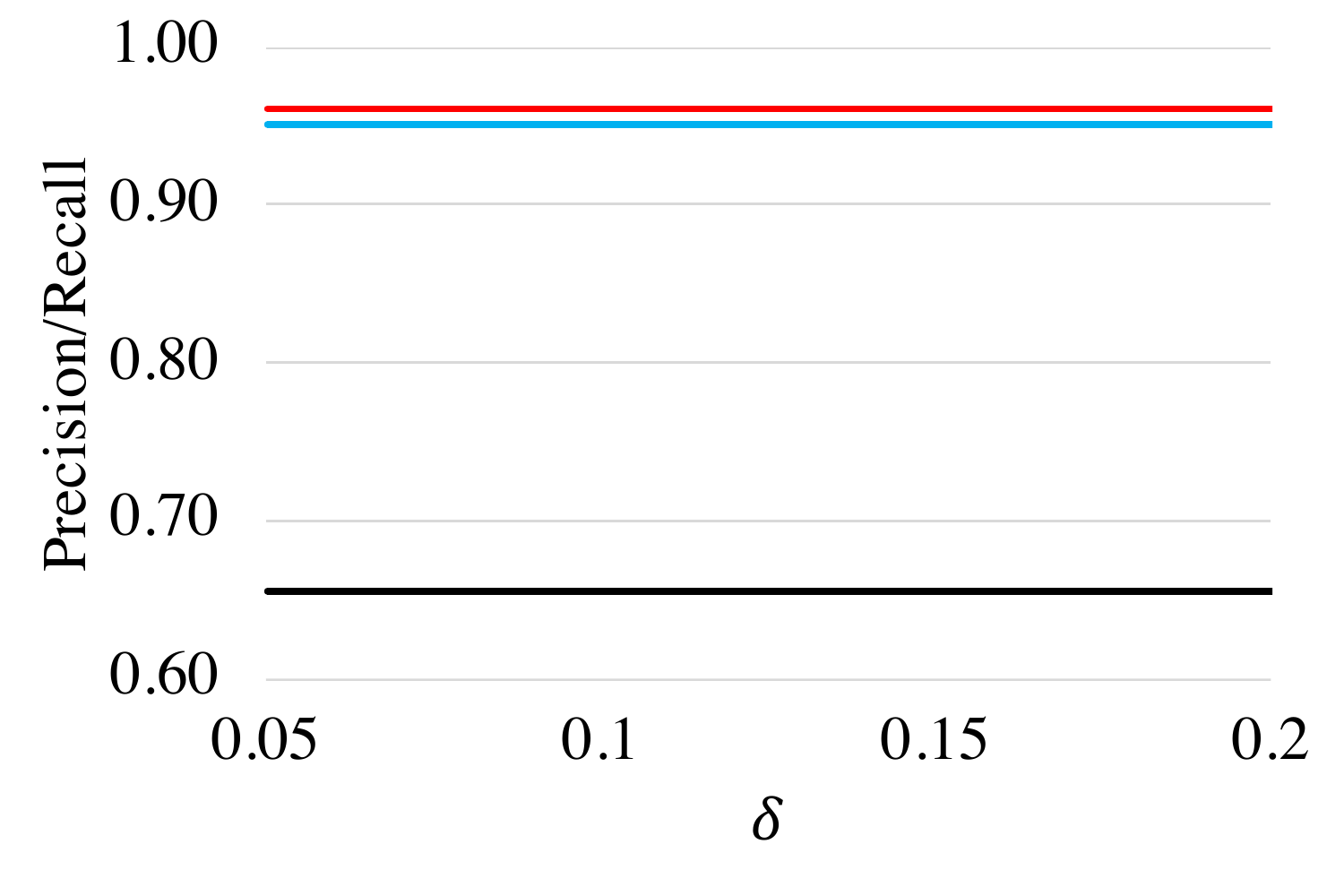} &
\includegraphics[width=0.3\textwidth]{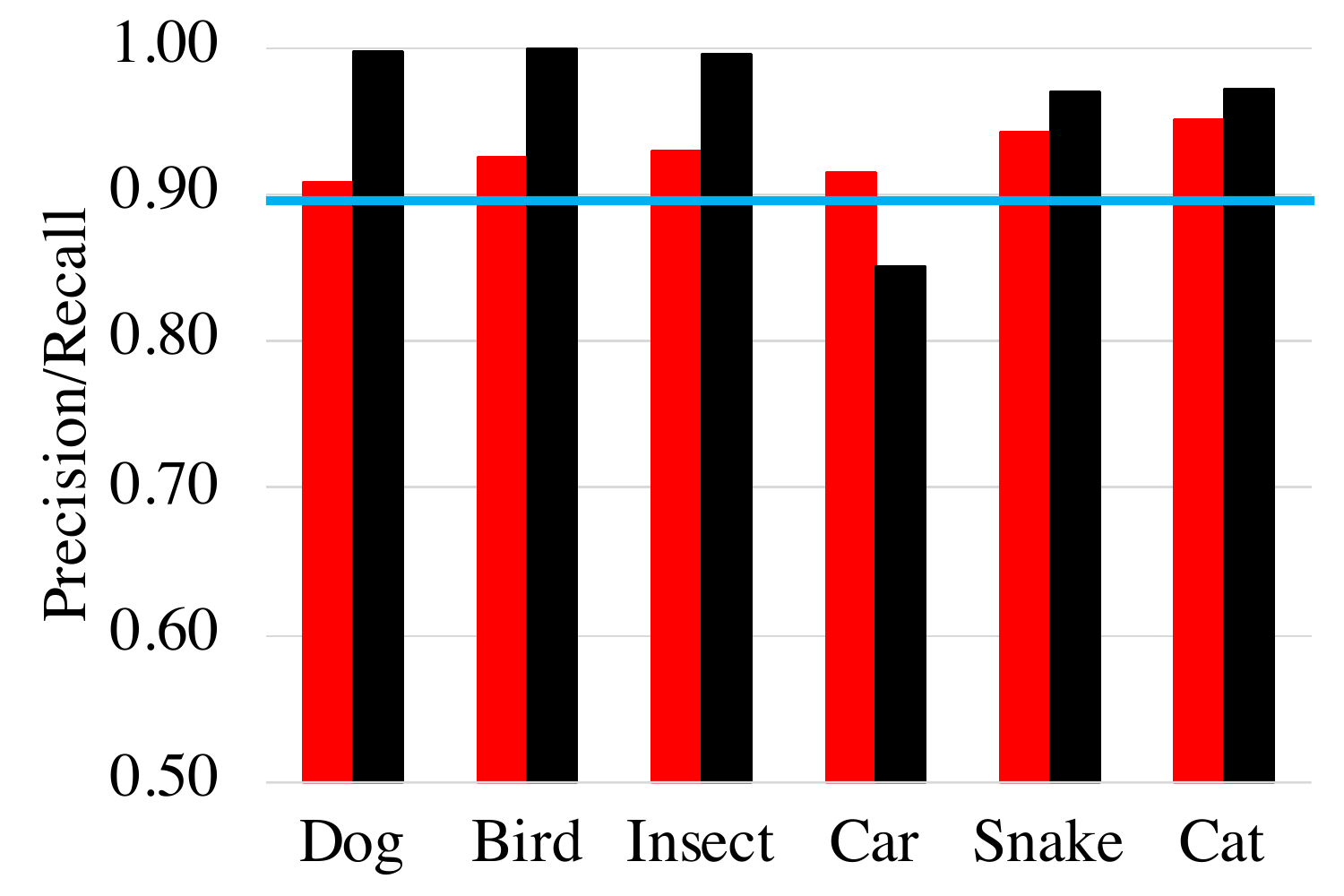}
\\
(a) & (b) & (c) \\
\includegraphics[width=0.3\textwidth]{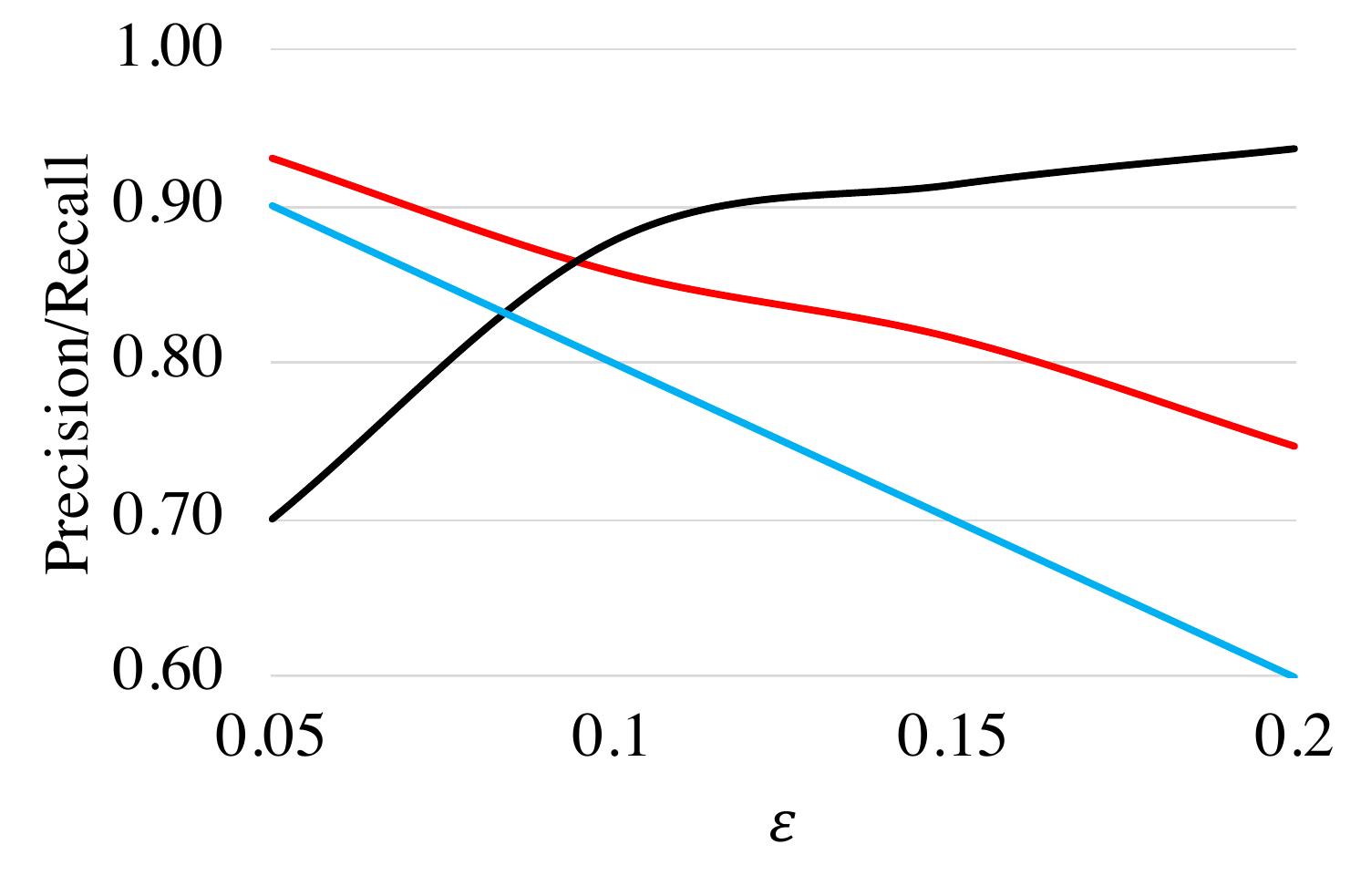} &
\includegraphics[width=0.3\textwidth]{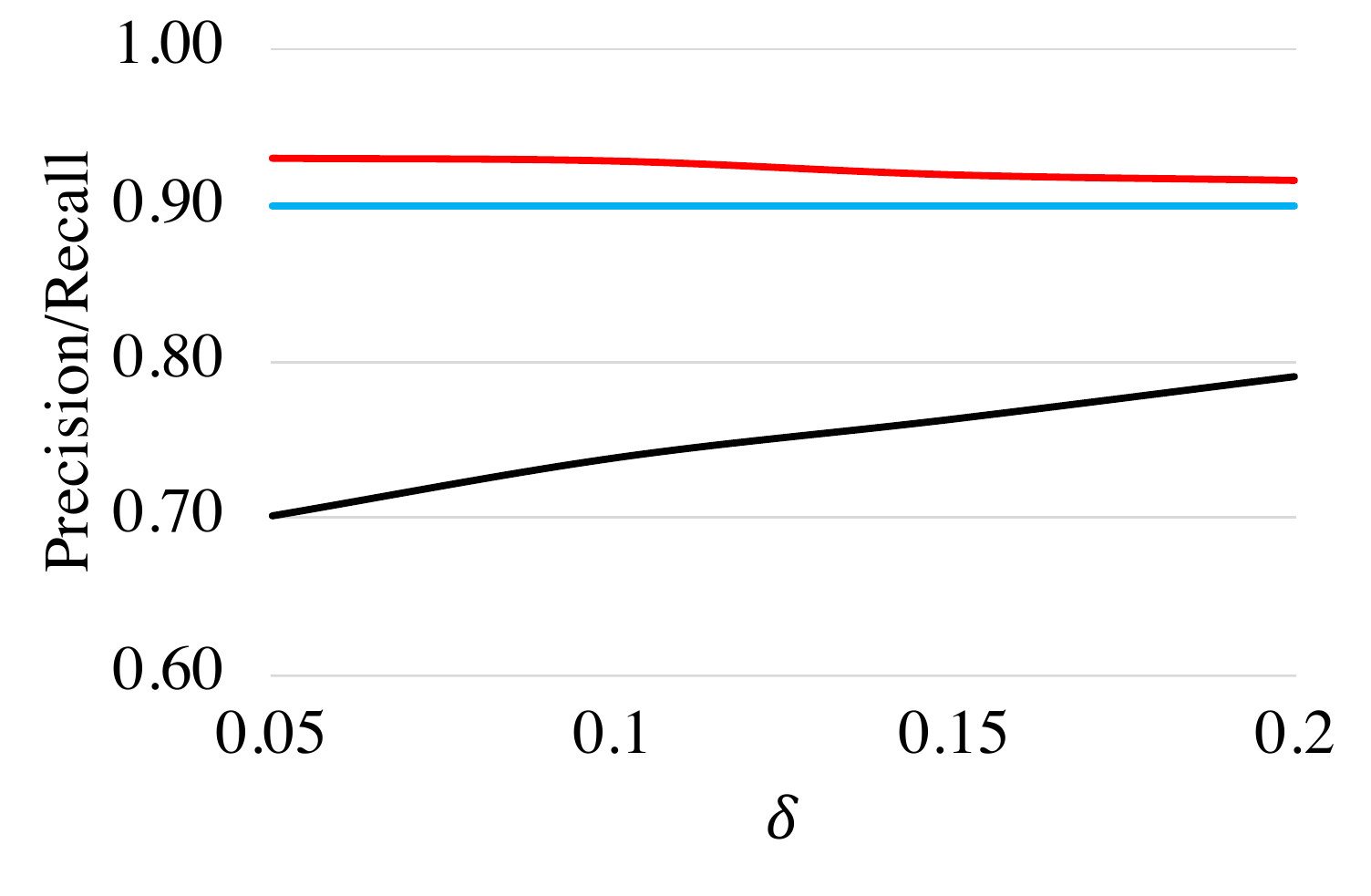} &
\includegraphics[width=0.3\textwidth]{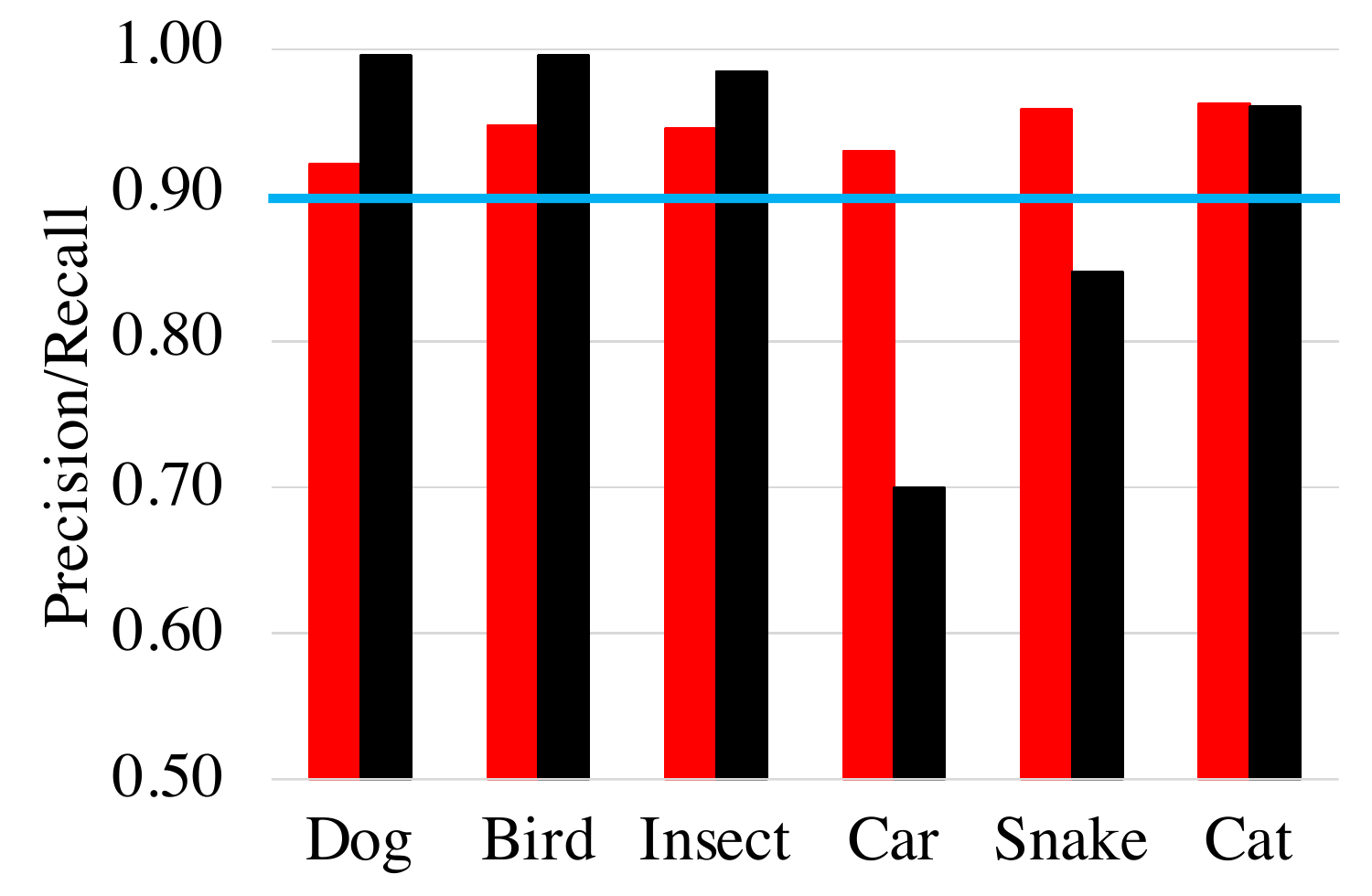} \\
(d) & (e) & (f) \\
\includegraphics[width=0.3\textwidth]{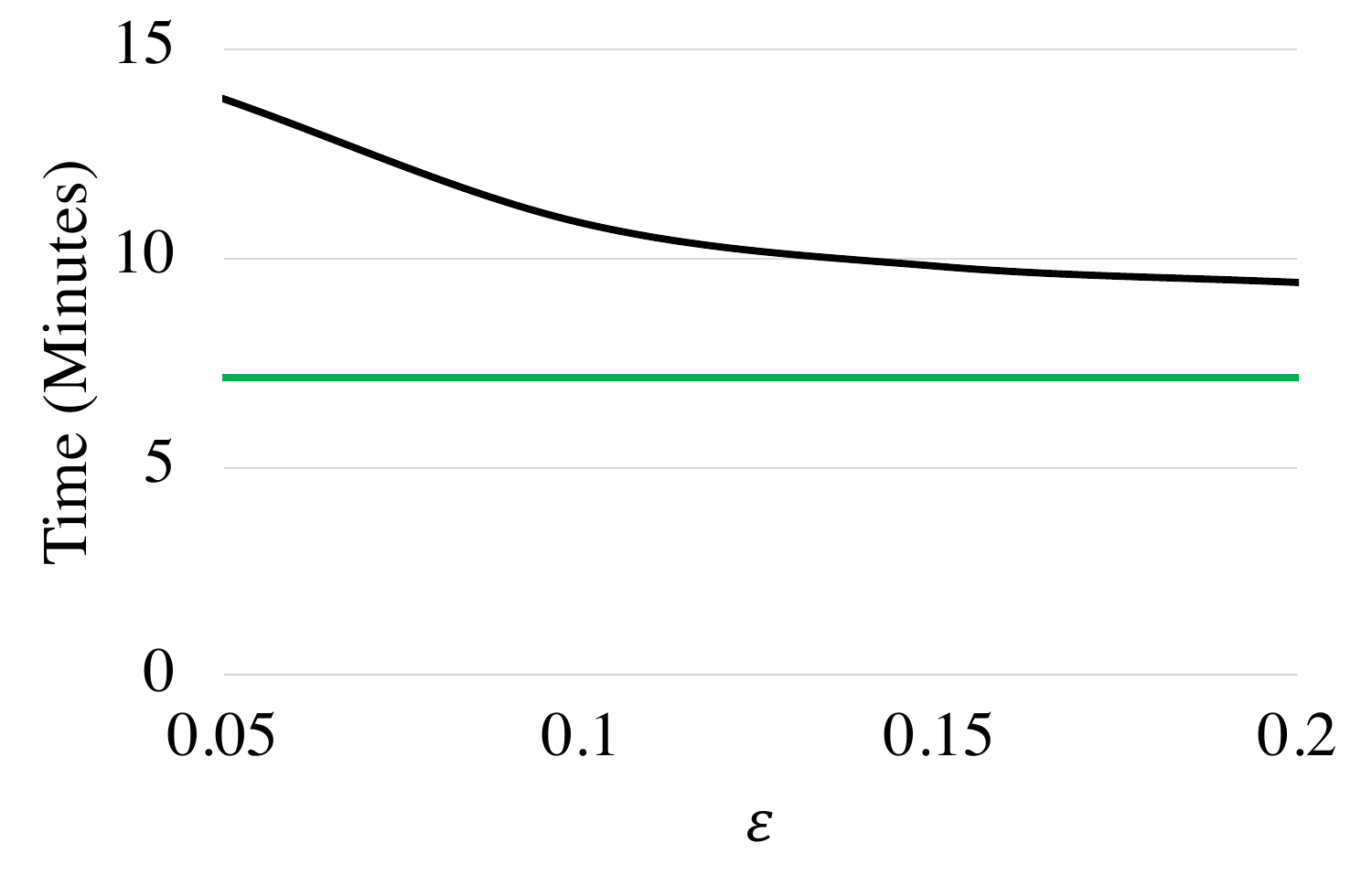} &
\includegraphics[width=0.3\textwidth]{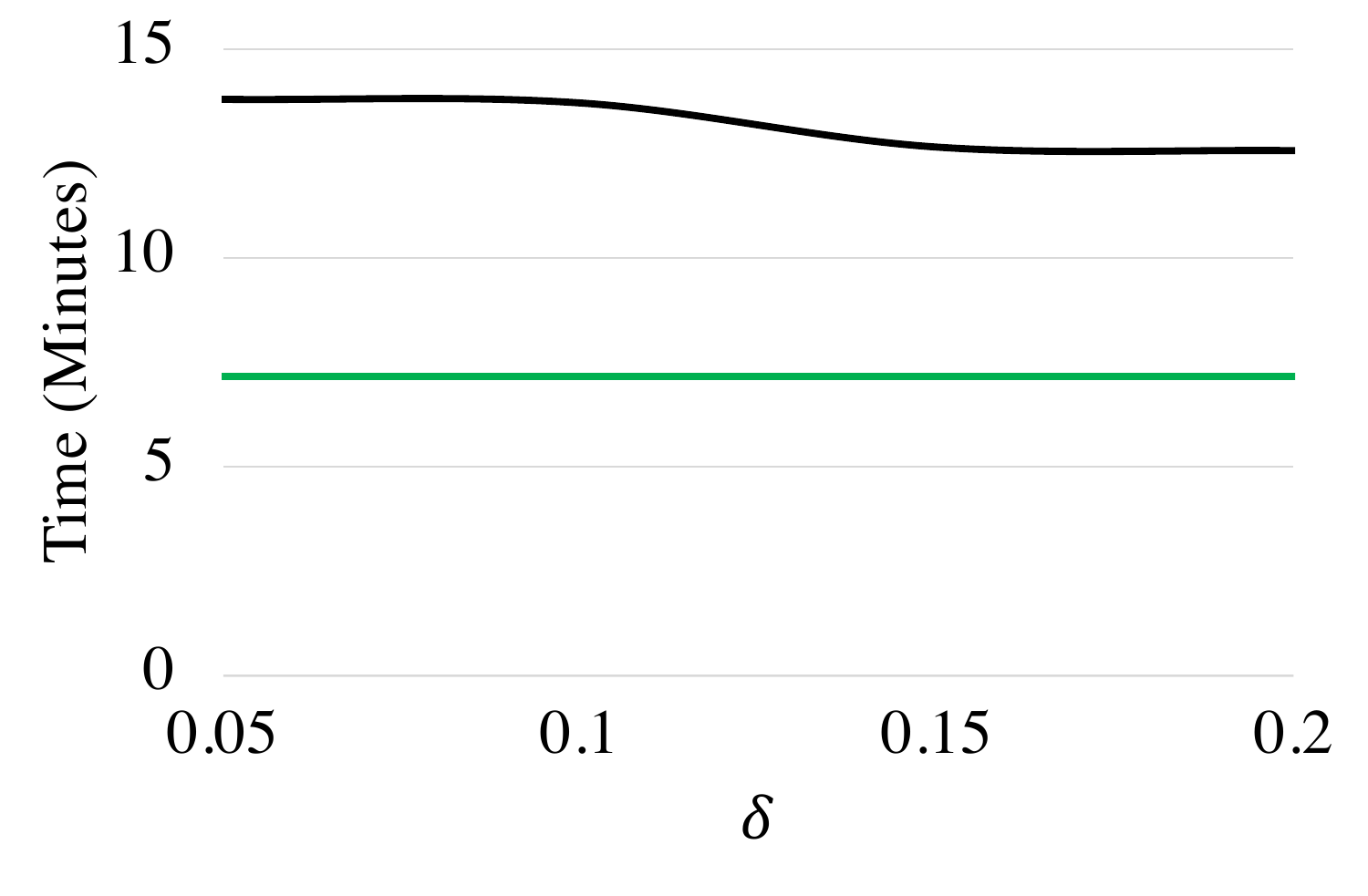} &
\includegraphics[width=0.3\textwidth]{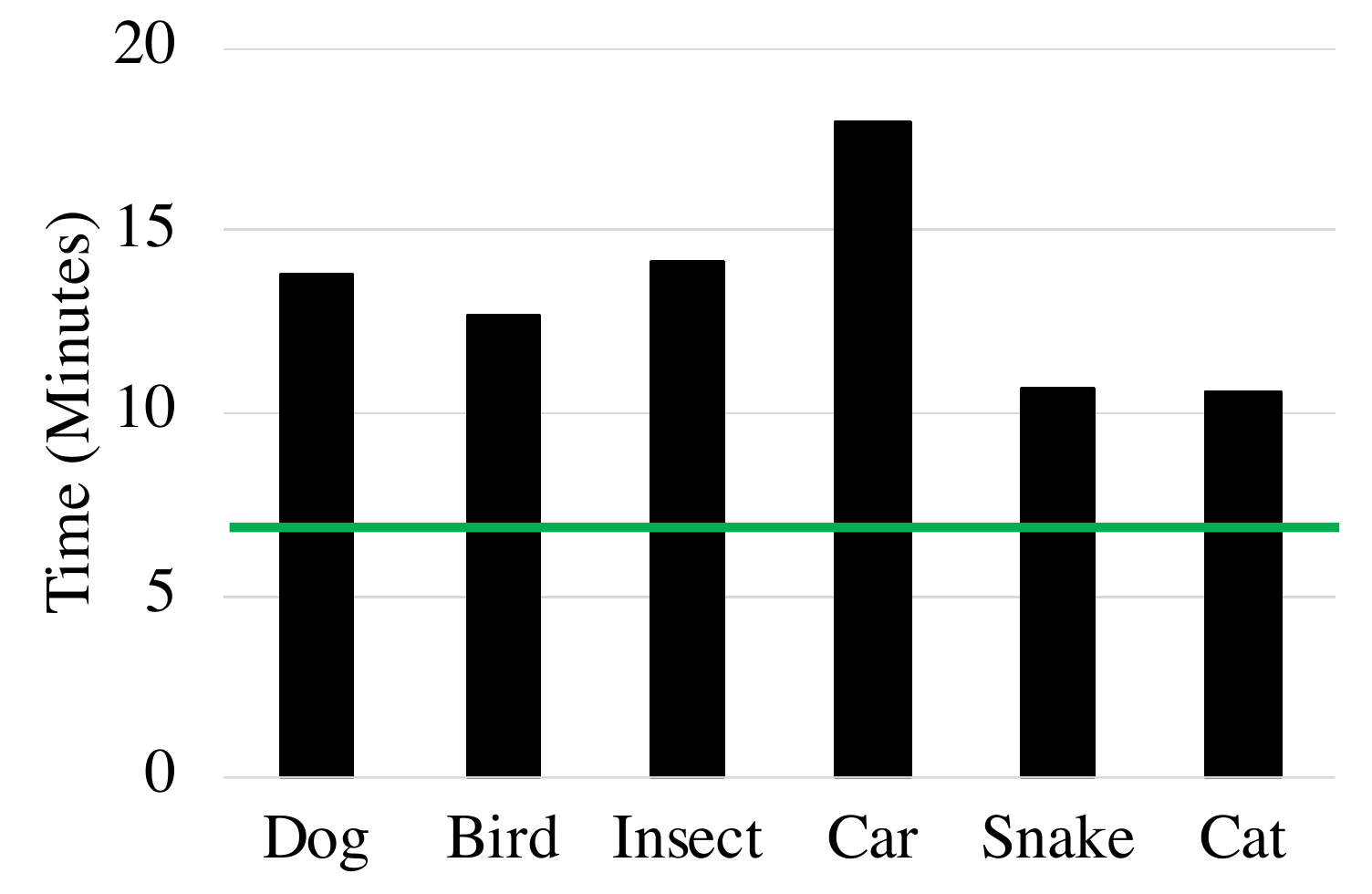} \\
(g) & (h) & (i) \\
& \includegraphics[width=0.3\textwidth]{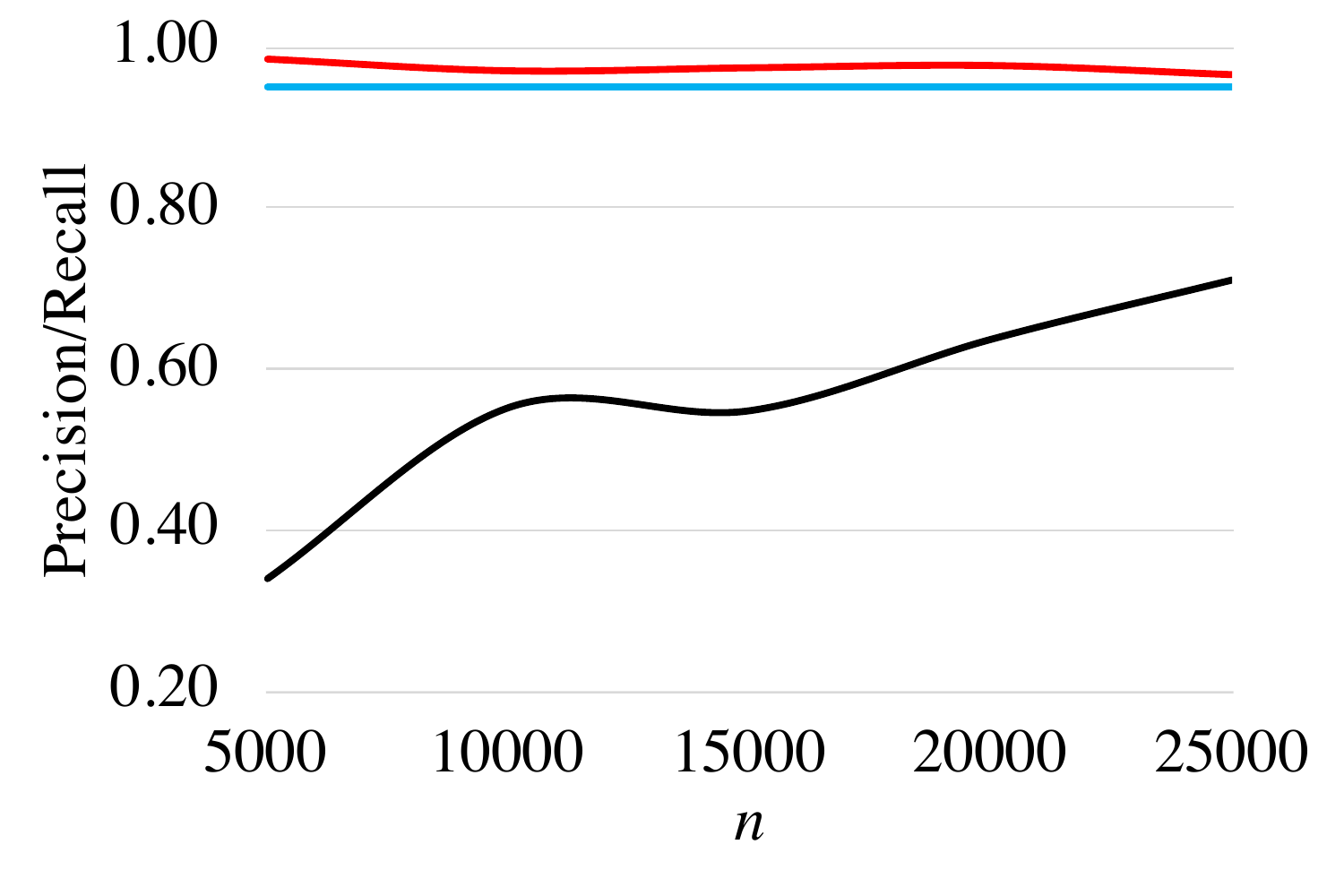} & \\
& (j) &
\end{tabular}
\caption{For ResNet alone, we show the recall (red), desired lower bound on the recall (blue), and precision (black) as a function of (a) $\epsilon$, (b) $\delta$, (c) $y\in\mathcal{Y}$, and (j) the number of synthesis examples $n$; the defaults are $\epsilon=\delta=0.05$, $n=25,000$, and $y=\text{``car''}$, except in (c) we use $\epsilon=0.1$ to facilitate the comparison with (f). We show the same values for ResNet$+$AlexNet as a function of (d) $\epsilon$, (e) $\delta$, and (f) $y\in\mathcal{Y}$. For ResNet$+$AlexNet (black) compared to AlexNet alone (green), we show the running time as a function of (g) $\epsilon$, (h) $\delta$, and (i) $y\in\mathcal{Y}$; we omit ResNet alone since its running time (82.6 minutes) is significantly above the scale.}
\label{fig:expclassification}
\end{figure*}

\subsection{Case Study 1: ImageNet Image Classification}
\label{sec:evalclassification}

\paragraph{Correctness.}

Consider program shown in Figure~\ref{fig:example}, which classifies images as ``person'' (returns \textsf{true}) or ``not person'' (returns \textsf{false}). The function \texttt{is\_person} takes as input an image $x$, and optionally the ground truth label $y^*$ (which is only used during sketching). The specification in \texttt{is\_person} says that the program should return \textsf{true} with high probability if the image is of a person (i.e., $y^*=1$). The predicate $\mathbbm{1}(1-f(x)\le c)$ is shown in blue, where the value of $c$ has been left as a hole \texttt{??1}, the specification $y^*=1$ is shown in green in the curly braces, and the value $\epsilon=0.05$ is shown in green in the square braces. We perform a case study in the context of this program (though for labels other than ``person''). We consider the ImageNet dataset~\cite{deng2009imagenet}, a large image classification benchmark with over one million images in 1000 categories, including various different animals and inanimate objects. We consider the ResNet-152 DNN architecture~\cite{he2016deep}, a state-of-the-art image classification model trained on ImageNet that achieves about 88\% accuracy overall. For both architectures, we use the implementation in PyTorch~\cite{paszke2017automatic}.

To use our system, we split the ImageNet validation set consisting of 50,000 held-out images into (at most) 25,000 for synthesis (i.e., the \emph{synthesis set}) and 25,000 for validation. Because ImageNet has so many labels, each object category has very few examples in the validation dataset (50 on average). Thus, we group the labels into larger, coarse-grained categories, focusing on ones that correspond to many fine-grained ImageNet labels. We consider ``dog'' (130 labels, 6,500 images) ``bird'' (59 labels, 2,950 images), ``insect'' (27 labels, 1,350 images), ``car'' (21 labels, 1050 images), ``snake'' (17 labels, 850 images), and ``cat'' (13 labels, 650 images). The default one we use is ``car''; this category contains vehicles such as passenger cars, bikes, busses, trolleys, etc. For the scoring function, given a coarse-grained category $Y\subseteq\mathcal{Y}$, we use the sum of the fine-grained label probabilities---i.e., $f(x)=\sum_{y\in Y}p(x,y)$, where $p(x,y)$ is the predicted probability of label $y$ according to ResNet-152.

Then, we use our sketching algorithm to synthesize $c$ to fill \texttt{??1}. We show results in the first and fourth rows of Figure~\ref{fig:expclassification}. Note that the red curves ideally equal the blue curves, but are slightly conservative to account for synthesis being based on finitely many samples. The value of $\epsilon$ has the biggest effect on performance, since it directly governs recall; as $\epsilon$ grows, recall drops (as desired) and precision substantially improves. In contrast, the performance does not vary significantly with $\delta$. These trends match sample complexity guarantees from learning theory relevant to our setting of $n=O(\log(1/\delta)/\epsilon)$ ~\cite{kearns1994introduction,vapnik2013nature}. Next, as $n$ grows larger, recall can more closely match the desired maximum, allowing precision to improve dramatically (the non-monotone effect is most likely due to random chance). Finally, the dependence on the target label is also governed by the number of synthesis images in each category.

\paragraph{Improving speed.}

Next, we describe how our framework can be used to compose $f$ with a second DNN $f_{\text{fast}}$, which is much faster than $f$ but has lower accuracy. Intuitively, we want to use $f_{\text{fast}}$ when we can guarantee its prediction is correct with high probability, and use $f$ otherwise. This approach has been used to reduce running time~\cite{teerapittayanon2016branchynet,bolukbasi2017adaptive}; our framework can be used to do so while providing rigorous accuracy guarantees.

The code for this approach is shown in \texttt{is\_person\_fast} in Figure~\ref{fig:example}. As before, the idea is to compute a threshold $c'$ such that the prediction $f_{\text{fast}}(x)\ge1-c'$ is correct with high probability. There are two differences. First, if we conclude that there might be a person in the image according to $f_{\text{fast}}$, then we return the prediction according to $f$ (instead of \textsf{true}). While $f_{\text{fast}}$ is guaranteed to detect 95\% images with people with high probability, it may have more false positives than $f$; calling $f$ after $f_{\text{fast}}$ reduces these false positives. Second, the correctness guarantee is with respect to the prediction $\hat{y}=\mathbbm{1}(f(x)\ge1-c)$ rather than $y^*$. We could use $y^*$, but there is no need---if $\hat{y}$ is incorrect, then it is not helpful for $f_{\text{fast}}$ to predict correctly since it falls back on $\hat{y}$.

For $f_{\text{fast}}$, we use AlexNet, which achieves about 57\% accuracy overall; in particular, we use $f_{\text{fast}}(x)=\sum_{y\in Y}p_{\text{fast}}(x,y)$, where $p_{\text{fast}}(x,y)$ is the predicted probability of label $y$ according to AlexNet. Then, we conclude that $x$ (may) have label $y$ if $f_{\text{fast}}(x)\ge1-c'$, where $c'$ is synthesized by our algorithm. We obtain results on an Nvidia GeForce RTX 2080 Ti GPU. We show results on the second and third rows of Figure~\ref{fig:expclassification}. All results shown are for the combined predictions (i.e., using both AlexNet and ResNet), and are estimated on the validation set. For running time, we omit results for ResNet since its running time is $82.6$ minutes, which is more than $4\times$ the running time of our combined model. For the ``dog'' category, our approach reduces running time $6\times$ from 82.6 minutes to 13.8 minutes without any sacrifice in precision or recall.

Thus, our approach significantly reduces running time while achieving the desired error rate. Furthermore, comparing to Figure~\ref{fig:expclassification} (d), the precision does not significantly decrease across most labels. It does suffer for the labels ``car'' and ``snake''. Intuitively, for these labels, there are relatively few examples in the synthesis set, so the synthesis algorithm needs to choose more conservative thresholds. Since the fast program has two thresholds whereas the original program only has one, it is more conservative in the latter case. This difference is reflected in the fact that Figure~\ref{fig:expclassification} (e) has higher recall than (d), especially for ``car'' and ``snake''.

Importantly, these results rely on the fact that we are tailoring our predictions to a single category---i.e., our system enables the user to tailor the predictions of pretrained DNN models such as ResNet and AlexNet to their desired task. For instance, it can focus on predicting cars rather than achieving good performance on all 1000 ImageNet categories.

\paragraph{Runtime monitoring.}

As described in Appendix~\ref{sec:monitoring}, our framework can monitor the synthesized program at runtime, which is useful since PAC guarantees are specific to the data distribution $p(x,y^*)$. Thus, if the program is executed on data from a different distribution, called \emph{distribution shift}~\cite{ben2007analysis,quionero2009dataset}, then our guarantees may not hold. Monitoring requires us to obtain ground truth labels $y^*$ for inputs $x$ encountered at run time; then, we use these ground truth labels to estimate the failure rate of the model and ensure it is below the desired value $\epsilon$.

We show how we can monitor the correctness of \texttt{is\_person\_fast}. In this case, we can easily obtain ground truth labels since the specification for \texttt{??2} can be obtained by evaluating $f(x)$. We want to avoid running $f$ on every input since this would defeat the purpose of using a fast DNN; instead, we might run it once every $N$ iterations for some large $N$. The function \texttt{monitor\_correctness} implements this check, generating a ground truth label once every $N=100$ iterations on average. Note that we formulate the check as a probabilistic assertion~\cite{sampson2014expressing}---i.e.,
\begin{align*}
\textsf{passert}~~1-f_{\text{fast}}(x)\le c'~~\{1-f(x)\le c\}_{0.05}^\mid,
\end{align*}
which has the semantics
\begin{align*}
\mathbb{P}_{p(x,y^*)}\big(1-f_{\text{fast}}(x)\le c'\mid1-f(x)\le c\big),
\end{align*}
which is the specification in \texttt{is\_person\_fast}. When our framework synthesizes a value $c'$ to fill \texttt{??2} in \texttt{is\_person\_fast}, it uses the same value to fill \texttt{??2} in \texttt{monitor\_correctness}. Then, at run time, it accumulates pairs $(x,\hat{y})$, where $\hat{y}=\mathbbm{1}(f(x)\le c)$, in calls to \texttt{monitor\_correctness} and uses them to check whether the probabilistic assertion in that function is true.

To evaluate whether monitoring can detect shifts, we select two subsets of the ``car'' category: (i) bikes, including motor bikes, and (ii) passenger cars, excluding busses, trucks, etc., with 6 fine-grained labels each. Then, we consider a shift from the car category to the bike category---i.e., if we imagine that bikes were instead labeled as cars, would the recall of our program continue to be above the desired threshold. First, we check whether it proves correctness when the data distribution does \emph{not} shift---i.e., using the test images labeled ``passenger car''. We run our verification algorithm on this property using the test set images labeled  As expected, our verification algorithm correctly concludes that both the recall and the running time are within the expected bounds. Then, we check whether it proves correctness when the data distribution shifts---i.e., using the test images labeled ``bike''. In this case, our verification algorithm concludes that recall is incorrect, but running time is correct. Indeed, the average running time is now lower---intuitively, $f_{\text{fast}}$ is incorrectly rejecting many ``car'' images, which reduces recall (undesired) as well as running time (desired).

As a side note, our framework can also be used to monitor quantitative properties. For instance, we can keep monitor how frequently the branch $f_{\text{fast}}(x)>c$ is taken---i.e., avoiding the need to evaluate $f(x)$. In Figure~\ref{fig:example}, \texttt{monitor\_running\_time} includes a probabilistic assertion
\begin{align*}
\textsf{passert}~~1-f_{\text{fast}}(x)>c'~~\{\text{true}\}_{\epsilon}^\mid
\end{align*}
to perform this check. This assertion says that $1-f_{\text{fast}}(x)>c'$ with probability at least $1-\epsilon$---i.e., the faster branch in \texttt{is\_person\_fast} should be taken at least $1-\epsilon$ fraction of the time according to $p(x,y^*)$. We might not know what is a reasonable value of $\epsilon$---i.e., the rate at which $f_{\text{fast}}$ predicts there is a person in the image. Thus, we leave it as a hole \texttt{??3}. Given training examples $\vec{z}$, our framework can be used to synthesize a value of $\epsilon$ to fill this hole.

\subsection{Case Study 2: Precision Medicine}

\begin{figure}
\centering\tiny
\begin{minipage}{0.52\textwidth}
$\begin{array}{l}
\texttt{def predict\_warfarin\_dose(x, y\_true=None):} \\
\qquad\texttt{y = argmax([(ys, f(x, y)) for y in [`low', `med', `high']])} \\
\qquad\texttt{if y == `low' and \blue{f(x, `low') >= ??1} \green{\{y\_true != `high'\} [|, 0.05]}:} \\
\qquad\qquad\texttt{return `low'} \\
\qquad\texttt{if y == `high' and \blue{f(x, `high') >= ??2} \green{\{y\_true != `low'\} [|, 0.05]}:} \\
\qquad\qquad\texttt{return `high'} \\
\qquad\texttt{return `med'}
\end{array}$
\end{minipage}
\begin{minipage}{0.47\textwidth}
$\begin{array}{l}
\texttt{def monitor\_correctness(x):} \\
\qquad\texttt{y = argmax([(ys, f(x, y)) for y in [`low', `med', `high']])} \\
\qquad\texttt{y\_true = obtain\_result(x)} \\
\qquad\texttt{if y == `low':} \\
\qquad\qquad\texttt{passert \blue{f(x, `low') >= ??1} \green{\{y\_true != `high'\} [|, 0.05]}} \\
\qquad\texttt{if y == `high':} \\
\qquad\qquad\texttt{passert \blue{f(x, `high') >= ??2} \green{\{y\_true != `low'\} [|, 0.05]}}
\end{array}$
\end{minipage}
\caption{A program that predicts the Warfarin dose for a patient with covariates $x$. Specifications are shown in green; curly brackets is the specification and square brackets is the value of $\epsilon$. The corresponding inequality with a hole in blue. Holes with the same number are filled with the same value.}
\label{fig:warfarinprog}
\end{figure}

\begin{figure*}
\centering
\begin{tabular}{ccc}
\includegraphics[width=0.3\textwidth]{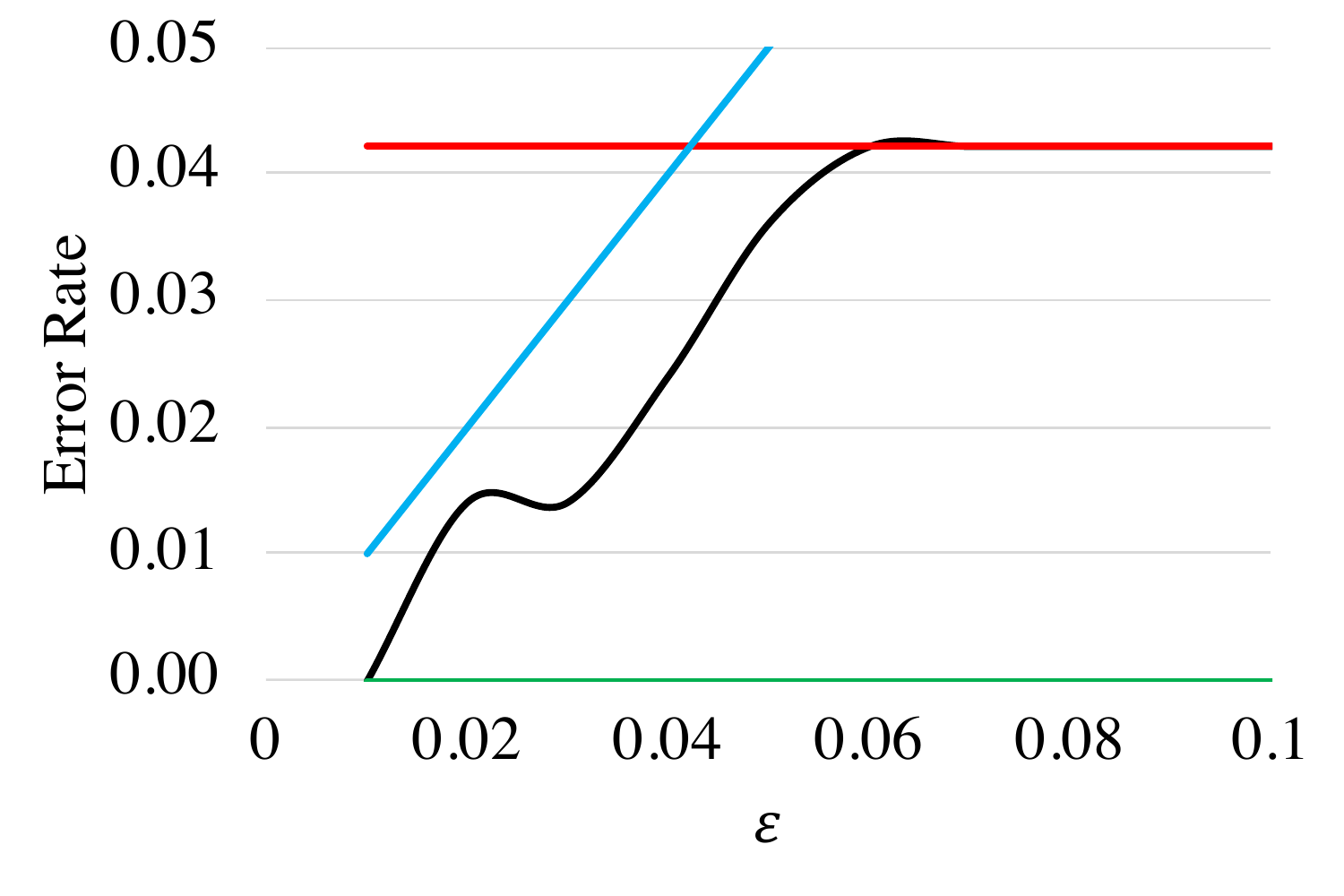} &
\includegraphics[width=0.3\textwidth]{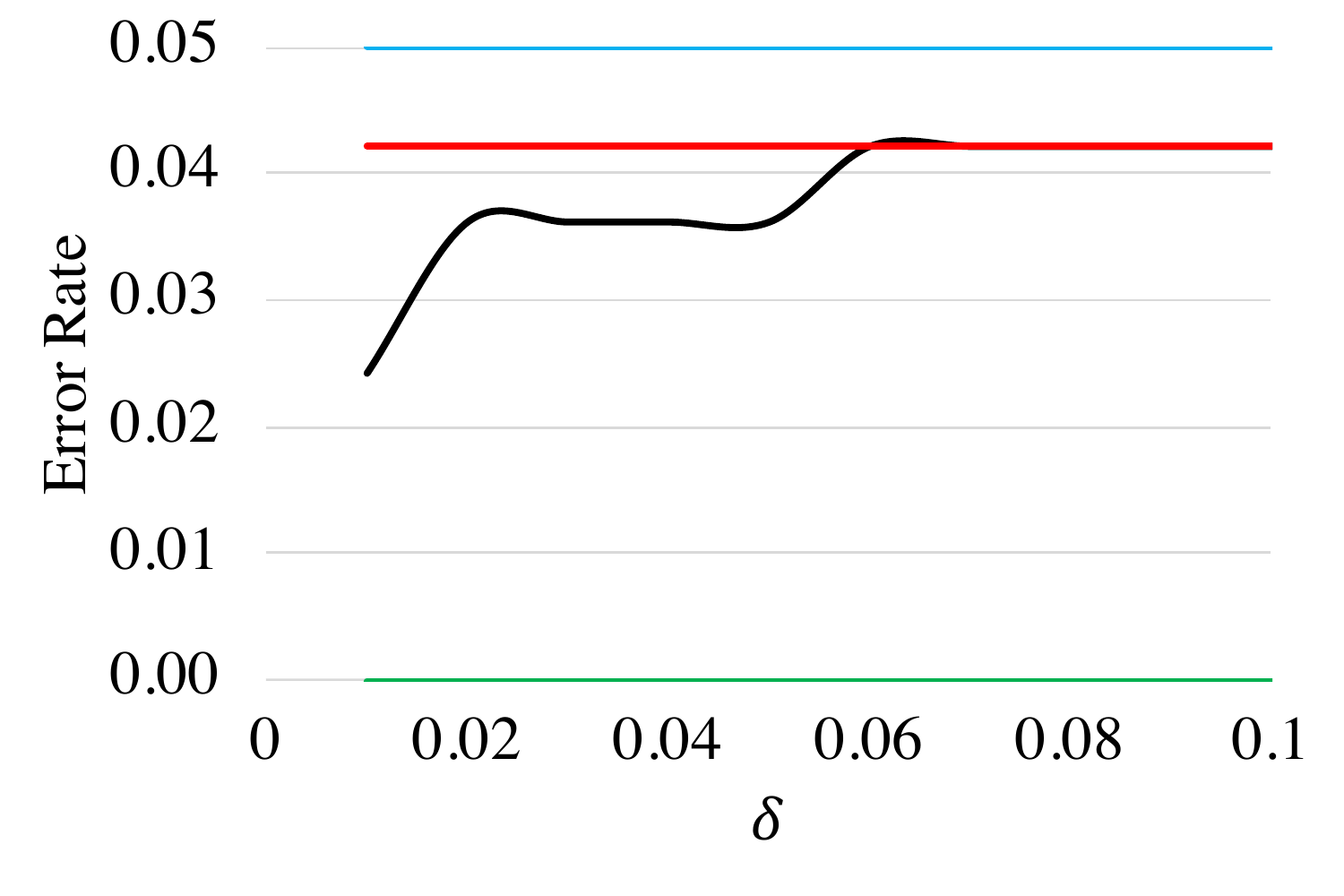} &
\includegraphics[width=0.3\textwidth]{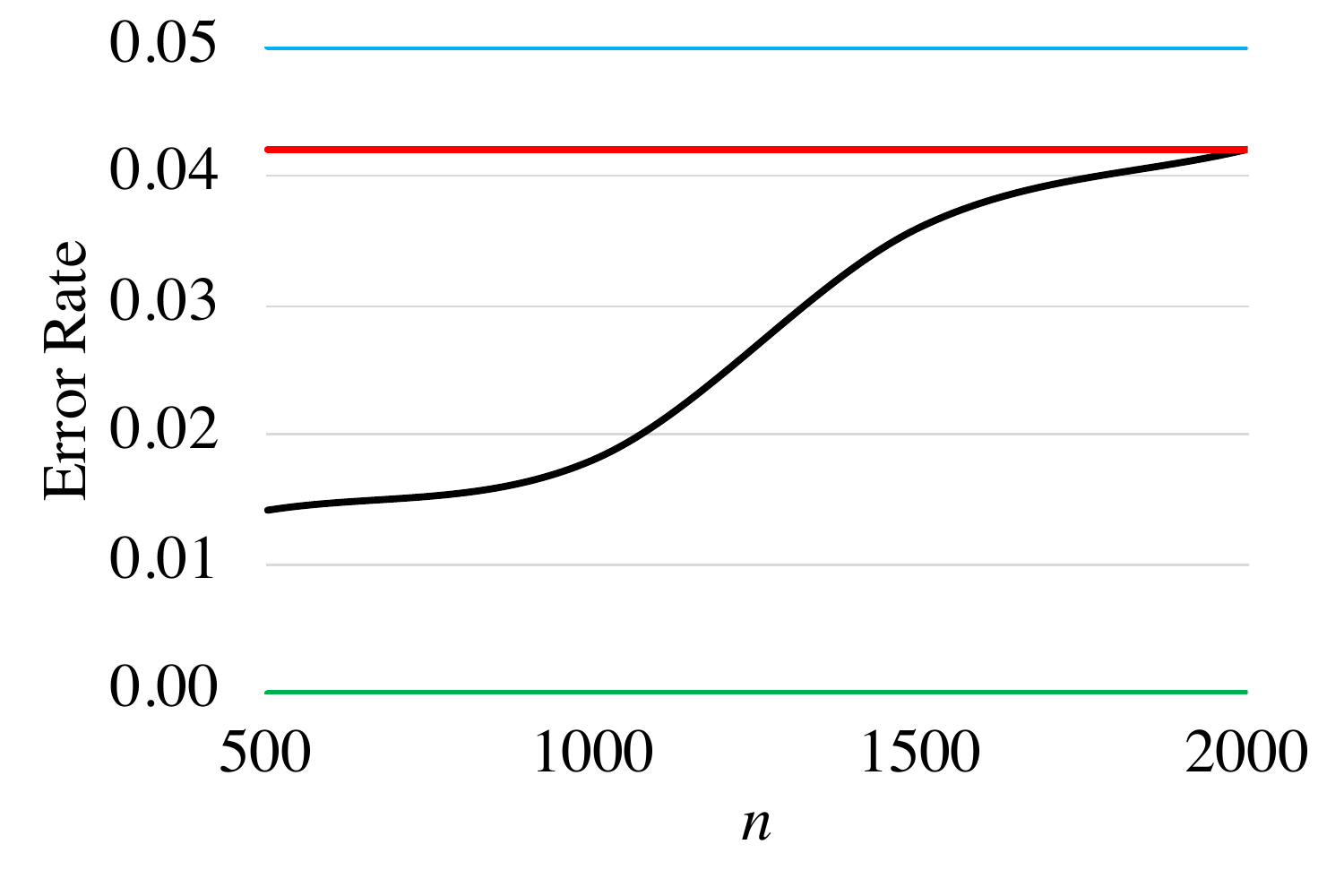} \\
(a) & (b) & (c) \\
\includegraphics[width=0.3\textwidth]{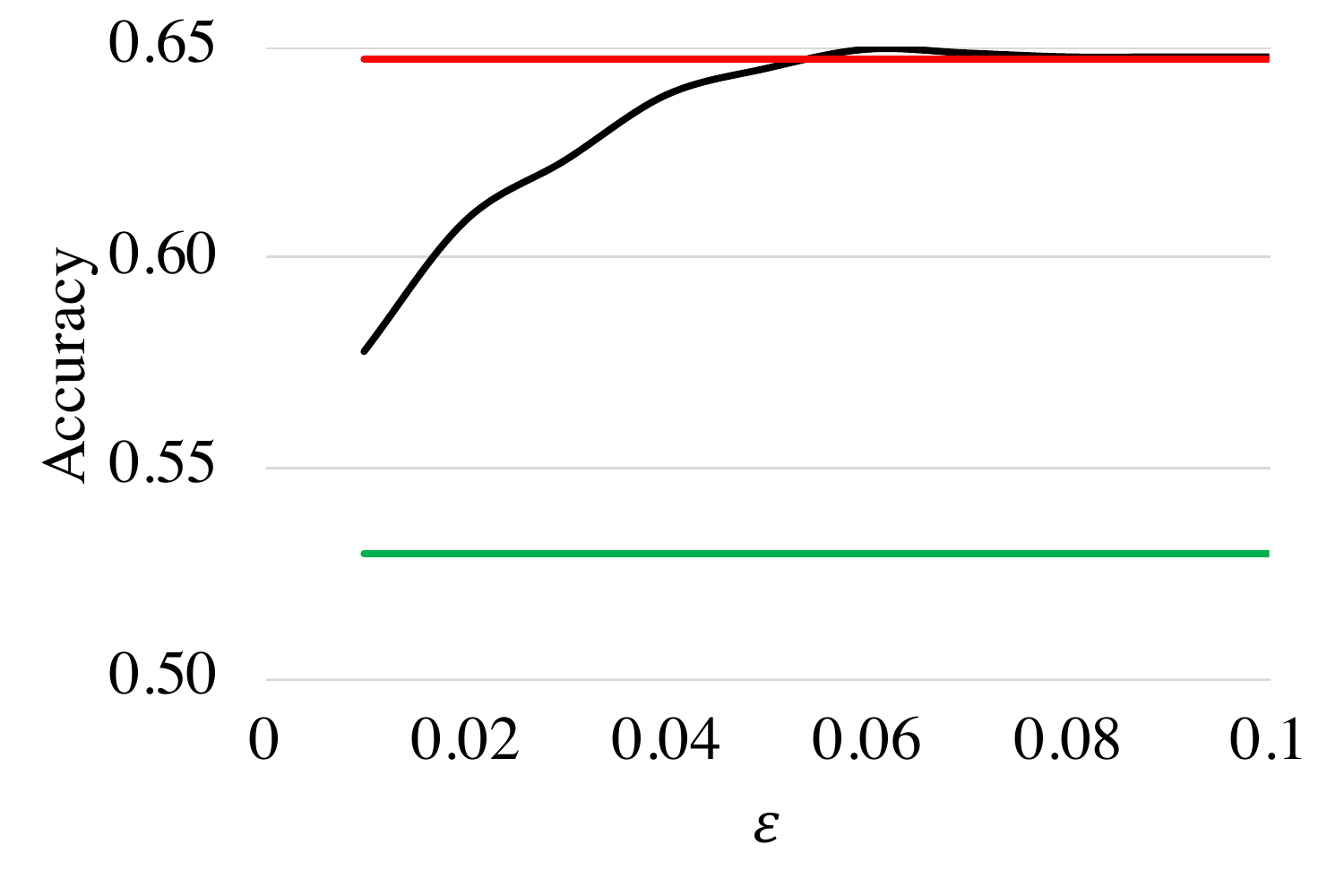} &
\includegraphics[width=0.3\textwidth]{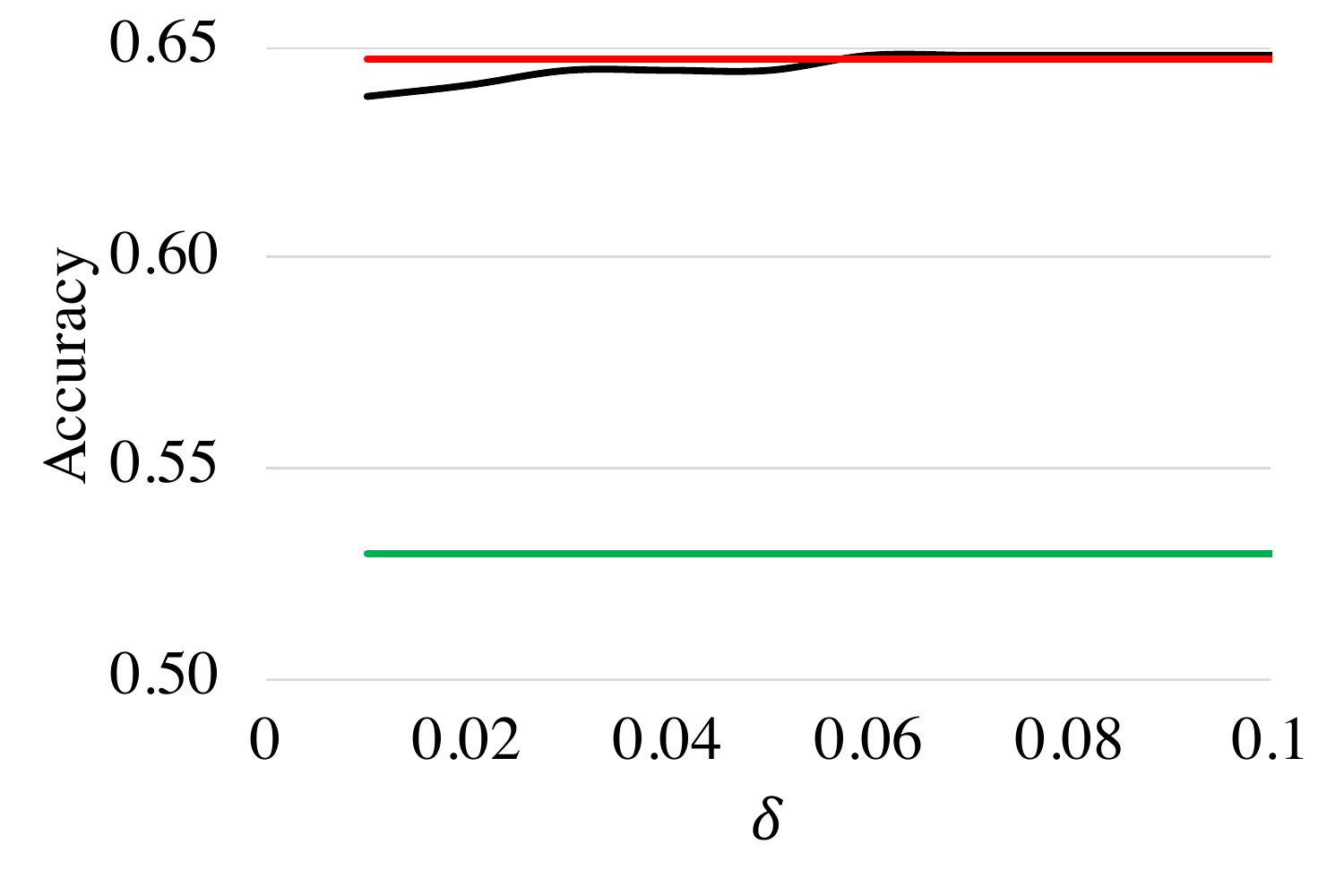} &
\includegraphics[width=0.3\textwidth]{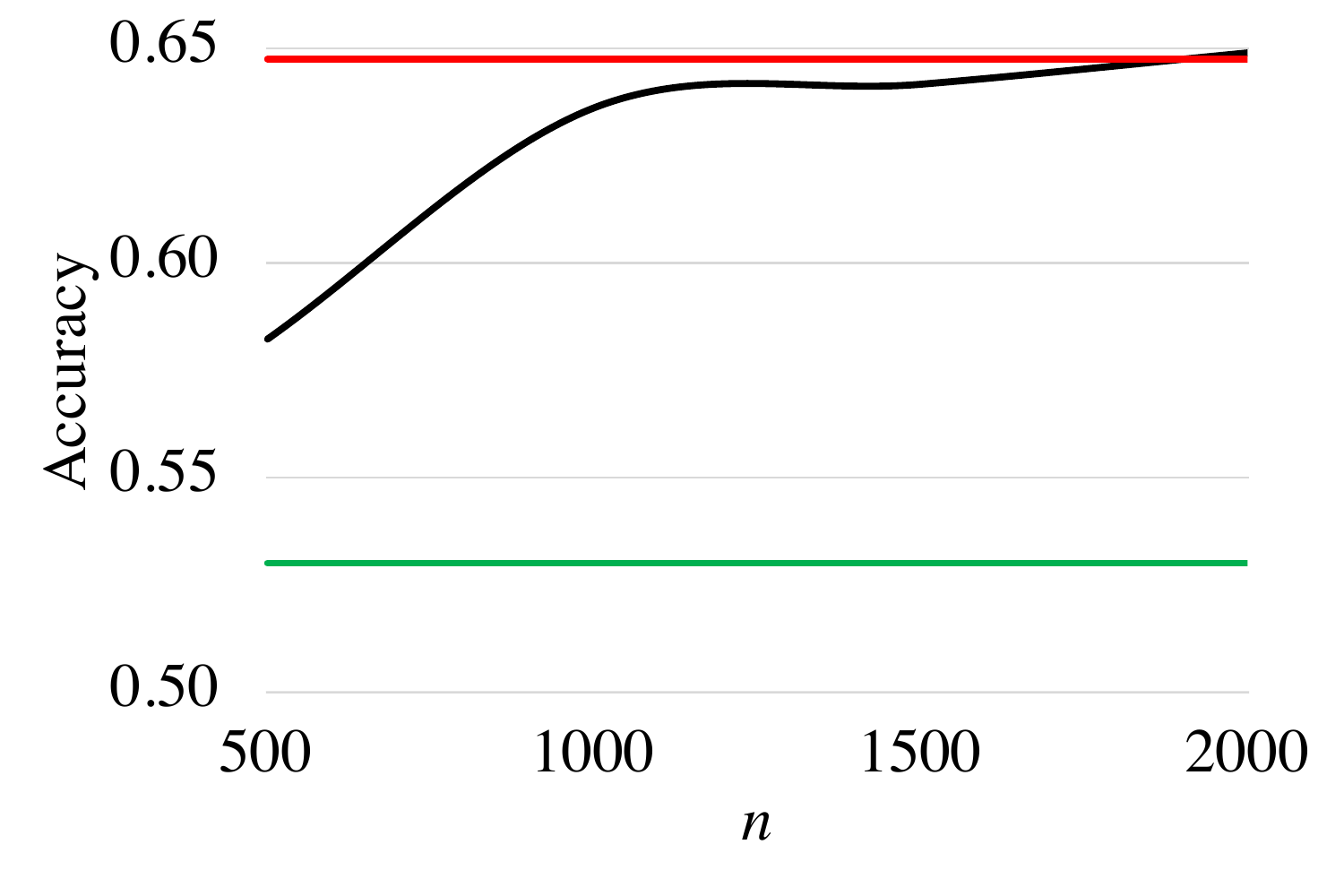} \\
(d) & (e) & (f)
\end{tabular}
\caption{We show the error rate (top) and accuracy (bottom) for our program (black), the random forest (red), always predicting ``medium'' (green) as a function of (a,d) $\epsilon$, (b,e) $\delta$, and (c,f) the number of synthesis examples $n$; for the top plots, we also show the desired upper bound on the error rate (blue).}
\label{fig:expwarfarin}
\end{figure*}

\paragraph{Warfarin dosing task.}

Next, we consider a task from precision medicine. In particular, we consider a random forest trained to predict dosing level for the Warfarin drug based on individual covariates such as genetic biomarkers~\cite{kimmel2013pharmacogenetic}. Personalized dosing can improve patient outcomes, but significant errors can lead to adverse events if not quickly corrected. The ideal dosage is a real-valued label. The goal is to train a model to predict this dosage as a decision support tool for physicians. For simplicity, we build on an approach that converts the problem into a classification problem by discretizing this value into labels $\mathcal{Y}=\{\text{high},\text{medium},\text{low}\}$ dose~\cite{bastani2015online}. Then, the goal is to maximize accuracy while ensuring that very few patients for whom a high dose is predicted but should have been assigned a low dose, and vice versa.

\paragraph{Experimental setup.}

We split the dataset (5,528 examples) into training (1,658 examples), synthesis (2,764 examples), and test (1,106 examples) sets. Then, we use scikit-learn~\cite{pedregosa2011scikit} to train a random forest $f:\mathcal{X}\times\mathcal{Y}\to\mathbb{R}$ with 100 trees on the training set, where $f(x,y)\in\mathbb{R}$ is the probability assigned to label $y\in\mathcal{Y}$, and use $f$ in conjunction with the program shown in Figure~\ref{fig:expwarfarin}. This program includes two thresholds $c_{\text{low}}$ and $c_{\text{high}}$, and only assigns a low dose to a patient with covariates $x$ if $f(x)\ge1-c_{\text{low}}$, and similarly for a high dose---i.e., it only assigns the riskier outcomes when $f$ is sufficiently confident in its prediction. Importantly, the specification on $c_{\text{low}}$ refers not to the error rate on predictions for patients for whom $y=\text{low}$, but for whom $y=\text{high}$---i.e., we want to choose $c_{\text{low}}$ to ensure precision specifically on patients for whom $y=\text{high}$, and conversely for $c_{\text{high}}$. We use our synthesis algorithm to synthesize values of $c_{\text{low}}$ and $c_{\text{high}}$ that satisfy these specifications.

\paragraph{Correctness.}

Figure~\ref{fig:expwarfarin} shows the results of our approach (black) compared to directly predicting the highest probability label according to the random forest $f$ (red), always predicting ``medium'' (green), and the desired error rate (blue), as a function of the maximum error rate $\epsilon$, the maximum failure probability $\delta$, and the number of synthesis examples $n$. The top plots show the error rate, which is the maximum of the rate at which patients with $y=\text{low}$ are assigned a high dose, and the rate at which patients $y=\text{high}$ are assigned a low dose; this value should be below the blue line. The bottom plot shows the overall accuracy of the program---i.e., how often its predicted dose equals the ground truth dose. All values are estimated on the held-out test set. As before, $\epsilon$ has the largest impact on performance since it directly controls the error rate; however, once it hits $\epsilon=0.06$, performance levels off since its accuracy now equals that of $f$, and the program never assigns a dose not predicted by $f$. Performance is flat as a function of $\delta$. Finally, performance increases quickly as $n$ goes from $500$ to $1000$, but plateaus thereafter, again once accuracy equals that of $f$.

\paragraph{Runtime monitoring.}

In the case of Warfarin dosing, the doctor administers an initial dose to the patient (possibly the predicted dose, depending on the doctor's judgement), and gradually adjusts it based on the patient response. Thus, we eventually observe the ground truth dose that should have been recommended, which we can use to monitor our program. This process is achieved by the \texttt{monitor\_correctness} subroutine; here, \texttt{obtain\_result} returns the true dose eventually observed for a patient with covariates $x$. We evaluate whether our runtime monitoring can detect shifts in the data distribution that lead to a reduction in performance. We consider a shift in terms of the ethnicity of the patients, which has recently been identified as an important challenge in algorithmic healthcare~\cite{obermeyer2019dissecting}. In particular, we consider a model trained using non-Hispanic White patients (2,969 examples), which we refer to as the ``majority patients'', and test it on Black, Hispanic, and Asian patients (2,559 examples), which we refer to as the ``minority patients''.

First, we check if it proves correctness when the data distribution does not shift---i.e., we train the random forest, and synthesize and verify the program on majority patients. As expected, it successfully verifies correctness. Next, we check if it proves correctness when there is a shift---i.e., we train the random forest and synthesize the program on majority patients, but verify the program on minority patients. As expected, it rejects the program as incorrect.

Finally, recall that whether verification is successful depends on how many test examples are provided; thus, we also evaluate how many test examples are needed in this setting. To make sure we have enough examples, we use all examples in this case. Then, we find that for 2,000 test examples, our verification algorithm successfully proves correctness, but for 500, 1,000, or 1,500 test examples, it fails. Intuitively, the number of test examples needed to verify correctness needs to be more than the number use to synthesize the parameters, or else the synthesized thresholds will be more precise (i.e., closer to their ``optimal'' value) and the verification algorithm will not have enough data to validate them. In this case, we use 1,000 synthesis examples, so about $2\times$ as many test examples are needed to verify correctness.

\section{Related Work}

\paragraph{Synthesizing machine learning programs.}

There has been work on synthesizing
programs that include DNN components~\cite{gaunt2017differentiable,valkov2018houdini,ellis2018learning,young2019learning,shah2020learning} and on synthesizing probabilistic programs~\cite{nori2015efficient,saad2019bayesian}; however, they do not provide guarantees on the synthesized program.
There has been work on synthesizing control policies that satisfy provable guarantees~\cite{verma2018programmatically,bastani2018verifiable,zhu2019inductive,anderson2020neurosymbolic}; however, they focus on the setting where the learner can interact with the environment, and are not applicable to our supervised learning setting. Finally, there has been work on synthesizing programs with probabilistic constraints~\cite{drews2019efficient}, but requires that the search space of programs has finite VC dimension.

\paragraph{Verified machine learning.}

There has been recent interest in verifying machine learning programs---e.g., verifying robustness~\cite{bastani2016measuring,katz2017reluplex,huang2017safety,gehr2018ai2,anderson2019optimization}, fairness~\cite{albarghouthi2017fairsquare,bastani2019probabilistic}, and safety~\cite{katz2017reluplex,ivanov2019verisig}. More broadly, there has been work verifying systems such as approximate computing~\cite{rinard2006probabilistic,carbin2012proving,carbin2013verifying,misailovic2014chisel} and probabilistic programming~\cite{sankaranarayanan2013static,sampson2014expressing}. The most closely related work is~\cite{o2018scalable,dreossi2019verifai,fremont2019scenic,kim2020programmatic}, which verify semantic properties of machine learning models by sampling synthetic inputs from a user-specified space. In contrast, our focus is on synthesizing machine learning programs.

\paragraph{Statistical verification.}

There has been work leveraging statistical bounds to verify stochastic systems~\cite{younes2002probabilistic,sen2004statistical,sen2005statistical}, probabilistic programs~\cite{sankaranarayanan2013static,sampson2014expressing}, and machine learning programs~\cite{bastani2019probabilistic}. Our verification algorithm in Appendix~\ref{sec:verification} relies on bounds similar to the ones used in these approaches~\cite{younes2002probabilistic}. To the best of our knowledge, we are the first to focus on synthesis; in contrast to verification, our approach relies on bounds from learning theory to provide correctness guarantees.

\paragraph{Conformal prediction.}

There has been work on \emph{conformal prediction}~\cite{shafer2008tutorial,balasubramanian2014conformal,tibshirani2019conformal,romano2019conformalized}, including applications of these ideas to deep learning~\cite{park2020pac,angelopoulos2020uncertainty,kivaranovic2020adaptive,park2020pac2}, which aim to use statistical techniques to provide guarantees on the predictions of machine learning models. In particular, they provide confidence sets of outputs that contain the true label with high probability. Our techniques are inspired by these approaches, extending them to a general framework of synthesizing machine learning programs that satisfy provable guarantees.

\section{Conclusion}

We have proposed algorithms for synthesizing machine learning programs that come with PAC guarantees. Our technique leverages novel statistical learning bounds to achieve these guarantees. We have empirically demonstrated how our approach can be used to synthesize list processing programs that manipulate images using DNN components while satisfying PAC guarantees, as well as on two case studies in image classification and precision medicine. A key direction for future work is how to extend these ideas to settings where the underlying data distribution may shift, and to settings beyond supervised learning such as reinforcement learning.

\bibliographystyle{ACM-Reference-Format}
\bibliography{paper}

\clearpage
\appendix

\section{Statistical Verification}
\label{sec:verification}

We describe our algorithm for verifying a complete program. Our algorithm (Algorithm~\ref{alg:verification}) takes as input a complete program $\bar{P}$, training valuations $\vec{\alpha}=(\alpha_1,...,\alpha_n)$, where $\alpha_1,...,\alpha_n\sim p$ are i.i.d. samples, and a confidence level $\delta\in\mathbb{R}_{>0}$, and outputs a value $A(\bar{P},\vec{\alpha})\in\{0,1\}$ indicating whether $\bar{P}$ is approximately complete, which is correct with probability at least $1-\delta$ with respect to $p(\vec{\alpha})$.

Our algorithm is based on \emph{statistical verification}~\cite{younes2002probabilistic,sen2004statistical,sen2005statistical}. These algorithms leverage \emph{concentration inequalities} from probability theory to provide high-probability correctness guarantees. Concentration inequalities are theorems that provide rigorous bounds on the rate of convergence of statistical estimators. For instance, consider a Bernoulli distribution $p=\text{Bernoulli}(\mu)$ with unknown mean $\mu$. Given samples $z_1,...,z_n\sim p$, Hoeffding's inequality~\cite{hoeffding1994probability} says that the empirical mean $\hat{\mu}(\vec{z})=n^{-1}\sum_{i=1}^nz_i$ converges to $\mu$:
\begin{align}
\label{eqn:hoeffding}
\mathbb{P}_{p(\vec{z})}\big(|\hat{\mu}(\vec{z})-\mu|\le\epsilon\big)\ge1-\delta
~\text{where}~
\delta=2e^{-2n\epsilon^2},
\end{align}
i.e., $\hat{\mu}(\vec{z})$ is a good approximation of $\mu$ with high probability.

In our setting, given training valuation $\alpha$ and a specification $E=\phi(\bar{P}',c)~\{Q\}_{\epsilon}^\omega$ in $\bar{P}$, we let $z_{\alpha}=\llbracket E\rrbracket_{\alpha}$ and $z_{\alpha}^*=\llbracket E\rrbracket_{\alpha}^*$. Then, $\epsilon$-approximate soundness of $E$ is equivalent to $\mu=\mathbb{P}_{p(\alpha)}(z_{\alpha}\mid z_{\alpha}^*)\ge1-\epsilon$ if $\omega=\;\mid$, or $\mu=\mathbb{P}_{p(\alpha)}(z_{\alpha}^*\Rightarrow z_{\alpha})\ge1-\epsilon$ if $\omega=\;\Rightarrow$.
That is, $\epsilon$-approximate soundness is equivalent to $\mu\ge1-\epsilon$, where $\mu$ is the mean of a Bernoulli random variable $z_{\alpha}$ that is a function of a random variable $\alpha$ with distribution $p(\alpha\mid z_{\alpha}^*)$ (if $\omega=\;\mid$) or
the mean of a Bernoulli random variable $z_{\alpha}^*\Rightarrow z_{\alpha}$ that is a function of $\alpha$ with distribution $p(\alpha)$ (if $\omega=\;\Rightarrow$). However, $z_{\alpha}$ is potentially a complicated function of $\alpha$ and $p(\alpha)$ is unknown, so $\mu$ is hard to compute directly. Instead, given i.i.d. samples $\alpha_1,...,\alpha_n\sim p(\alpha)$, we can construct the samples $z_{\alpha_1},...,z_{\alpha_n}$ and $z_{\alpha_1}^*,...,z_{\alpha_n}^*$ and use them estimate $\mu$:
\begin{align*}
\hat{\mu}(\vec{\alpha})=\begin{cases}
\frac{\sum_{i=1}^nz_{\alpha_i}\wedge z_{\alpha_i}^*}{\sum_{i=1}^nz_{\alpha_i}^*}&\text{if}~\omega=\;\mid \\
\sum_{i=1}^nz_{\alpha_i}^*\Rightarrow z_{\alpha_i}&\text{if}~\omega=\;\Rightarrow.
\end{cases}
\end{align*}
Then, we can use (\ref{eqn:hoeffding}) to bound the error of $\hat{\mu}(\vec{\alpha})$---e.g., if $\hat{\mu}(\vec{\alpha})\ge1-\frac{\epsilon}{2}$ and $|\hat{\mu}(\vec{\alpha})-\mu|\le\frac{\epsilon}{2}$ with probability at least $1-\delta$, then $\mu\ge1-\epsilon$ with probability at least $1-\delta$. However, this approach is inefficient since Hoeffding's inequality is not tight for our setting. Instead, our verification algorithm (Appendix~\ref{sec:verificationalgo}) leverages a concentration inequality tailored to our setting (Appendix~\ref{sec:concentration}). Finally, we disucss how our approach can be used in the context of runtime monitoring (Appendix~\ref{sec:monitoring}).

\subsection{A Concentration Bound}
\label{sec:concentration}

\paragraph{Problem formulation.}

Consider a Bernoulli distribution $p=\text{Bernoulli}(\mu)$ with unknown mean $\mu\in[0,1]$. Given $\epsilon\in\mathbb{R}_{>0}$, our goal is to determine whether $\mu\ge1-\epsilon$. For instance, a sample $z\sim p$ may indicate a desired outcome (e.g., a correctly classified input), in which case $\mu$ is the correctness rate and $\epsilon$ is a desired bound on the error rate; then, our goal is to check whether the $\mu$ meets the desired error bound. More precisely, we want to compute $\psi\in\{0,1\}$ such that
\begin{align}
\label{eqn:epsverify}
\psi\Rightarrow(\mu\ge1-\epsilon).
\end{align}
That is, $\psi$ is a sound overapproximation of the property $\mu\ge1-\epsilon$ (i.e., $\psi=1$ implies $\mu\ge1-\epsilon$).

To compute such a $\psi$, we are given a training set of examples $\vec{z}=(z_1,...,z_n)\in\{0,1\}^n$, where $z_1,...,z_n\sim p$ are $n$ i.i.d. samples from $p$. An \emph{estimator} is a mapping $\hat{\psi}:\mathbb{R}^n\to\mathbb{R}$. We say such an estimator is \emph{approximately correct} if it satisfies the condition (\ref{eqn:epsverify})---i.e., $\hat{\psi}(\vec{z})\Rightarrow(\mu\ge1-\epsilon)$.

In general, we cannot guarantee $\hat{\psi}(\vec{z})$ is approximately correct due to the randomness in the training examples $\vec{z}$.\footnote{Note that $\hat{\psi}$ is a deterministic function; the randomness of $\hat{\psi}(\vec{z})$ is entirely due to the randomness in the training data $\vec{z}$.} Thus, we allow a probability $\delta\in\mathbb{R}_{>0}$ that $\hat{\psi}(\vec{z})$ is not approximately correct. 
\begin{definition}
\rm
Given $\epsilon,\delta\in\mathbb{R}_{>0}$, $\hat{\psi}$ is \emph{$(\epsilon,\delta)$-PAC} if $\mathbb{P}_{p(\vec{z})}\big(\hat{\psi}(\vec{z})\Rightarrow(\mu\ge1-\epsilon)\big)\ge1-\delta$.
\end{definition}
In other words, $\hat{\psi}(\vec{z})$ is approximately correct with probability at least $1-\delta$ according to the randomness in $p(\vec{z})$. Our goal is to construct an $(\epsilon,\delta)$-PAC estimator $\hat{\psi}(\vec{z})$.

\paragraph{Estimator.}

Given $\epsilon,\delta\in\mathbb{R}_{>0}$, consider the estimator
\begin{align}
\label{eqn:psihat}
\hat{\psi}(\vec{z})=\mathbbm{1}(L(\vec{z})\le k)
\quad\text{where}\quad
k=\max\left\{h\in\mathbb{N}\Biggm\vert\sum_{i=0}^h\binom{n}{i}\epsilon^i(1-\epsilon)^{n-i}\le\delta\right\}
\end{align}
and where $L(\vec{z})=\sum_{z\in\vec{z}}^n(1-z)$. Intuitively, $L(\vec{z})$ counts the number of errors, so we conclude the desired property holds as long as $L(\vec{z})$ is below a threshold $k$. This threshold is chosen so $\hat{\psi}$ is $(\epsilon,\delta)$-PAC---in particular, $\delta$ upper bounds the CDF of the binomial distribution evaluated at $k$.

To compute the solution $k$ in (\ref{eqn:psihat}), we start with $h=0$ and increment it until it no longer satisfies the condition. To ensure numerical stability, this computation is performed using logarithms. Note that $k$ does not exist if the set inside the maximum in (\ref{eqn:psihat}) is empty; in this case, we choose $\hat{\psi}(\vec{z})=0$, which trivially satisfies the PAC property. We have the following; see Appendix~\ref{sec:thmbinomproof} for a proof:
\begin{theorem}
\label{thm:binom}
The estimator $\hat{\psi}$ is $(\epsilon,\delta)$-PAC.
\end{theorem}

\begin{algorithm}[t]
\begin{algorithmic}
\Procedure{Verify}{$\bar{P},\vec{\alpha},\delta$}
\State $m\gets|\Phi(P)|$
\For{$\phi(\bar{P}',c)~\{Q\}_{\epsilon}^\omega\in\Phi(\bar{P})$}
\State Compute $\vec{z}_{\vec{\alpha}}$ according to (\ref{eqn:verification})
\State Compute $\hat{\psi}(\vec{z}_{\vec{\alpha}})$ according to (\ref{eqn:psihat}) with $(\epsilon,\delta/m)$
\If{$\neg\hat{\psi}(\vec{z}_{\vec{\alpha}})$}
\State\Return {\bf false}
\EndIf
\EndFor
\State\Return {\bf true}
\EndProcedure
\end{algorithmic}
\caption{Use statistical verification to check if $P$ is approximately correct.}
\label{alg:verification}
\end{algorithm}

\subsection{Verification Algorithm}
\label{sec:verificationalgo}

\paragraph{Problem formulation.}

A verification algorithm $A:\bar{\mathcal{P}}\times\mathcal{A}^n\to\{0,1\}$ takes as input a complete program $\bar{P}\in\bar{\mathcal{P}}$, and a set of test valuations $\vec{\alpha}=(\alpha_1,...,\alpha_n)\in\mathcal{A}^n$, where $\alpha_1,...,\alpha_n\sim p$ are i.i.d. samples from a distribution $p(\alpha)$. For example, $p(\alpha)$ may be the distribution of input images to an image classifier encountered while running in production, that have been manually labeled using crowdsourcing. Then, $A(\bar{P},\vec{\alpha})\in\{0,1\}$ should indicate whether $\bar{P}$ is approximately sound---i.e., whether every expression $\phi(\bar{P}',c)~\{Q\}_{\epsilon}^\omega\in\Phi(\bar{P})$ is approximately sound. We say that $A$ is \emph{sound} if $A(\bar{P},\vec{\alpha})\Rightarrow\bar{P}\in\bar{\mathcal{P}}^*$. As before, we cannot guarantee that $A$ is sound; instead, given $\delta\in\mathbb{R}_{>0}$, we want this property to hold with probability at least $1-\delta$ according to $p(\vec{\alpha})$.
\begin{definition}
\rm
A verification algorithm $A:\bar{\mathcal{P}}\times\mathcal{A}^n\to\{0,1\}$ is $\delta$-probably approximately sound if for all $\bar{P}\in\bar{\mathcal{P}}$,
$\mathbb{P}_{p(\vec{\alpha})}\big(A(\bar{P},\vec{\alpha})\Rightarrow\bar{P}\in\bar{\mathcal{P}}^*\big)\ge1-\delta$.
\end{definition}

\paragraph{Algorithm.}

Our verification algorithm is shown in Algorithm~\ref{alg:verification}. It check approximate correctness of $\bar{P}$ by checking approximate soundness of each $\phi(\bar{P}',c)~\{Q\}_{\epsilon}^\omega\in\Phi(\bar{P})$. It does so by allocating a $\delta/m$ probability of failure for each expression, where $m=|\Phi(\bar{P})|$ is the number of such expressions.

Next, we describe how our algorithm checks approximate soundness for a single expression $\phi(\bar{P}',c)~\{Q\}_{\epsilon}^\omega$. Given a single test valuation $\alpha\sim p$, consider the indicators
\begin{align*}
z_{\alpha}=\llbracket\phi(\bar{P}',c)~\{Q\}_{\epsilon}^\omega\rrbracket_{\alpha}
\qquad\text{and}\qquad
z_{\alpha}^*=\llbracket\phi(\bar{P}',c)~\{Q\}_{\epsilon}^\omega\rrbracket_{\alpha}^*.
\end{align*}
That is, $z_{\alpha}$ indicates whether $\phi(\bar{P}',c)$ holds, and $z_{\alpha}^*$ indicates whether $Q$ holds. Then, $\phi(\bar{P}',c)~\{Q\}$ is approximately sound if and only if
\begin{align*}
\mathbb{P}_{p(\alpha)}(z_{\alpha}\mid z_{\alpha}^*)\ge1-\epsilon&\quad\text{if}~\omega=\;\mid
\qquad\text{or}\qquad
\mathbb{P}_{p(\alpha)}(z_{\alpha}^*\Rightarrow z_{\alpha})\ge1-\epsilon\quad\text{if}~\omega=\;\Rightarrow.
\end{align*}
Next, note that $z_{\alpha}\in\{0,1\}$ is a Bernoulli random variable with mean $\mu=\mathbb{P}_{p(\alpha)}(z_{\alpha}\mid z_{\alpha}^*)$ (if $\omega=\;\mid$) or $\mu=\mathbb{P}_{p(\alpha)}(z_{\alpha}^*\Rightarrow z_{\alpha})$ (if $\omega=\;\Rightarrow$). Thus, given $\vec{\alpha}=(\alpha_1,...,\alpha_n)$, where $\alpha_1,...,\alpha_n\sim p$ are i.i.d. samples,
\begin{align}
\label{eqn:verification}
\vec{z}_{\vec{\alpha}}=\begin{cases}
\{z_{\alpha}\mid\alpha\in\vec{\alpha}\wedge z_{\alpha}^*\}&\text{if}~\omega=\mid \\
\{z_{\alpha}^*\Rightarrow z_{\alpha}\mid\alpha\in\vec{\alpha}\}&\text{if}~\omega=\Rightarrow
\end{cases}
\end{align}
is a vector of i.i.d. samples from $\text{Bernoulli}(\mu)$. Then, the estimator $\hat{\psi}(\vec{z}_{\vec{\alpha}})$ in (\ref{eqn:psihat}) with parameters $(\epsilon,\delta/m)$ indicates whether $\mu\ge1-\epsilon$ with high probability---i.e., if $\hat{\psi}(\vec{z}_{\vec{\alpha}})=1$, then $\mu\ge1-\epsilon$ holds with probability at least $1-\delta/m$ according to $p(\vec{\alpha})$. The following guarantee follows from Theorem~\ref{thm:binom} by a union bound over expressions in $\Phi(\bar{P})$:
\begin{theorem}
Algorithm~\ref{alg:verification} is $\delta$-probably approximately sound.
\end{theorem}

\subsection{Runtime Monitoring}
\label{sec:monitoring}

One challenge is that the specifications considered by our framework depend on the distribution of the data. As a consequence, if this distribution changes, then our correctness guarantees may no longer hold. This potential failure mode, called \emph{distribution shift}~\cite{ben2007analysis,quionero2009dataset}, is a major challenge for machine learning components. A key feature of our framework is that it can be used not only to sketch or verify the program before it is deployed, but also to continuously re-sketch the program based on feedback obtained in production to account for potential distribution shift. The primary requirement for using this approach is the need for feedback---i.e., continuing to collect labeled examples in production. In some settings, this kind of feedback is naturally available; otherwise, a solution is to manually label a small fraction of examples---e.g., using crowdsourcing~\cite{deng2009imagenet}.

Given ground truth labels for the input examples encountered in production, our verification algorithm can be straightforwardly adapted to the runtime setting. In particular, our system collects examples during execution; once it collects at least $N$ examples, it re-runs verification or sketching. It can do so after every subsequent example, or every $K$ examples. Finally, we may want to discard an examples after $T$ steps, both for computational efficiency and to account for the fact that the data distribution may be shifting over time so older examples are less representative. Here, $K,N,T\in\mathbb{N}$ are hyperparameters. Finally, we note that our statistical sketching algorithm can similarly be adapted to the runtime setting.

\section{Proofs}

\subsection{Proof of Theorem~\ref{thm:binom}}
\label{sec:thmbinomproof}

It suffices to show that if $\mu<1-\epsilon$, then $\mathbb{P}_{p(\vec{z})}(\hat{\psi}(\vec{z}))<\delta$. First, note that since $z_1,...,z_n$ are i.i.d. Bernoulli random variables with mean $\mu$, then $1-z_1,...,1-z_n$ are i.i.d. Bernoulli random variables with mean $\nu=1-\mu$. Their sum $L(\vec{z})$ is a binomial random variable---i.e., $L(\vec{z})\sim\text{Binomial}(n,\nu)$. Also, note that the condition $\mu<1-\epsilon$ is equivalent to $\nu>\epsilon$. Thus, we have
\begin{align*}
\mathbb{P}_{p(\vec{z})}(\hat{\psi}(\vec{z}))
&=\sum_{i=0}^{k}\binom{n}{i}\nu^i(1-\nu)^{n-i} \\
&<\sum_{i=0}^{k}\binom{n}{i}\epsilon^i(1-\epsilon)^{n-i} \\
&\le\delta,
\end{align*}
where the first inequality follows by standard properties of the CDF of the Binomial distribution. The claim follows. $\qed$

\subsection{Proof of Theorem~\ref{thm:pac}}
\label{sec:thmpacproof}

First, define
\begin{align*}
t_{\epsilon}^0=\inf_{t\in\mathbb{R}}\mathcal{T}_{\epsilon}.
\end{align*}
Intuitively, $t_{\epsilon}^0\in\mathbb{R}$ is the threshold that determines whether $t$ is $\epsilon$-approximately correct. In particular, it is clear that $t\in\mathcal{T}_{\epsilon}$ for all $t>t_{\epsilon}^0$ and $t\not\in\mathcal{T}_{\epsilon}$ for all $t<t_{\epsilon}^0$; in general, $t_{\epsilon}^0\in\mathcal{T}_{\epsilon}$ may or may not hold. Thus, it suffices to show
\begin{align*}
\mathbb{P}_{p(\vec{z})}(\hat{t}(\vec{z})\le t_{\epsilon}^0)<\delta.
\end{align*}
To this end, note that the constraint $L(t;\vec{z})\le k$ in (\ref{eqn:that}) implies
\begin{align*}
\sum_{z\in\vec{z}}\mathbbm{1}(z>\hat{t}(\vec{z})-\gamma(\vec{z}))\le k.
\end{align*}
Thus, on event $\hat{t}(\vec{z})\le t_{\epsilon}^0$, we have $\hat{t}(\vec{z})-\gamma(\vec{z})\le t_{\epsilon}^0-\gamma(\vec{z})$, so
\begin{align*}
k\le\sum_{z\in\vec{z}}\mathbbm{1}(z>\hat{t}(\vec{z})-\gamma(\vec{z}))\le\sum_{z\in\vec{z}}\mathbbm{1}(z>t_{\epsilon}^0-\gamma(\vec{z})).
\end{align*}
As a consequence, we have
\begin{align*}
\mathbb{P}_{p(\vec{z})}(\hat{t}(\vec{z})\le t_{\epsilon}^0)
\le\mathbb{P}_{p(\vec{z})}\left(\sum_{z\in\vec{z}}\mathbbm{1}(z> t_{\epsilon}^0-\gamma(\vec{z}))\ge k\right).
\end{align*}
Next, since $t_{\epsilon}^0-\gamma(\vec{z})<t_{\epsilon}^0$, we have $t_{\epsilon}^0-\gamma(\vec{z})\not\in\mathcal{T}_{\epsilon}$---i.e.,
\begin{align*}
\epsilon<\mathbb{P}_{p(z)}(z> t_{\epsilon}^0-\gamma(\vec{z}))=\mathbb{E}_{p(z)}(\mathbbm{1}(z> t_{\epsilon}^0-\gamma(\vec{z}))).
\end{align*}
In other words, the random variables $\mathbbm{1}(z> t_{\epsilon}^0-\gamma(\vec{z}))$ for $z\in\vec{z}$ are i.i.d. Bernoulli random variables with mean $\nu>\epsilon$. Thus, we have
\begin{align*}
\mathbb{P}_{p(\vec{z})}\left(\sum_{z\in\vec{z}}\mathbbm{1}(z> t_{\epsilon}^0-\gamma(\vec{z}))\ge k\right)
&=\sum_{i=0}^{k}\mathbb{P}_{p(\vec{z})}\left(\sum_{z\in\vec{z}}\mathbbm{1}(z>t_{\epsilon}^0+\gamma(\vec{z}))=i\right) \\
&=\sum_{i=0}^{k}\text{Binomial}(i;n,\nu) \\
&<\sum_{i=0}^{k}\text{Binomial}(i;n,\epsilon) \\
&=\sum_{i=0}^{k}\binom{n}{i}\epsilon^i(1-\epsilon)^{n-i} \\
&\le\delta,
\end{align*}
where the first inequality follows by standard properties of the CDF of the Binomial distribution. The claim follows. $\qed$

\subsection{Proof of Theorem~\ref{thm:binommean}}
\label{sec:binommeanproof}

First, we have the following classical inequality~\cite{hoeffding1994probability}:
\begin{theorem}
\label{thm:hoeffding}
(Hoeffding's inequality) We have
\begin{align*}
\mathbb{P}_{p(\vec{z})}\Big(\mu-\hat{\mu}(\vec{z})\ge t\Big)\le e^{-2nt^2}.
\end{align*}
\end{theorem}
Now, letting $t=\sqrt{\frac{\log(1/\delta)}{2n}}$, we have
\begin{align*}
\mathbb{P}_{p(\vec{z})}\Big(\mu\ge\hat{\nu}(\vec{z})\Big)
\le\mathbb{P}_{p(\vec{z})}\Big(\mu-\hat{\mu}(\vec{z})\ge t\Big)
\le e^{-2nt^2}
\le\delta,
\end{align*}
where the second-to-last inequality follows from Theorem~\ref{thm:hoeffding}. The claim follows. $\qed$

\section{Additional Case Study: Object Detection}
\label{sec:appendixdetection}

\begin{figure}
\scriptsize
$\begin{array}{l}
\texttt{def detect\_ppl(x):} \\
\qquad\texttt{y\_hat = f(x)} \\
\qquad\texttt{return [d.box for d in y\_hat if check\_det(x, d)]} \vspace{5pt} \\
\texttt{def check\_det(x, d, d\_true=None)} \\
\qquad\texttt{return \blue{d.score > ??1} \green{\{IOU(d.box, d\_true) >= 0.5\} [|, 0.05]}} \vspace{5pt} \\
\texttt{def detect\_ppl\_fast(x, y\_true=None):} \\
\qquad\texttt{y\_hat = f\_fast(x)} \\
\qquad\texttt{no\_ppl\_score = 1.0 - max([d.score for d in y\_hat])} \\
\qquad\texttt{if \blue{no\_ppl\_score > ??2} \green{\{len(detect\_ppl(x)) != 0\} [|, 0.05]}:} \\
\qquad\qquad\texttt{return []} \\
\qquad\texttt{else:} \\
\qquad\qquad\texttt{return detect\_ppl(x)} \vspace{5pt} \\
\texttt{def monitor\_correctness(x):} \\
\qquad\texttt{if np.random.uniform() <= 0.99:} \\
\qquad\qquad\texttt{return} \\
\qquad\texttt{y\_hat = f\_fast(x)} \\
\qquad\texttt{no\_ppl\_score = 1.0 - max([d.score for d in y\_hat])} \\
\qquad\texttt{passert \blue{no\_ppl\_score > ??1} \green{\{ len(detect\_ppl(x)) != 0\} [??3]}} \\
\texttt{def monitor\_speed(x):} \\
\qquad\texttt{y\_hat = f\_fast(x)} \\
\qquad\texttt{no\_ppl\_score = 1.0 - max([d.score for d in y\_hat])} \\
\qquad\texttt{passert \blue{no\_ppl\_score > ??1} \green{\{true\} [??3]}}
\end{array}$
\caption{A program used to detect people in a given image $x$. Specifications are shown in green; curly brackets is the specification and square brackets is the value of $\epsilon$. The corresponding inequality with a hole in blue. Holes with the same number are filled with the same value.}
\label{fig:objdet}
\end{figure}

\begin{figure*}
\centering
\begin{tabular}{ccc}
\includegraphics[width=0.3\textwidth]{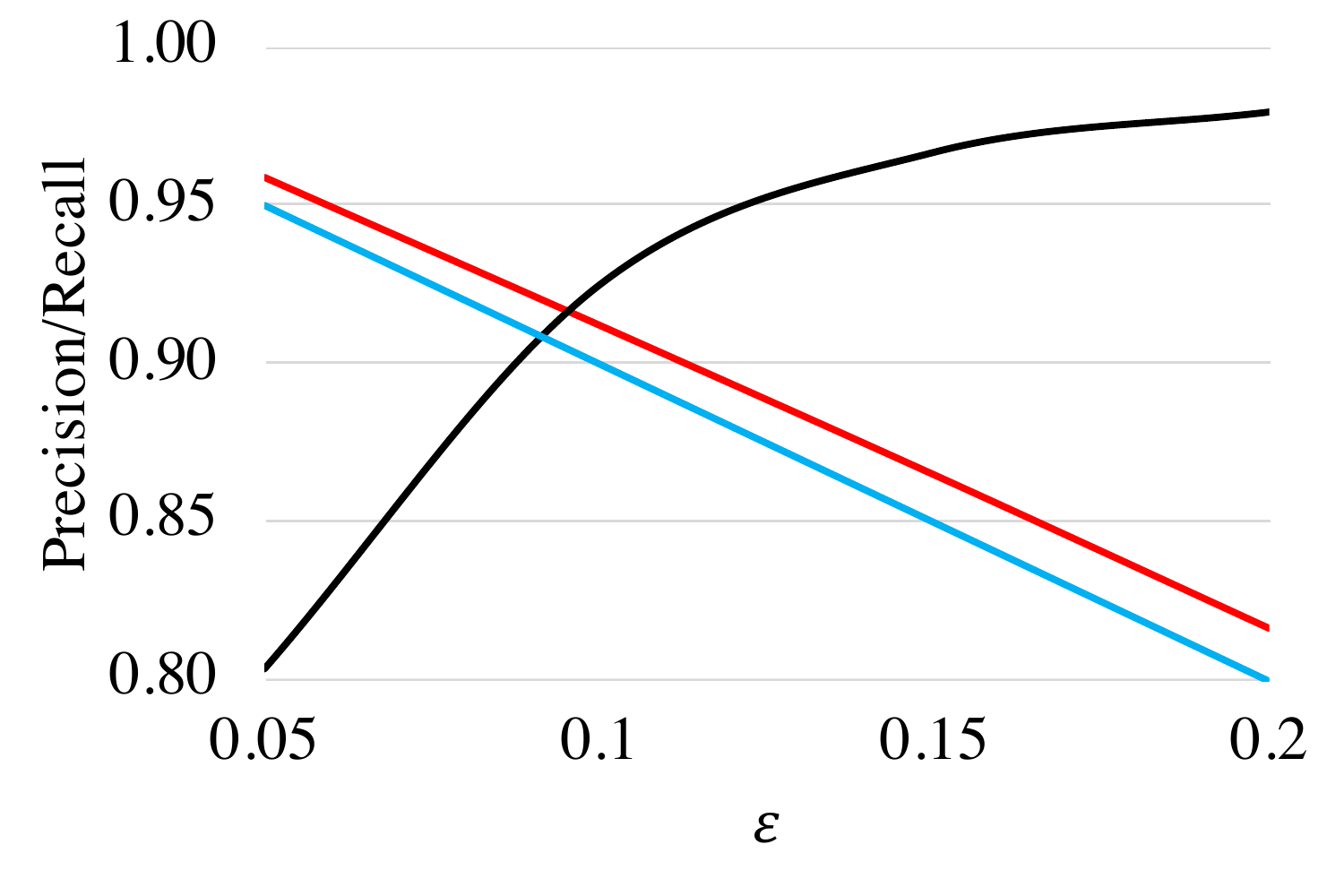} &
\includegraphics[width=0.3\textwidth]{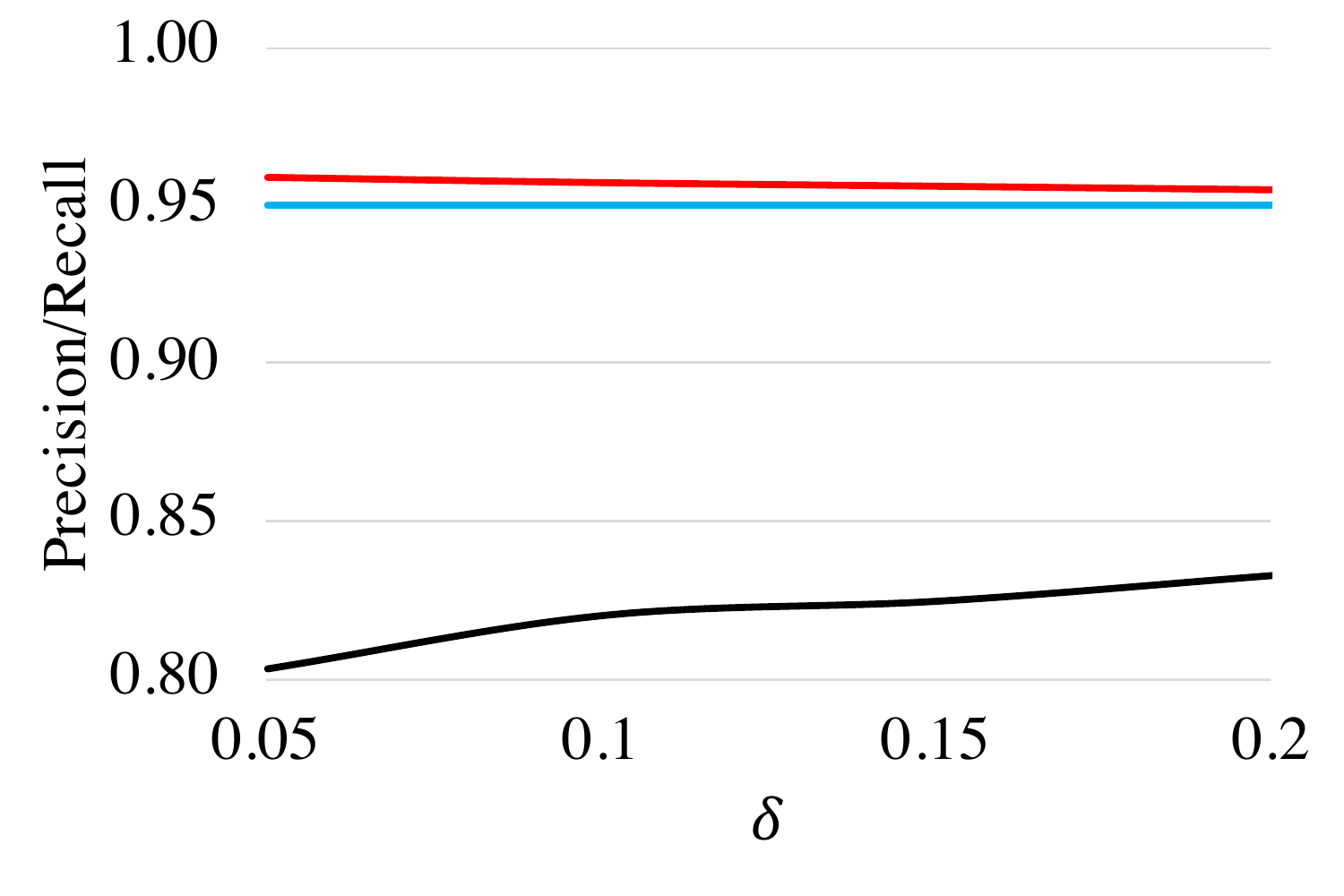} &
\includegraphics[width=0.3\textwidth]{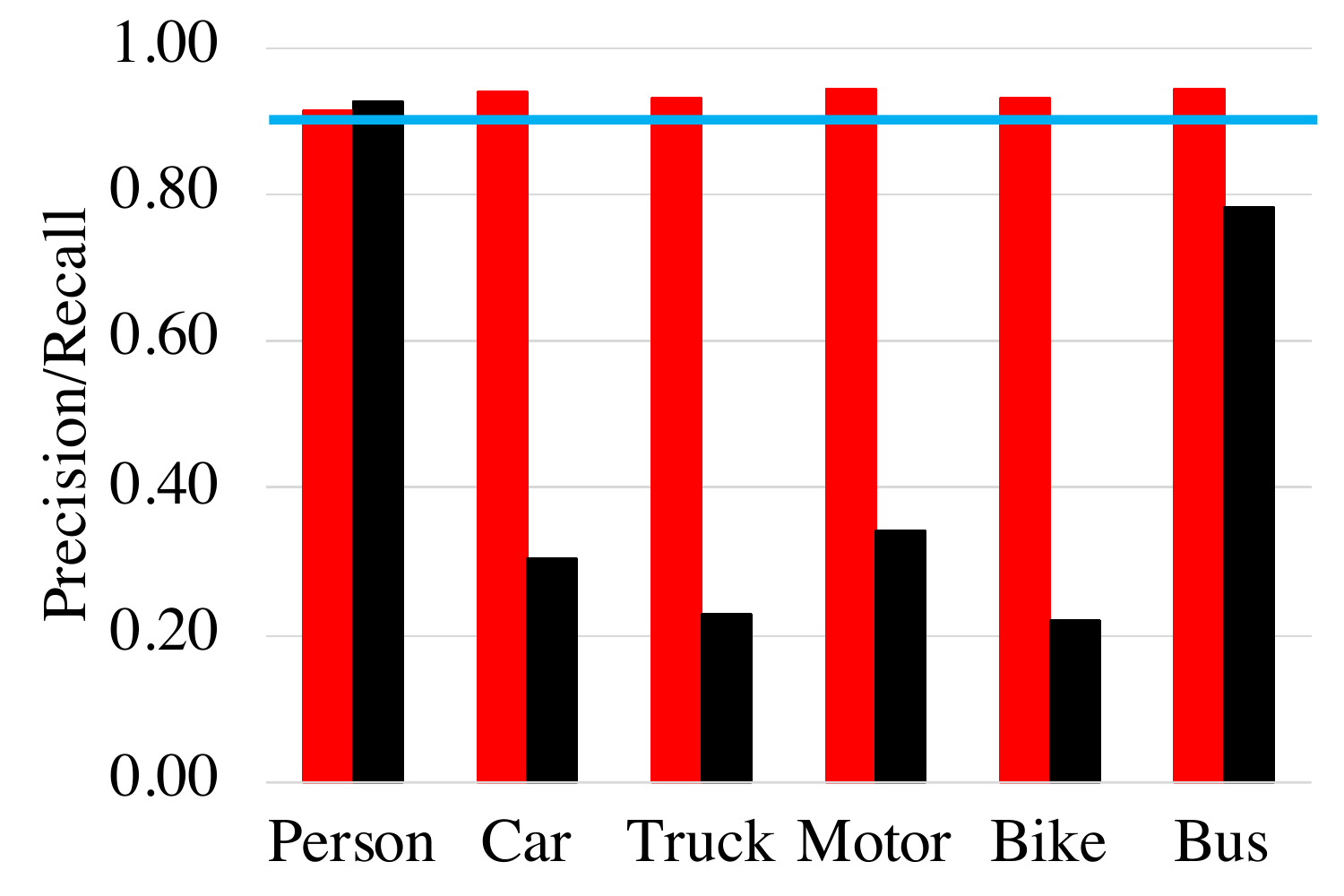} \\
(a) & (b) & (c) \\
\includegraphics[width=0.3\textwidth]{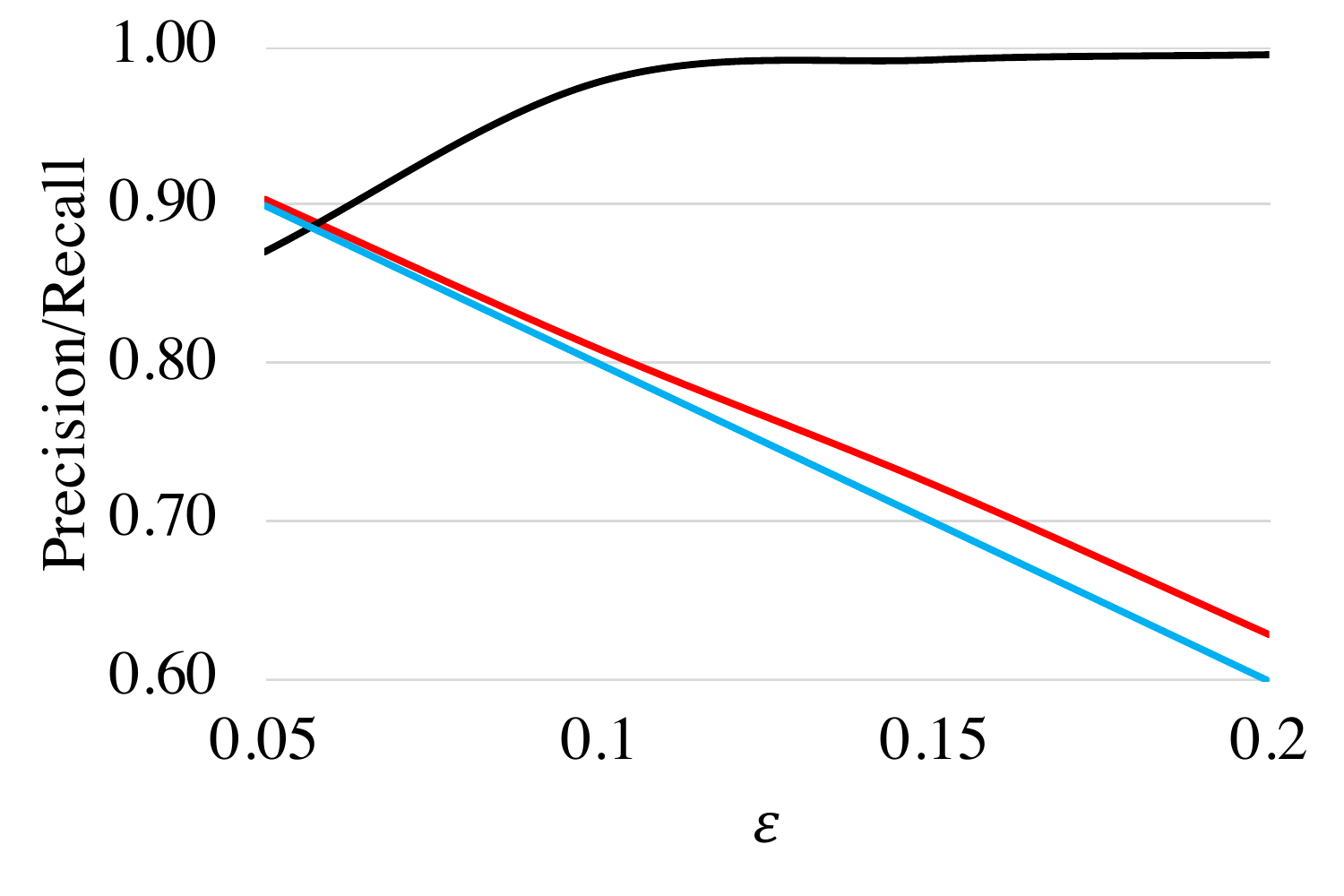} &
\includegraphics[width=0.3\textwidth]{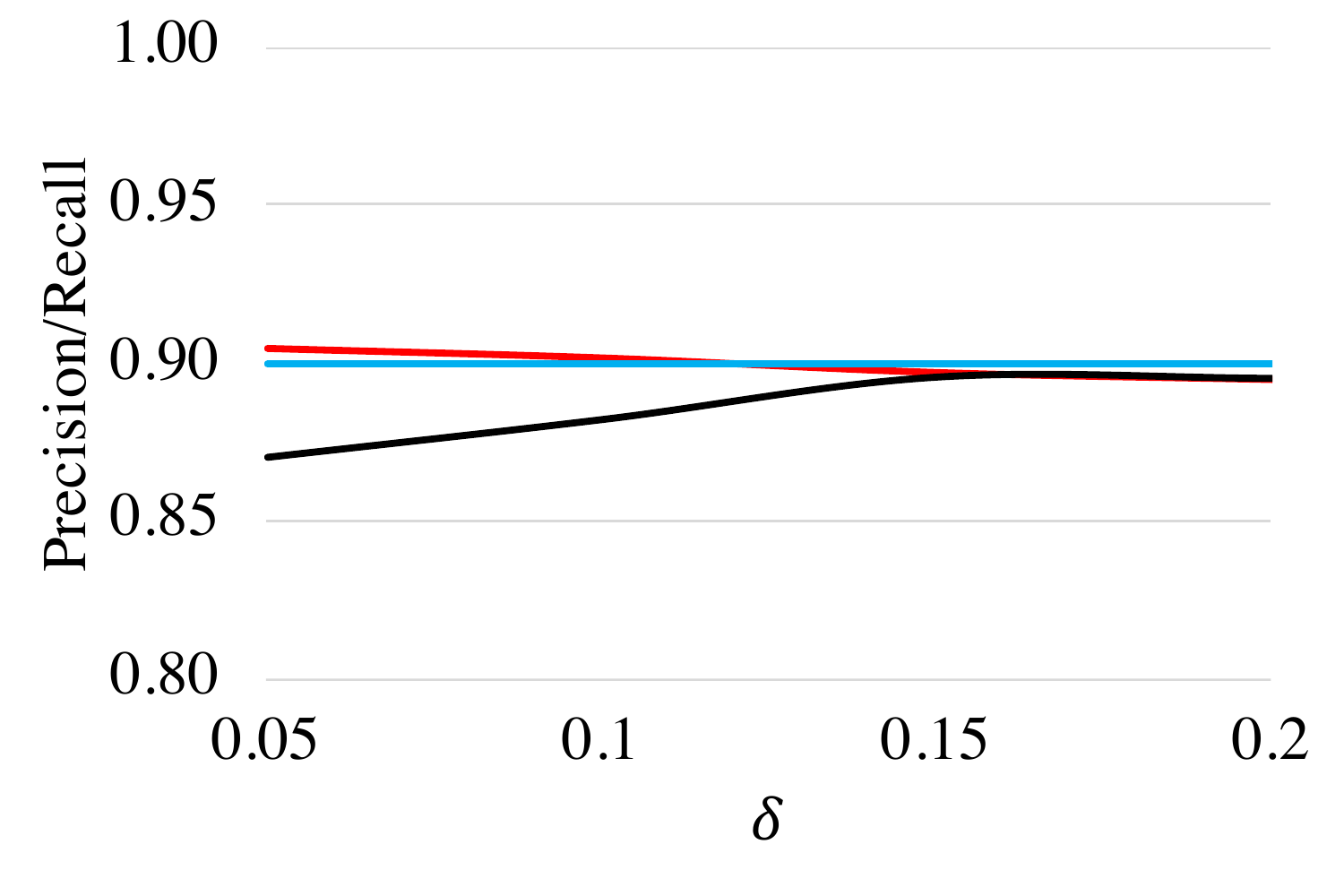} &
\includegraphics[width=0.3\textwidth]{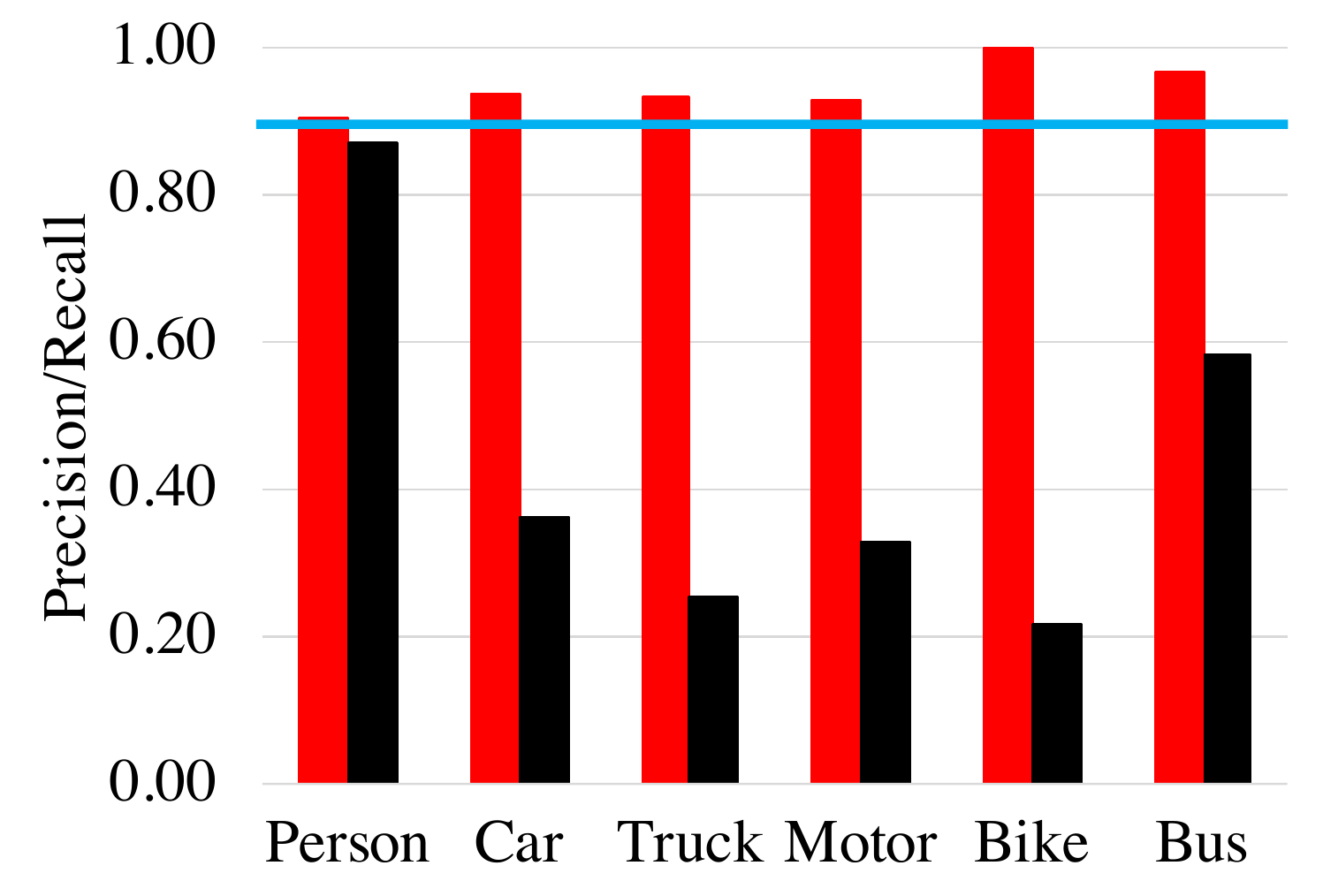} \\
(d) & (e) & (f) \\
\includegraphics[width=0.3\textwidth]{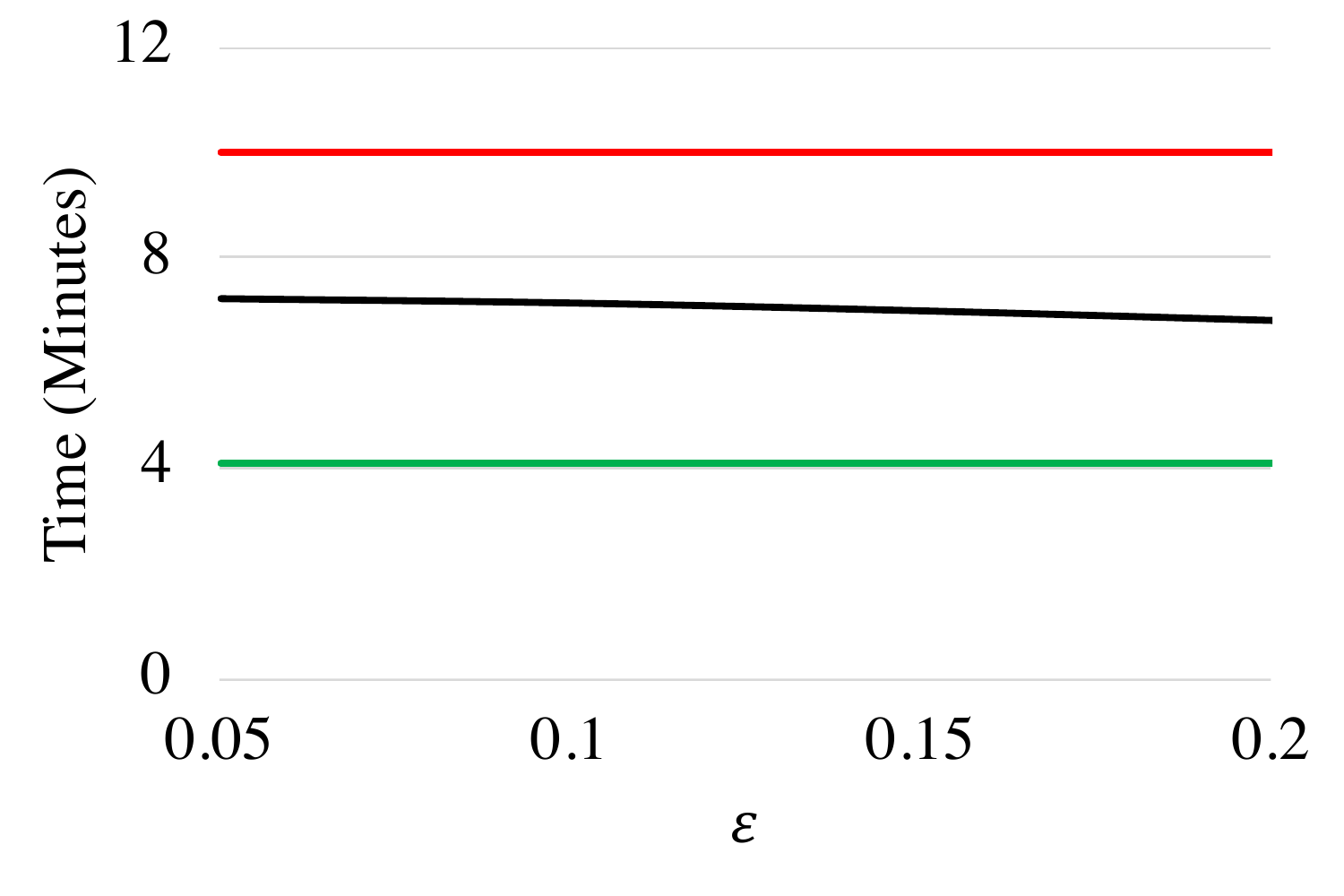} &
\includegraphics[width=0.3\textwidth]{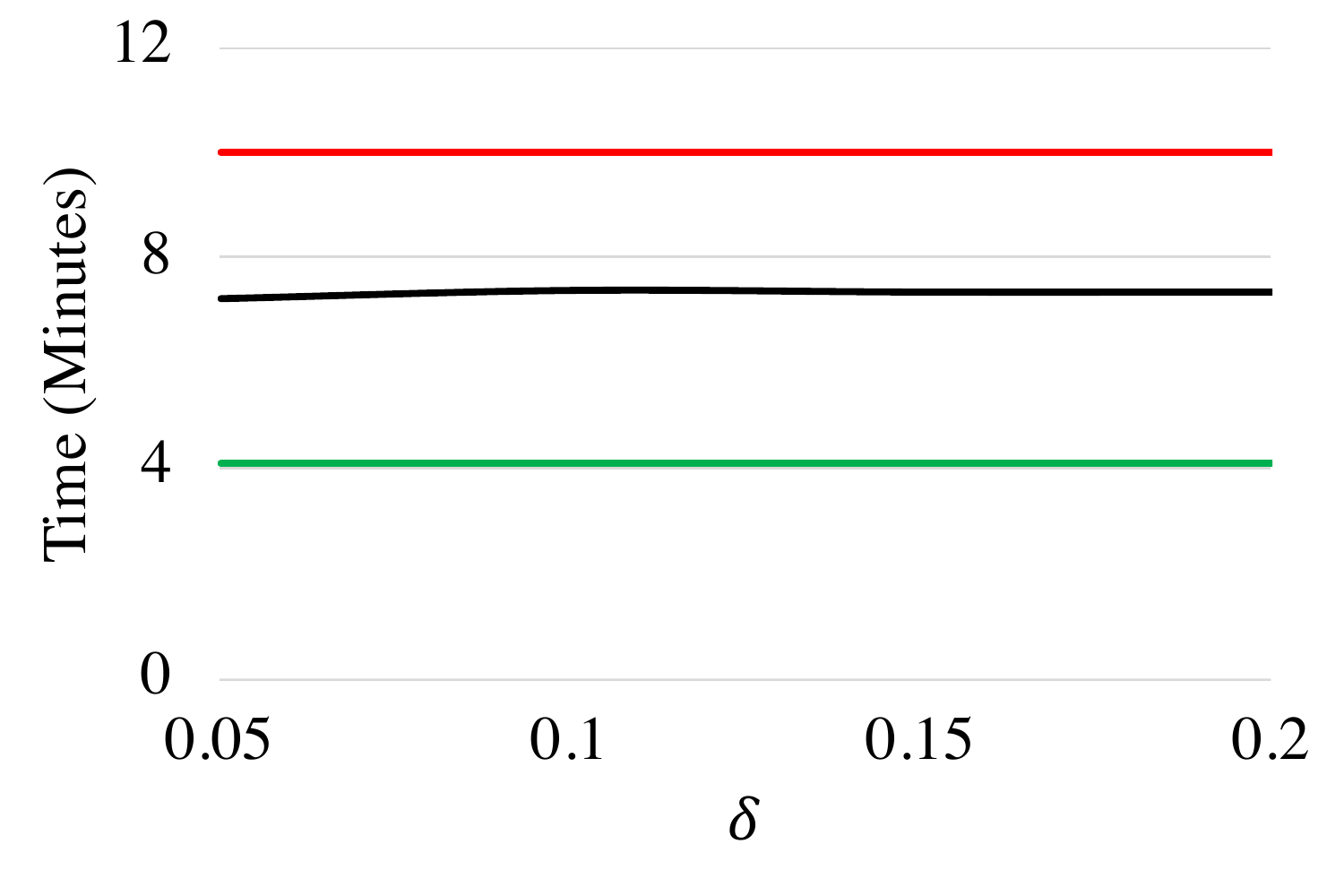} &
\includegraphics[width=0.3\textwidth]{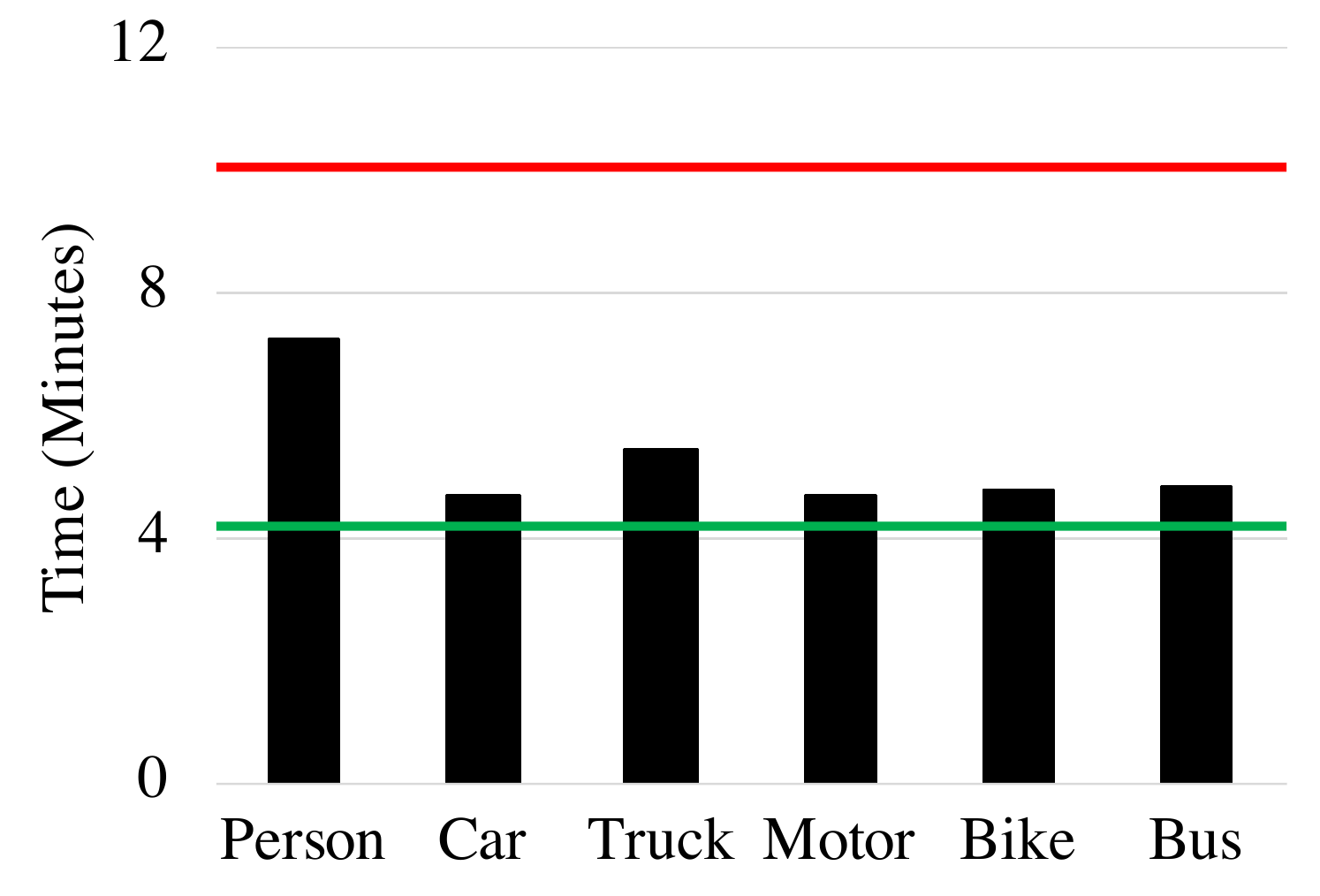} \\
(g) & (h) & (i)
\end{tabular}
\caption{For the slow model alone (top) and slow$+$fast model (middle), we show recall (red), the desired lower bound on recall (blue), and precision (black) as a function of (a,d) $\epsilon$, (b,e) $\delta$, and (c,f) the object category $y$. For slow$+$fast (black), slow alone (red), and fast alone (green), we show running time as a function of (g) $\epsilon$, (h) $\delta$, and (i) the object category $y$.}
\label{fig:expdetection}
\end{figure*}

\paragraph{Object detection.}

We assume given a DNN component $f$ that given an image $x$, is designed to detect people in $x$. Our formulation of object detection in this section is slightly different than the previous setup. In particular, $d\in f(x)$ is a list of \emph{detections}, which is a pair $d=(b,p)$ including a \emph{bounding box} $b\in\mathbb{R}^4$ that encodes the center, width, and height of a rectangular region of $x$, and a value $p\in[0,1]$ that is the predicted probability that $b$ exists. In addition, the ground truth label $y^*$ for an image $x$ is a list of bounding boxes $b\in y^*$. In general, we cannot expect to get a perfect match between the predicted bounding boxes and the ground truth ones. Typically, two bounding boxes $b,'$ \emph{match} if have significant overlap---in particular, their intersection-over-union satisfies $\text{IOU}(b,b')\ge\rho$ for some threshold $\rho\in[0,1]$; we use a standard choice of $\rho=0.5$. We denote that $b$ and $b'$ match in this sense by $b\cong b'$. Finally, $b$ approximately matches a bounding box in $y^*$ it $b\cong b'$ for some $b'\in y^*$, which we denote by $b\approxin y^*$.

\paragraph{Experimental setup.}

We use a pretrained state-of-the-art object detector called Faster R-CNN~\cite{ren2016faster} available in PyTorch~\cite{paszke2017automatic}, tailored to the COCO dataset~\cite{lin2014microsoft}. There are multiple variants of Faster R-CNN; we use the most accurate one, termed X101-FPN with $3\times$ learning rate schedule. For each predicted bounding box, this model additionally outputs a predicted object category (e.g., ``person''), as well as the size of the bounding box (``small'', ``medium'', and ``large''). For most of our evaluation, we use ``person'' and ``large''. We specify alternative choices when we used them; in particular, we additionally consider 6 of the 91 object categories: ``person'' (10777 bounding boxes), ``car'' (1918 bounding boxes), ``truck'' (414 bounding boxes), ``motorcycle'' (367 bounding boxes), ``bike'' (314 bounding boxes), and ``bus'' (283 bounding boxes). We split the COCO validation set into 2000 synthesis images and 3000 test images.

\paragraph{Correctness.}

Our goal is to detect a majority of people. In particular, we consider synthesizing a threshold $c$ and selecting all bounding boxes with probability above $c$---i.e.,
\begin{align*}
f(x,c)=\{b\mid (b,p)\in f(x)\wedge p\ge c\}.
\end{align*}
This task is more challenging to specify than our examples so far since $f(x)$ is a structured output. In particular, we are not reasoning about whether $f(x)$ is correct with high probability with respect to $p(x,y^*)$, but whether bounding boxes $(b,p)\in f(x)$ are correct. Thus, we need a distribution $p(b\mid x)$ over bounding boxes $b$ in an image $x$. Given such a distribution, our goal is to choose $c$ so
\begin{align}
\label{eqn:person}
\mathbb{P}_{p(x,y^*),p(b\mid x)}\left(b\approxin f(x,c)\mid b\approxin y^*\right)\ge1-\epsilon,
\end{align}
where $p(x,y^*)$ is the data distribution. Intuitively, this property says that $f(x,c)$ contains at least a $1-\epsilon$ fraction of ground truth bounding boxes. A reasonable choice for $p(b\mid x)$ is the uniform distribution over $f(x,0)$---i.e., the set of all bounding boxes predicted by $f$. One issue is when a ground truth bounding box $b\in y^*$ is completely missing from $f(x,0)$; in this case, $b$ would not occur in $p(b\mid x)$, so (\ref{eqn:person}) would not count it as an error even though it is missing from $f(x,c)$ for any $c$. To handle this case, we simply add $(b,0)$ to $f(x)$ during synthesis for such bounding boxes $b$---i.e., $f$ predicts $b$ occurs with probability zero.

The program for achieving this goal is shown in the subroutine \texttt{detect\_ppl} in Figure~\ref{fig:objdet}. We use our algorithm in conjunction with the synthesis examples to synthesize the parameter \texttt{??1} for this program, using the default values $\epsilon=\delta=0.05$ and the object category ``person''. In Figure~\ref{fig:expdetection}, we show the recall (red), desired lower bound on recall (blue), and precision (black) as a function of (a) $\epsilon$, (b) $\delta$, and (c) the object category $y$. The trends are largely similar to before---e.g., performance varies significantly with $\epsilon$ and the object category, but not very much with $\delta$. For (c), we use $\epsilon=0.1$ to facilitate comparison to our fast program described below.

\paragraph{Improving speed.}

We use a similar approach to improve speed as before---i.e., given a fast object detector $f_{\text{fast}}$, we want to use it to check the image, and only send it to the slow object detector $f$ if necessary. A challenge compared to image classification is that the object detection model does not operate at the level of bounding boxes, which is the level at which we defined correctness, but at the level of images. Thus, we cannot decide whether we want to run the slow model independently for each detection $d\in f_{\text{fast}}(x)$; instead, we have to make such a decision for an image $x$ as a whole. Intuitively, we check whether the fast model returns \emph{any} detections in the given image $x$. To this end, we compute the maximum score $p$ across all detections $(b,p)\in f_{\text{fast}}(x)$---i.e.,
\begin{align*}
\tilde{f}_{\text{fast}}(x)=\max_{(b,p)\in f_{\text{fast}}(x)}p.
\end{align*}
Then, we want to guarantee that $y^*=\varnothing$ if this score is below some threshold that ensures that $y^*=\varnothing$; this property is equivalent to its contrapositive
\begin{align}
\label{eqn:objdetfast}
(y^*\neq\varnothing)\Rightarrow(1-\tilde{f}_{\text{fast}}(x)\le c),
\end{align}
where the right-hand side of the implication is equivalent to $f_{\text{fast}}(x)\ge1-c$---i.e., the score is above the threshold $1-c$. As before, we cannot ensure this property holds with probability one, so instead we use the high-probability variant
\begin{align*}
\mathbb{P}_{p(x,y^*)}(1-\tilde{f}_{\text{fast}}(x)\le c\mid y^*\neq\varnothing)\ge1-\epsilon.
\end{align*}
This approach is shown in the \texttt{detect\_ppl\_fast} subroutine in Figure~\ref{fig:objdet}. We note that this approach does not provide guarantees as strong as the ones for image classification---in particular, there is a chance that the false negative images $x$ of $f_{\text{fast}}$ (i.e., $x$ does not satisfy (\ref{eqn:objdetfast})) will contain larger numbers of ground truth bounding boxes compared to true positive images. Then, the recall at the level of bounding boxes may be less than $1-2\epsilon$. However, we find that it works well in practice; intuitively, $f_{\text{fast}}$ is more likely to have false negative images that contain \emph{fewer} ground truth bounding boxes.

For $f_{\text{fast}}$, we use a variant of Faster R-CNN termed R50-FPN with $3\times$ learning rate schedule, which is the fastest variant available. Then, we synthesize the parameters of \texttt{??1} and \texttt{??2} in Figure~\ref{fig:objdet} using the synthesis examples. As before, all results are run on an Nvidia GeForce RTX 2080 Ti GPU. In Figure~\ref{fig:expdetection}, we show the recall (red), desired lower bound on recall (blue), and precision (black) of our approach as a function of (d) $\epsilon$, (e) $\delta$, and (f) the object category $y$. Similarly, we show the running time (on the entire test set) of the combined program slow$+$fast (black), fast alone (green), and slow alone (red). As can be seen, our approach reduces running time by more than $2\times$ except in the case of ``person'' (28\% reduction) and ``truck'' (45\% reduction). The person speedup is relatively small because so many of the images in the COCO dataset contain people. Compared to the image classification setting, we obtain a smaller speedup since the gap between the fast and slow models is not as large, and also because we can only avoid using the slow model for images that contain zero detections. Furthermore, comparing Figure~\ref{fig:expdetection} (c) and (f) (i.e., slow alone vs. slow$+$fast, respectively), for categories ``car'' and ``truck'', we suffer no loss in precision, though we suffer a small loss in precision for the others.

Finally, we note that in Figure~\ref{fig:expdetection} (e), for $\delta=0.15$ and $\delta=0.2$, the estimated recall falls slightly below the desired lower bound on recall. This result is most likely due to random chance, either because of randomness in the synthesis set or because these values are estimates based on a random test set. In particular, $0.15$ is a fairly high failure probability (note that the results across $\delta$ are correlated, since we are using the same synthesis and test sets across all $\delta$).

\paragraph{Runtime monitoring.}

We use runtime monitors to check that our program meets the desired bounds both in terms of error rate (the subroutine \texttt{monitor\_correctness} in Figure~\ref{fig:objdet}) and running time (the subroutine \texttt{monitor\_speed} in Figure~\ref{fig:objdet}). These approaches are the same as for image classification---the correctness monitor checks that the error rate (i.e., $f_{\text{fast}}(x)$ concludes there are no detections but $f(x)\neq\varnothing$) is below the desired rate $\epsilon$, and the running time monitor checks that $f$ is not called too often (i.e., $f_{\text{fast}}(x)$ concludes there are no detections sufficiently frequently).

To evaluate these monitors, we consider a shift from the default ``large'' bounding boxes we use to ``small'' and ``medium''. Intuitively, the smaller bounding boxes correspond to objects farther in the background, which are harder to detect but also tend to be less important (e.g., an autonomous car may not care as much about detecting far-away pedestrians). The trends are as before. First, we find that the monitors correctly prove correctness when there is no shift. Second, we find that the running time does not increase due to the shift, so the running time monitor continues to prove correctness. Finally, our correctness monitor rejects correctness for the shift to ``small'' bounding boxes; interestingly, it proves correctness for ``medium'' bounding boxes, which suggests that our synthesized program generalizes to this case.

\end{document}